\documentclass[twocolumn]{aastex63}

\usepackage{graphicx}	
\usepackage{amsmath}	
\usepackage{amssymb}	
\usepackage{enumitem}
\usepackage{hyperref}

\defcitealias{lehner13}{L13}
\defcitealias{rafelski12}{R12-DLA}
\defcitealias{lehner18}{CCC\,I}
\defcitealias{wotta19}{CCC\,II}
\defcitealias{lehner19}{CCC\,III}
\defcitealias{wotta16}{W16}
\defcitealias{lehner16}{KODIAQ-Z\,I}
\defcitealias{fumagalli16}{HD-LLS}
\defcitealias{haardt96}{HM05}
\defcitealias{haardt12}{HM12}

\newcommand{\ssz}{202}

\newcommand{\logU}{\ensuremath{\log U}}
\def\nodata{ ~$\cdots$~ }

\newcommand{\zabs}{$z_{\rm abs}$}
\newcommand{\mzabs}{z_{\rm abs}}

\newcommand{\zem}{$z_{\rm em}$}
\newcommand{\mzem}{z_{\rm em}}

\newcommand{\cmm}{cm$^{-2}$}
\newcommand{\cmmm}{cm$^{-3}$}

\newcommand{\lya}{Ly$\alpha$}

\newcommand{\lyb}{Ly$\beta$}
\newcommand{\lyg}{Ly$\gamma$}
\newcommand{\lyd}{Ly$\delta$}
\newcommand{\nhi}{$N_{\rm H\,I}$}
\newcommand{\nh}{$N_{\rm H}$}
\newcommand{\nnh}{$n_{\rm H}$}

\newcommand{\mnnh}{n_{\rm H}}
\newcommand{\mnh}{N_{\rm H}}
\newcommand{\mnhi}{N_{\rm H\,I}}
\newcommand{\mnhii}{N_{\rm H\,II}}
\newcommand{\lnhi}{$\log N_{\rm H\,I}$}

\newcommand{\mlnhi}{\log N_{\rm H\,I}}
\newcommand{\mlnh}{\log N_{\rm H}}
\newcommand{\mlnnh}{\log n_{\rm H}}

\newcommand{\km}{${\rm km\,s}^{-1}$}
\newcommand{\kms}{${\rm km\,s}^{-1}$}

\newcommand{\xh}{\ensuremath{{\rm [X/H]}}}
\newcommand{\ah}{\ensuremath{{\rm [\alpha/H]}}}
\newcommand{\ca}{\ensuremath{{\rm [C/\alpha]}}}

\newcommand{\hi}{\ensuremath{\mbox{\ion{H}{1}}}}
\newcommand{\hii}{\ensuremath{\mbox{\ion{H}{2}}}}

\newcommand{\alii}{\ensuremath{\mbox{\ion{Al}{2}}}}
\newcommand{\aliii}{\ensuremath{\mbox{\ion{Al}{3}}}}

\newcommand{\cii}{\ensuremath{\mbox{\ion{C}{2}}}}
\newcommand{\ciii}{\ensuremath{\mbox{\ion{C}{3}}}}
\newcommand{\civ}{\ensuremath{\mbox{\ion{C}{4}}}}

\newcommand{\oi}{\ensuremath{\mbox{\ion{O}{1}}}}

\newcommand{\ovi}{\ensuremath{\mbox{\ion{O}{6}}}}

\newcommand{\siii}{\ensuremath{\mbox{\ion{Si}{2}}}}
\newcommand{\siiii}{\ensuremath{\mbox{\ion{Si}{3}}}}
\newcommand{\siiv}{\ensuremath{\mbox{\ion{Si}{4}}}}

\newcommand{\feii}{\ensuremath{\mbox{\ion{Fe}{2}}}}
\newcommand{\feiii}{\ensuremath{\mbox{\ion{Fe}{3}}}}

\shortauthors{Lehner et al.}
\shorttitle{KODIAQ-Z: Metals and Baryons in the cool CGM/IGM}

\begin{document}

\title{KODIAQ-Z: Metals and Baryons in the Cool Intergalactic and Circumgalactic Gas at $2.2\la \lowercase{z} \la 3.6$}


\author[0000-0001-9158-0829]{Nicolas Lehner}
\affiliation{Department of Physics, University of Notre Dame, Notre Dame, IN 46556}

\author[0000-0001-5158-1966]{Claire Kopenhafer}
\affiliation{ Department of Physics and Astronomy, Department of Computational Mathematics, Science, and Engineering, Michigan State University, East Lansing, MI 48824}

\author[0000-0002-7893-1054]{John M. O'Meara}
\affiliation{W.M. Keck Observatory 65-1120 Mamalahoa Highway Kamuela, HI 96743}

\author[0000-0002-2591-3792]{J. Christopher Howk}
\affiliation{Department of Physics, University of Notre Dame, Notre Dame, IN 46556}

\author[0000-0001-6676-3842]{Michele Fumagalli}
\affiliation{Dipartimento di Fisica G. Occhialini, Universit\`a degli Studi di Milano Bicocca, Piazza della Scienza 3, 20126 Milano, Italy}
\affiliation{INAF - Osservatorio Astronomico di Trieste, via G. B. Tiepolo 11, 34143 Trieste, Italy}

\author[0000-0002-7738-6875]{J. Xavier Prochaska}
\affiliation{UCO/Lick Observatory, Department of Astronomy \& Astrophysics, University of California Santa Cruz, 1156 High Street, Santa Cruz, CA 95064}
\affiliation{Kavli Institute for the Physics and Mathematics of the Universe (WIP), 5-1-5 Kashiwanoha, Kashiwa, 277-8583, Japan}

\author[0000-0003-4804-7142]{Ayan Acharyya}
\affiliation{Space Telescope Science Institute, 3700 San Martin Drive, Baltimore, MD 21218}
\affiliation{Department of Physics \& Astronomy, Johns Hopkins University, 3400 N.\ Charles Street, Baltimore, MD 21218}
\affiliation{Research School of Astronomy and Astrophysics, Australian National University, Weston Creek, ACT 2611, Australia}

\author[0000-0002-2786-0348]{Brian W.\ O'Shea}
\affiliation{Department of Computational Mathematics, Science, and Engineering, Department of Physics and Astronomy, National Superconducting Cyclotron Laboratory,  Michigan State University, East Lansing, MI 48824}

\author[0000-0003-1455-8788]{Molly S.\ Peeples}
\affiliation{Space Telescope Science Institute, 3700 San Martin Drive, Baltimore, MD 21218}
\affiliation{Department of Physics \& Astronomy, Johns Hopkins University, 3400 N.\ Charles Street, Baltimore, MD 21218}

\author[0000-0002-7982-412X]{Jason Tumlinson}
\affiliation{Space Telescope Science Institute, 3700 San Martin Drive, Baltimore, MD 21218}
\affiliation{Department of Physics \& Astronomy, Johns Hopkins University, 3400 N.\ Charles Street, Baltimore, MD 21218}

\author[0000-0002-3817-8133]{Cameron B.\ Hummels}
\affiliation{Department of Astronomy, California Institute of Technology, Pasadena, CA 91125}

\begin{abstract}
We present the KODIAQ-Z survey aimed to characterize the cool, photoionized gas at $2.2 \la z \la 3.6$ in 202 \hi-selected absorbers with $14.6 \le \mlnhi <20$, i.e., the gaseous interface between galaxies and the intergalactic medium (IGM).  We find that the $14.6 \le \mlnhi <20$ gas at $2.2 \la z \la 3.6$ can be metal-rich gas ($-1.6 \la \xh \la -0.2$) as  seen in damped \lya\ absorbers (DLAs); it can also be very metal-poor ($\xh<-2.4$) or even pristine gas ($\xh <-3.8$) not observed in DLAs, but commonly observed in the IGM. For $16< \mlnhi <20$ absorbers, the frequency of pristine absorbers is about 1\%--10\%, while for  $14.6 \le \mlnhi \le 16$ absorbers it is 10\%--20\%, similar to the diffuse IGM. Supersolar gas is extremely rare ($<1\%$) in this gas. The factor of several thousand spread from the lowest to highest metallicities and large metallicity variations (a factor of a few to $>100$) between absorbers separated by less than $\Delta v < 500$ \km\ imply that the metals are poorly mixed in $14.6 \le \mlnhi <20$ gas. We show that these photoionized absorbers contribute to about 10\% of the cosmic baryons and  30\% of the cosmic metals at $2.2 \la z \la 3.6$. We find the mean metallicity increases with \nhi, consistent with what is found in $z<1$ gas. The metallicity of gas in this column density regime has increased by a factor $\sim$8 from  $2.2 \la z \la 3.6$ to $z<1$, but the contribution of the  $14.6 \le \mlnhi <19$ absorbers to the total metal budget of the universe at $z<1$ is half that at $2.2 \la z \la 3.6$, indicating a substantial shift in the reservoirs of metals between these two epochs. We compare the KODIAQ-Z results to FOGGIE cosmological zoom simulations. The simulations show an evolution of \xh\ with \nhi\ similar to our observational results. Very metal-poor absorbers with $\xh <-2.4$ at $z\sim 2$--3 in these simulations are excellent tracers of inflows, while higher metallicity absorbers are a mixture of inflows and outflows.
\end{abstract}


\section{Introduction}\label{s-intro}
The intergalactic medium (IGM) and the circumgalactic medium (CGM) are massive reservoirs of baryons \citep[e.g.,][]{fukugita98,mcquinn16,macquart20} and major fuel sources for star formation in galaxies \citep[e.g.,][]{tumlinson17}. They play a vital role in the formation and evolution of galactic and large-scale structures in the universe. The empirical and theoretical characterizations of the CGM and IGM before, during, and after the peak of cosmic star formation ($z \sim 2$) are therefore of prime importance to understanding their role in the evolution of galaxies. Gathering data that can probe both the low and high redshift universe is critical for robustly constraining the evolution of these cosmic structures. This is also the best route to fully test always-improving cosmological simulations. Modern numerical simulations are now reaching new milestones in fidelity. Notably a new generation of simulations boasts unprecedented high spatial resolution, even in the more diffuse regions of the simulation boxes (e.g., \citealt{vandevoort18,hummels19,peeples19,suresh19}). This allows for far more realistic modeling of the CGM and IGM and their inter-relationships with galaxies. 

Over the last few years, our group has engaged in several \hi-selected surveys at both low and high redshifts to target absorbers in the \hi\ column density range $15 \la \mlnhi \la 19$. This column density interval spans a range of physical environments from the diffuse IGM (Ly$\alpha$ forest, hereafter LYAF) to the denser portions of the CGM and the edges of galaxy disks. Following \citet{lehner18}, we describe absorbers as strong \lya\ forest systems (SLFSs, $\mlnhi = 14.5$--16.2), as partial Lyman systems (pLLSs, $\mlnhi =16.2$--17.2), or as Lyman limit systems (LLSs, $\mlnhi =17.2$--19). Below this range is the LYAF ($\mlnhi \la 14.5$) and above it are the super-LLSs (SLLSs, $19.0 - 20.3$, a.k.a. the sub-damped \lya\ absorbers) and the damped \lya\ absorbers (DLAs, $\ge 20.3$). With overdensities between the diffuse IGM (probed by the LYAF) and galaxies (probed by DLAs), the SLFSs, pLLSs, and LLSs spanning $15 \la \mlnhi \la 19$ should be at the heart of the exchange of matter between galaxies, their CGM, and the diffuse IGM.

One of the main goals of our \hi\ absorption surveys is to provide a census of the chemical enrichment of the absorbers with $14.6 \la \mlnhi \la 20$ over cosmic time (e.g., the COS CGM compedium, CCC, at low redshift---\citealt{lehner18,lehner19,wotta19}, and see also \citealt{lehner13,wotta16}, the HD-LLS survey---\citealt{prochaska15,fumagalli16}, the KODIAQ-Z survey---\citealt{lehner16}; this paper). The metallicity of the absorbers is a direct measure of their metal enrichment and a key diagnostic of their origins. The metallicity provides direct information on how efficient (or not) galaxies are at enriching their immediate surroundings and beyond and on the level of metal mixing in the diffuse regions of the universe. The \hi\ selection of our surveys provides a relatively unbiased way to obtain a census of the metallicity since selecting on hydrogen is sensitive to both very low and very high metallicities. 

While pLLSs and LLSs have been found often in the CGM of galaxies at both low and high redshift (e.g., \citealt{lehner09,lehner13,prochaska17}; M. Berg et al. 2021, in prep.), there is growing evidence that at least some of the pLLSs and LLSs with very low metallicities may probe the denser IGM (very low metallicity being $\xh \la -1.4$ at $z\la 1$, M. Berg et al. 2021, in prep., and $\xh \la -3$ at $2\la z\la 3.5$, \citealt{fumagalli11b,fumagalli16a,crighton16,lofthouse20})\footnote{For the metallicity, we use conventional squared-bracket notation $\xh \equiv \log N_{\rm X}/N_{\rm H} - \log ({\rm X/H})_\sun $, where X is an $\alpha$-element, unless otherwise stated.}. There are, however, also counter-examples at both low and high redshifts where very low metallicity absorbers are found in vicinity of galaxies (e.g., \citealt{ribaudo11,fumagalli16a}). Fully characterizing the metallicity distributions of the SLFSs, pLLSs, and LLSs at low and high $z$ directly probes the metallicity enrichment (or lack of it) relative to galaxies and their inner regions probed by DLAs and SLLSs. This technique can also assess the amount of pristine gas in these overdense regions of the universe and how it compares to the more diffuse IGM probed by LYAF absorbers \citep{fumagalli11b}. 

In previous papers, our group presented a sample of 261 $z < 1$ absorbers with $15 \la \mlnhi \la 19$ (\citealt{lehner18,lehner19,wotta19}, hereafter, \citetalias{lehner18,lehner19,wotta19}). This sample includes an unexpectedly large fraction of metal poor absorbers, many of which have metallicities $\xh <-2$ implying little chemical enrichment since $z\sim 2$--3. 

The finding that metal-poor absorbers with $\xh \le -1.4$ represents about half of the population of $z\la 1$ absorbers with $15 \la \mlnhi \la 19$ was unanticipated. This is because neither the LYAF (owing, in retrospect, to a lack of sensitivity) nor the SLLSs/DLAs  hinted at the presence and importance of such low-metallicity gas at low redshift. Although not fully appreciated at the time, there was, however, already some evidence of very low metallicity gas at low redshift based on individual studies of LLSs (\citealt{cooksey08,ribaudo11}, and see also \citealt{tripp05} for a rare example of a primitive SLLS at $z\sim 0.06$). CCC also reveals that at $z<1$ there is no evidence of pristine gas, implying $>99\%$ gas with $15 \la \mlnhi \la 19$ has been polluted by $z\la 1$, even if it is only at very low level. Low-redshift photoionized gas with $15 \la \mlnhi \la 19$ has a wide range of metallicities observed ($-3\la \xh\la +0.4$) and can show large metallicity variations (factor up to 25) over small redshift/velocity separation ($\Delta v<300$ \km), implying these absorbers sample a  highly inhomogeneous medium probed by these absorbers. The CCC results beg the questions: are these properties unique to $z\la 1$? how do they evolve from earlier times, across the peak of cosmic star formation at $z\sim 2.5$. 

To address these questions, we present here the KODIAQ-Z survey (Keck Database of Ionized Absorbers toward Quasars), a survey designed to identify \hi\ absorbers with $\mlnhi \ga 14.5$ at $z>2$ in  Keck HIRES spectra available in the KODIAQ database \citep{omeara15,omeara17} and determine their metallicities. Owing to the higher sensitivity of the Keck HIRES data than the Hubble Space Telescope (HST) COS Origins Spectrograph data, we can probe low \nhi\ absorbers at high redshift and still be sensitive to low metallicity gas. In fact, due to the new sensitivity gained with 8--10 m ground-based telescopes, early studies  on the metal enrichment at high redshift largely focused mostly on IGM absorbers with $\mlnhi \la 14.5$ (e.g., \citealt{songaila98,ellison00,schaye03,aguirre04,simcoe04}). Our sample of \hi-selected absorbers consists of 155 SLFSs, 24 pLLSs, 16 LLSs, and 7 SLLSs at $2.2 \la z \la 3.6$, totaling \ssz\ absorbers, 196 of them with $14.5 \la \mlnhi \la 19$. On account of the nature of the Keck HIRES data, the sample of pLLSs and LLSs has only marginally increased by a factor 1.3 compared to our earlier pilot survey  (\citealt{lehner16}). However, in this paper, we undertake a systematic search and characterization of \hi-selected SLFSs.\footnote{\citet{simcoe04} surveyed the metallicity of the IGM at $2.2 \la z \la 2.8$. While they focused on absorbers with $\mlnhi <14.5$, they also included about 30--40 SLFSs as part of their survey.} In our analysis, we include the \hi-selected absorbers from the HD-LLS survey \citep{prochaska15,fumagalli16}, which adds another 46 LLSs, increasing the total sample to 241 absorbers with $14.5 \la \mlnhi \la 20$. The HD-LLS survey is directly complementary to KODIAQ-Z since most of the HD-LLS absorbers have $\mlnhi \ga 17.5$ while KODIAQ-Z largely samples gas with lower \nhi. 

The main goals of the present paper are to determine the metallicity distributions of these absorbers, assess how their metallicity changes with \nhi\ and $z$, and provide a robust limit on the amount of pristine gas at high redshift. We also present and discuss the physical properties of these absorbers and use these results to estimate the metal and baryon budgets of the cool photoionized CGM at $z \ga 2.2$--3.6. Our paper is organized as follows. In \S\ref{s-present}, we describe the survey design of KODIAQ-Z, the search of the \hi-selected absorbers, and the sample of SLFSs, pLLSs, and LLSs. In \S\ref{s-estimate-col}, we present the methods to derive the column densities of the metals and \hi\ for each absorber, while in \S\ref{s-immediate-results} we present basic statistical properties of our sample ($z$, \nhi, and the Doppler parameter $b$). In \S\ref{s-ion-model}, we present  ionization modeling of the KODIAQ-Z absorbers and assumptions for estimating their metallicity. Our main results on the metallicities of the \hi-selected absorbers are presented in \S\ref{s-metallicity}. In \S\ref{s-phys-cond}, we derive the physical properties of the absorbers, including their density (\nnh), total column density (\nh), temperature, and linear scale ($l\equiv \mnh/\mnnh$), and we model the neutral fraction as a function of \nhi\ in the range $14.6\le \mlnhi \le 20$. In \S\ref{s-budget}, we estimate  the cosmic metal and baryon budgets of these absorbers. In \S\ref{s-disc-sim}, we  compare the KODIAQ-Z results to the properties of the gas in cosmological zoom simulation from the Figuring Out Gas \& Galaxies in Enzo (FOGGIE) project \citep{peeples19,corlies20} and use these simulations to gain insight on the connection of the $14.5 \la \mlnhi \la 20$ gas depending on its metallicity with galaxies. In \S\ref{s-disc}, we discuss the results, and in \S\ref{s-sum}, we summarize our main conclusions.

\section{Survey Design, Database, and Observations}\label{s-present}
\subsection{Sample Selection Criteria}\label{s-criteria}

Our survey is based on a search for \hi\ absorption  in the KODIAQ DR2 database \citep{omeara17} of  300 QSO spectra observed with HIRES. The QSOs useful for our survey lie at $2 \la \mzem\ \la 4.5$, limiting the sample size to 235 QSOs. We also searched in the spectra of higher redshift ($\mzem >4.5$) QSOs, but the LYAF is too dense to allow us to reliably determine the \hi\ {\it and}\ metal-line column densities at these redshifts. The lower redshift-threshold $z\ga 2$ is required to have wavelength coverage of at least \lya\ and \lyb.\footnote{The Keck HIRES spectra are normalized and therefore we cannot rely on the break at the Lyman limit to estimate \nhi\ for the pLLSs and LLSs, relying solely on the Lyman series, and hence limiting the sample we can estimate accurately \nhi.}   In each of the QSO spectra, we search for absorbers with \hi\ column density $\mlnhi \ga 14.5$. This contrasts with our  \citetalias{lehner16} survey where our search of \hi\ absorbers was limited to absorbers with $\mlnhi \ga 16.2$ (to match our initial lower $z$ survey, \citetalias{lehner13}) and was by no means a systematic and complete examination of the KODIAQ DR1 database \citep{omeara15}. 

A major difference (besides the redshift probed) with the low redshift survey CCC \citepalias{lehner18} is that all the HIRES spectra are flux-normalized for the coaddition of the individual exposures and echelle orders   \citep{omeara15,omeara17}. This makes determination of large-scale flux decrements (e.g., at the Lyman limit  break or damping wings of DLAs) unreliable. Therefore, to estimate the \hi\ column density we can rely only on the absorption observed in the Lyman-series transitions; specifically, the break at the Lyman limit for absorbers with $\mlnhi \ga 16.5$ cannot be used. This limits the sample size of absorbers with $\mlnhi \ga 17.2$ because it requires enough weak Lyman transitions that are not too contaminated to derive \nhi\ (see \S\ref{s-estimate-hi}). For absorbers with $\mlnhi \ga 18$, damping wings start to be observable in \lya\ absorption, and therefore this transition can be used to derive \nhi\ depending on the level of contamination of the absorption from the LYAF.  On the other hand, while CCC is limited to absorbers with $\mlnhi \ga 15.3$ owing  to the need to be sensitive to metallicities $\xh \la -1$, the typically higher signal-to-noise ratios (SNRs) and higher spectral resolution of the Keck HIRES spectra allow us to probe lower \nhi, down to the LYAF threshold of about $ \mlnhi \simeq 14.5$,  and still be sensitive to metallicity with $\xh \la -2$.  Although with very high SNR Keck HIRES spectra, a strong limit ($\xh \la -2$) on the metallicities can be placed down to $ \mlnhi \simeq 13.6$ (e.g., \citealt{simcoe04}), our lower limit on \nhi\ was selected to correspond to the limit between the diffuse IGM and denser IGM/diffuse CGM. Following \citet{schaye01}, a 14.5 dex limit on the \hi\ column density corresponds to an overdensity of $\delta \ga 5$ at $z\sim 2.8$ (about the average redshift of the absorbers in our sample, see \S\ref{s-res-nhi-z-dist}). 

\subsection{Search for \hi\ Absorbers in the KODIAQ database}\label{s-search}

\begin{figure*}[tbp]
\epsscale{0.8}
\plotone{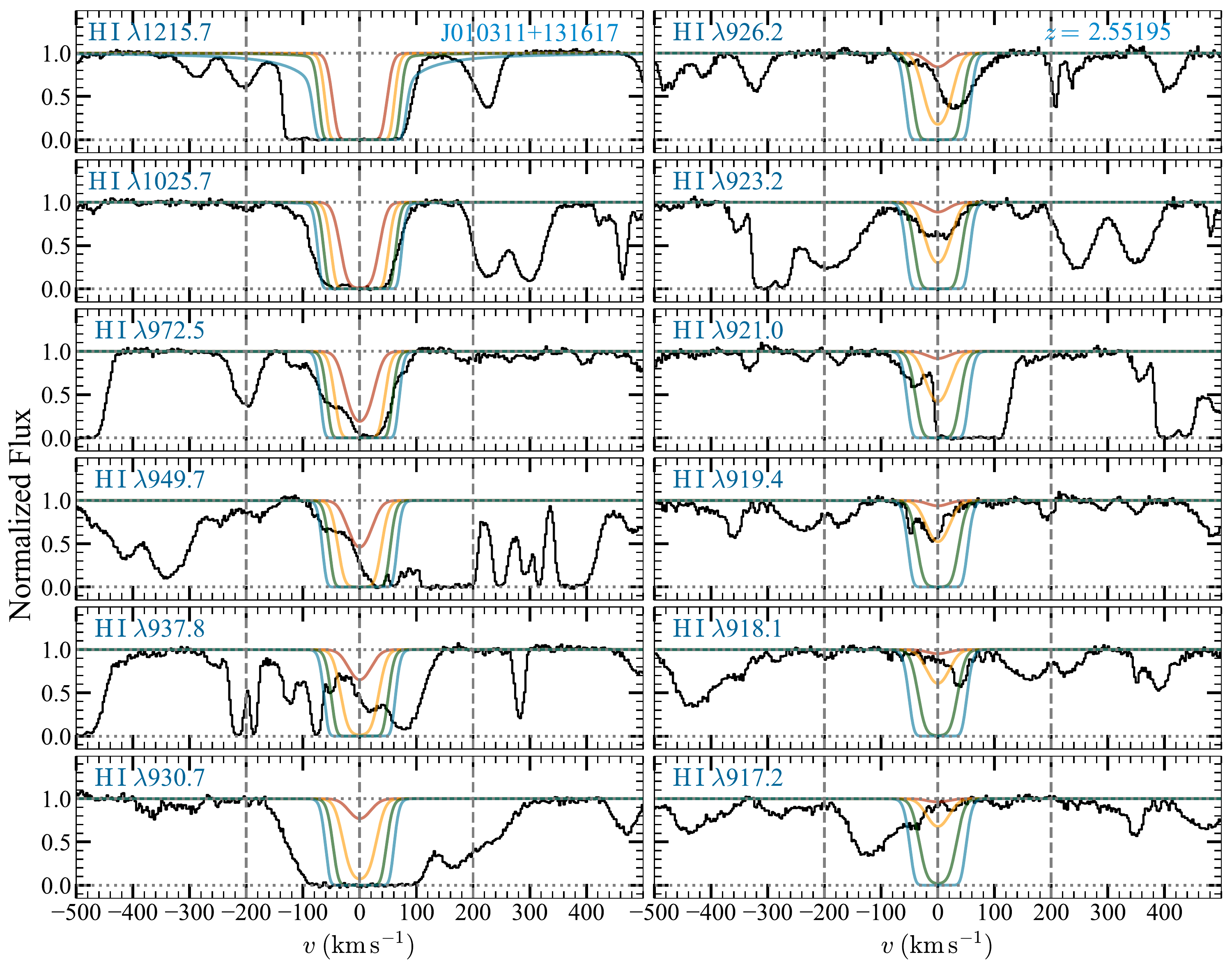}
\caption{Example of an absorber identified by our automatic searching tool. For each identified candidate-absorber, a similar figure is created that shows the normalized spectra (in black) of the \hi\ transitions from the 1215 to 917 \AA\ transitions  as function of the velocity in the redshift rest-frame. The colored profiles are Voigt profiles for an absorber with $b=25$ \km\ and $\mlnhi = 15,16,17,18$ and are used to help assess the level of contamination in each transition. Ultimately for this candidate two absorbers were identified at $z=2.55125$  and $2.55216$ (corresponding to $v\simeq -50$ and $+17$ \km\ in this figure).
\label{f-search-ex}}
\end{figure*}

To identify the strong \hi\ absorption, we developed an automated search tool to help identify the Lyman series transitions from 1215 \AA\ (\lya) down to 917 \AA. In Fig.~\ref{f-search-ex}, the  colored profiles show single-component Voigt models for absorbers with a Doppler parameter $b = 25$ \km\ and $\mlnhi = 15,16,17,18$. That $b$-value is selected based on previous profile fitting of strong \hi\ absorbers at similar redshifts (e.g., \citetalias{lehner16}); as we will see below that choice is close {\it a posteriori}\ to the mean $b$ value in our sample. We use these \hi\ models to set the following criteria for our automatic search: 1) \lya\ must have flux $F_\lambda \le 0$ $\ga  0.4$ \AA\ in contiguous pixels; 2) at the average velocity/redshift ($v_c$) of the identified \lya, the flux in the \lyb\ absorption must also reach $F_\lambda \le 0$; 3) at $v_c$, we also require there is substantial absorption in at least \lyg\ (and \lyd\ if there is wavelength coverage of that transition). Additional conditions were added to ensure that when a DLA or a SLLS is present, the search-program would not continue to look for additional absorbers near the redshift of the DLA or SLLS. For this search, we therefore require for each QSO wavelength coverage of at least \lya, \lyb, and \lyg. However, since the damping wings  of \lya\ can be used to determine \nhi\ if $\mlnhi \ga 18.5$, we also undertook a separate search that relies solely on \lya\ where the flux $f_\lambda \simeq 0$ over a wavelength width of $\ga 2$ \AA. {\it A posteriori}, only a few of these very strong \hi\ absorbers ended up in our sample owing to the fact that the KODIAQ spectra are normalized prior to coaddition, and the normalization was often not reliable enough to model the damping wings for these absorbers, especially the strong SLLSs and DLAs (DLAs being not an issue since they are not part of our survey). 

\begin{figure}[tbp]
\epsscale{1.15}
\plotone{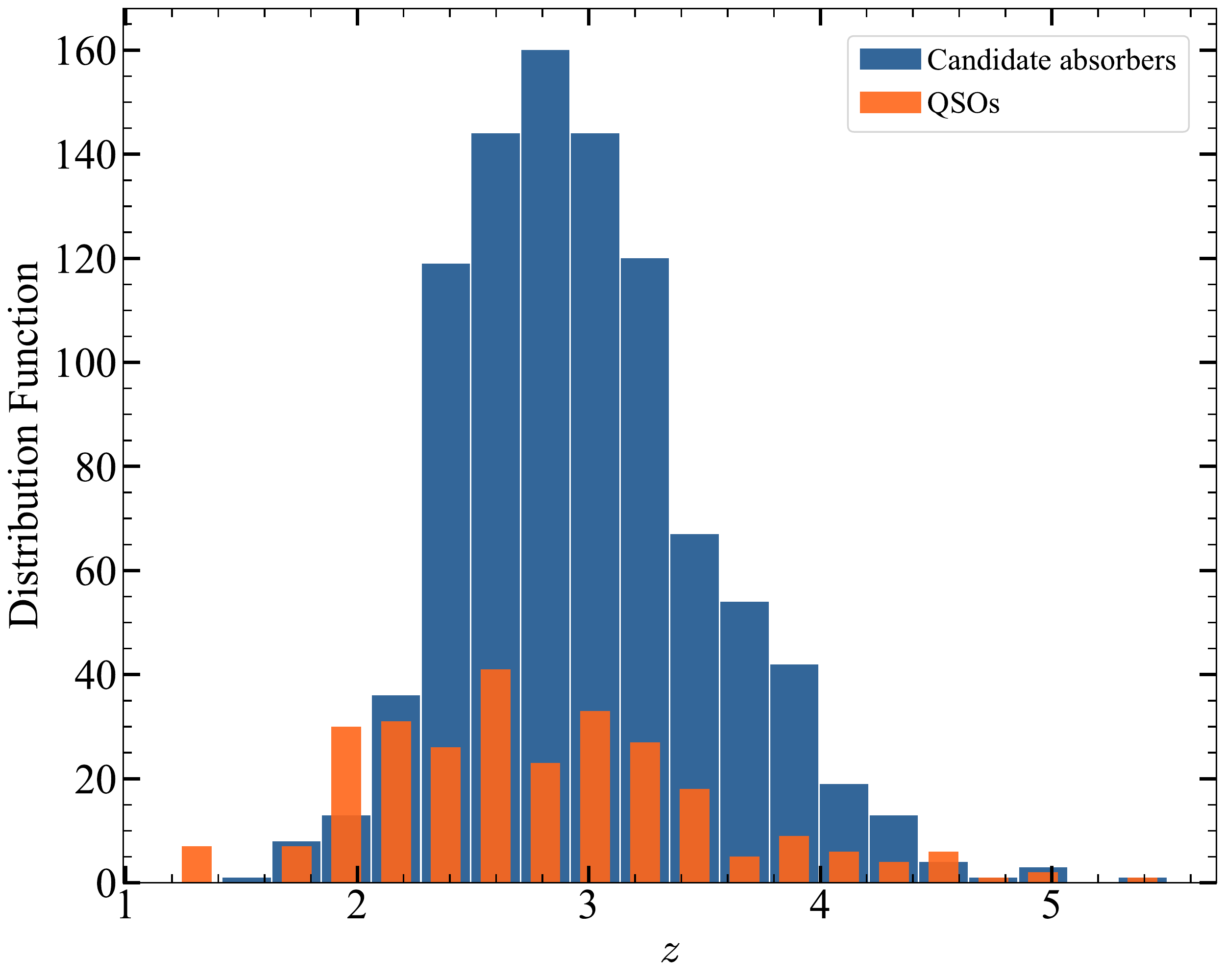}
\caption{Distribution of the redshifts for the QSOs and the {\it candidate}\ absorbers from the automatic search in KODIAQ DR2. 
\label{f-zdist-candidates}}
\end{figure}

For each candidate absorber, our search tool created a stack of \hi\ transitions as shown on Fig.~\ref{f-search-ex}. Over 3000 candidates were created, many of these being false positives owing to the large contamination from the LYAF and other metal lines. That number is still small enough that a quick look to the stack of \hi\ profiles could easily weed out most of the false positives using the Voigt models as a guide. This vetting led the sample to be reduced by more than a third to 949 candidate absorbers, for which we show their redshift distribution in Fig.~\ref{f-zdist-candidates}. In this figure, we also show the redshift distribution of the QSOs. About 83\% of the candidate absorbers are at $2.2 \le \mzabs \le 3.6$; candidate absorbers at $\mzabs \le 2$ are detected solely based on \lya. 

To assemble our final sample for metallicity determinations, we considered both the ability to estimate \nhi\ and the coverage of metal ions (in order to be able to estimate the metallicity).  For each candidate absorber, we therefore created  another set of stack of velocity profiles with \lya, \lyb, \lyd, and some key metal transitions used in the estimation of the metallicity (including \siii, \siiii, \siiv, \cii, \ciii, \civ, \alii, \aliii, \feii, \oi, and \ovi) to assess if there is enough metal-ion coverage to determine the metallicity. After all the measurements are done (detailed in \S\ref{s-estimate-col}), the final sample size sample consists of \ssz\ absorbers, i.e., $\approx 7\%$ of the candidates from the initial automatic search were absorbers where we could place a robust estimate on \nhi\ and metal-line column densities (including strong upper limits). The final KODIAQ-Z sample includes 155 SLFSs, 24 pLLSs, 16 LLSs, and 7 SLLSs.

\subsection{Comparison Samples}
\label{s-sample-comp}

One of the goals of our survey is to determine how the metallicity changes with \nhi\ at similar redshifts and between low and high $z$. For the low redshift, we use results from our CCC survey, including the compilation of \hi\ absorbers at $\mlnhi\ge 19$ (see \citetalias{lehner18,wotta19,lehner19} and references therein). At $z\ga 2$, we use two surveys: 1) the HD-LLS survey \citep{prochaska15,fumagalli16}, a survey of absorbers with $17.3 \le \mlnhi < 20.3$ at $z =2$--4.39, $\sim$95\% of the absorbers having \hi\ column densities in the range $18 \le \mlnhi < 20.3$ (which is directly complementary to KODIAQ-Z; HD-LLS also follows the same \hi-selection); and 2) the DLA survey by \citet{rafelski12}, hereafter the R12-DLA survey. The ease of  access to the data while working in the early stage of KODIAQ-Z, including the metal and \hi\ column densities for the HD-LLS, mostly dictated our use of these surveys. We note that the Magellan DLA survey by \citet{jorgenson13} has essentially the same metallicity distribution as R12-DLA in our redshift of interest $2.2 \la z\la 3.6$ (see below). The recent XQ-100 survey of SLLSs by \citet{berg21} shows also the same metallicity distributions as in HD-LLS. 

For the R12-DLA, we adopt the metallicities from the $\alpha$-elements (or Zn if not available). For the HD-LLS, we re-estimate all the metallicities using the same EUVB adopted in CCC in order to avoid mismatch in the comparisons of the metallicities although as we will show in \S\ref{s-uncertainties}, the systematic error arising from the EUVB in the photoionization modeling of the absorbers is much smaller (in fact negligible) at $z>2$ than at lower redshift, especially for the strong LLSs and SLLSs that are part of the HD-LLS survey.  

\section{Estimation of the Column Densities}\label{s-estimate-col}

Our analysis follows closely that undertaken in CCC to estimate the column densities of \hi\ and metal ions.  Throughout we assume the atomic parameters (central wavelengths and $f$-values) listed in \citet{morton03}. For each absorber, we first attempt to estimate the \hi\ column density since that step is necessary to estimate the metallicity of the absorbers. For \hi, we use a combination of different methods (apparent optical depth--AOD--method \citealt{savage91}, curve-of-growth--COG--method, and Voigt profile fitting--PF--method), while for the metal ions, we use solely the AOD method. We employ several methods to estimate the \hi\ column densities on the account that the contamination, blending, and saturation effects are more severe than for the metal lines (see Fig.~\ref{f-search-ex}).  

\subsection{Metal Column Densities}\label{s-metal-col}
Although we only estimate the column densities of the metal species  once the \hi\ transitions were modeled (and hence the candidate absorber became a selected absorber), we first describe the metal lines as we use only the AOD method to extract the properties of the absorption. In this method, the absorption profiles are converted into apparent optical depth per unit velocity, $\tau_a(v) = \ln[F_{\rm c}(v)/F_{\rm obs}(v)]$,  where $F_c(v)$ and $F_{\rm obs}(v)$ are the modeled continuum and observed fluxes as a function of velocity. The AOD, $\tau_a(v)$, is related to the apparent column density per unit velocity, $N_a(v)$, through the relation $N_a(v) = 3.768 \times 10^{14} \tau_a(v)/(f \lambda(\mbox{\AA})$)  ${\rm cm}^{-2}\,({\rm km\,s^{-1}})^{-1}$, where $f$ is the oscillator strength of the transition and $\lambda$ is the wavelength in \AA. The total column density is obtained by integrating the profile over the pre-defined velocity interval, $N = \int_{v_{\rm min}}^{v_{\rm max}} N_a(v) {\rm d}v $, where  $[v_{\rm min},v_{\rm max}]$ are the boundaries of the absorption. We computed the  average line centroids through the first moment of the AOD $v_a = \int v \tau_a(v) {\rm d}v/\int \tau_a(v){\rm d}v$ \km. The velocity range over which the line was integrated was determined from weak \hi\ transitions and metal line. 

\begin{figure*}[tbp]
\epsscale{0.8}
\plotone{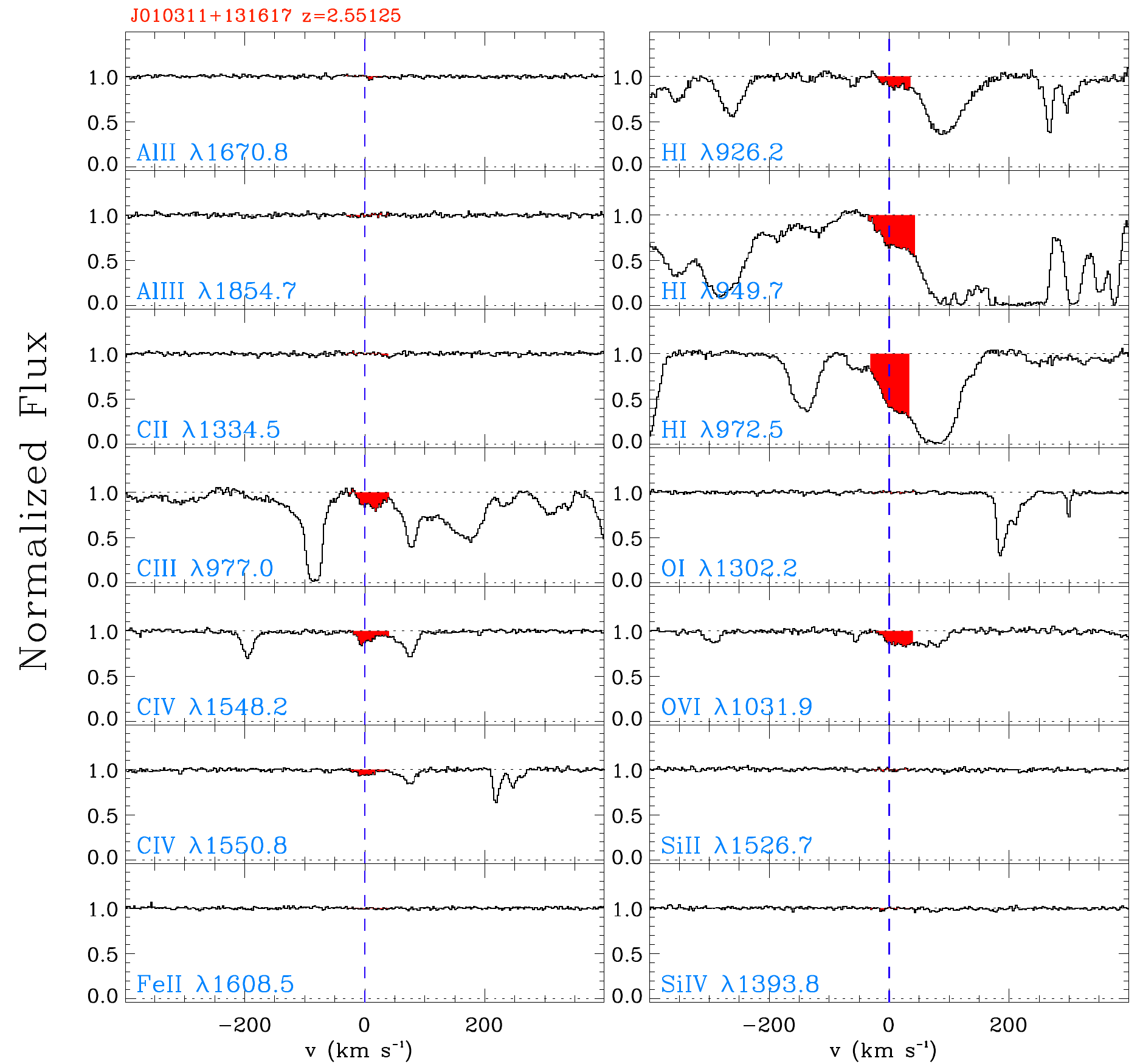}
\caption{Example of normalized spectra against the rest-frame velocity of the absorber at $z=2.55125$ toward J010311+131617 that show detection of \hi, \ciii, \civ, and \ovi\ (the integration range of the profiles are shown in red). Another absorber separated by 65 \km\ is found at $z=2.55216$; in this case the column densities can be estimated reliably in each component/absorber (see also Fig.~\ref{f-fit-ex})
\label{f-spectra-ex}}
\end{figure*}

 We emphasize that, as much as possible, the integration range ($[v_{\rm min},v_{\rm max}]$ corresponds to a single absorbing \hi\ complex as shown, e.g., in Fig.~\ref{f-spectra-ex}, but this does not mean necessarily the metal lines consist solely of a single component (often they are, but not always). For the absorbers considered in this survey, the main \hi\ components can typically be separated when the difference between central velocities between two \hi\ components are at least separated by $|\Delta v| \ga 40$ \km\ and $\mlnhi \la 18$.  As we will see below the mean Doppler parameter of the \hi\ components is around 27 \km, corresponding to a temperature of $T <4 \times 10^4$ K, typical for photoionized gas. This means the thermal broadening is important and hence the \hi\ lines are broader than the metal lines, i.e., there are \hi\ transitions that can appear as a single component (at the observed spectral resolution) despite the metal ions showing multiple components over a velocity width of $\sim 50$ \km. In that case, we integrate the metal ions over a similar velocity width as the \hi\ when that information is available from weak \hi\ transitions; if the latter is not available (e.g., SLLS where weak \hi\ transitions are not covered, which is the case for $\la 5\%$ of the sample), we guide our analysis using the low ions or \oi\ if it is detected, which are better tracers of denser regions probed by the strongest \hi\ absorbers. In this latter case, we integrate over the entire profiles of the low ions that may have a single or several components; this velocity integration range dictates the integration range of the higher ions, even though they may extend well beyond the absorption seen in the low ions (see, e.g., \citealt{lehner14}). 

As illustrated in Figs.~\ref{f-search-ex} and \ref{f-spectra-ex}, even deep in the LYAF, the original continuum placement (see \citealt{omeara15,omeara17}) is quite good and can be directly used in many cases to estimate the physical parameters of the absorption lines accurately. We therefore took full advantage that the Keck HIRES spectra are already normalized to use the AOD method in a fully automated way.  For each identified absorber, we made typically two iterations to estimate the physical parameters of the lines ($v_a$, $N_a$, and equivalent width $W_\lambda$). In the first iteration, the velocity range is estimated based on the stack of velocities profiles used to preview the data. Our program created figures that show for each transition the normalized and apparent column density profiles (such as Fig.~3 in \citetalias{lehner18}), allowing us to refine the integration velocity range and determine if the continuum needed any local refinement. In the second iteration, the refined velocity range is applied to the original or re-normalized spectra. If we determine that the continuum model needed to be re-estimated, we either use the automated Legendre polynomial fitting described in \citetalias{lehner18} or selected by hand the continuum region to be fitted by low-order Legendre polynomials. 

We integrate the equivalent widths using the same  $[v_{\rm min},v_{\rm max}]$ integration range adopted for the $N_a(v)$ profiles. The main use of the equivalent width for the metal lines in our survey is to determine if the absorption is detected at the $\ge 2\sigma$ level. If it is not, we quote a 2$\sigma$ upper limit on the column density, which is simply defined as twice the 1$\sigma$ error derived for the column density assuming the absorption line lies on the linear part of the curve of growth. The 1$\sigma$ error is determined by integrating the spectrum over a similar velocity interval to that of a detected ion or very weak \hi\ transition or otherwise $\pm 20$ \km\ based on the typical smallest velocity intervals in other absorbers with weak detection of metals.

When an absorption feature is detected at $\ge 2\sigma$ significance, the next step is to assess if there are any contamination or saturation issues. The main ions detected in this work are \cii, \ciii, \civ, \siii, \siiii, \siiv, and \ovi\ (see Fig.~\ref{f-spectra-ex} and Appendix). Other ions are also considered such as \feii, \feiii, \alii, \aliii\ or atoms such as \oi, but in most cases these are not detected (Fig.~\ref{f-spectra-ex} and Appendix~\ref{a-data}). For \civ\ $\lambda \lambda$1548, 1550, \siiv\ $\lambda \lambda$1393, 1402, and \ovi\ $\lambda \lambda$1031, 1037, \siii\ $\lambda \lambda$1193, 1260, 1304, 1526, these are either doublets or ions with several transitions, and both contamination and saturation can be readily checked by comparing their $N_a(v)$ profiles.  Ions with $\lambda \la 1215$ \AA\ are more often likely to be contaminated than ions with transitions at longer wavelength owing to the dense LYAF at these redshifts. For example, for the absorber shown in Fig.~\ref{f-spectra-ex}, the \civ\ transitions are not contaminated, but the weak transition of \ovi\ is.  In contrast, the weakest \hi\ absorber  at about $-55$ \km\ seen in Fig.~\ref{f-spectra-ex} (that is not analyzed as part of this work because its \hi\ column density is $\mlnhi <14.5$) shows absorption in both transitions of each doublet of \civ\ and \ovi. The \cii\ ion has two transitions at $1334$ and $1036$ \AA, and we always check the latter transition if $\lambda$1334 is detected; more  often than not it is contaminated by the LYAF. 

On the other hand, ions such as \ciii\ $\lambda$977 and \siiii\ $\lambda$1206 are  in the LYAF and have only a single transition. In this case, our strategy for assessing whether a given transition is contaminated by another unrelated absorber is based on a combination of visual inspection and comparison of the relevant quantities (e.g., mean velocity and apparent optical depth) with other transitions. Strong contamination can often be diagnosed visually, comparing both the central velocities of the peak optical depth and shape of the velocity structure of the absorption profiles across transitions from the same or similar ions.\footnote{Of course if the investigated ion is not in the same gas-phase than an ion that is uncontaminated (e.g., \ovi\ in some cases), these criteria cannot be used.}  The mean velocities of the absorption profiles estimated from the integration of the absorption profiles can often point to strong contamination if the mean velocity from one transition is significantly different than others that are aligned. In the case of obvious contamination, the absorption feature is discarded. The only exceptions are as follows. If the contamination is very mild, only occurring in the wing of a relatively strong absorption, we changed slightly the velocity integration range to avoid the contaminated portion for that absorption feature. In cases where we are not sure that the absorption arises solely from the understudied metal ion (e.g., detection of \ciii\ but no coverage or detection of \civ\ or other ions), but the feature could very likely be produced by that ion based on the same velocity as that of \hi\ and narrow absorption (as expected from a metal, low ion), we treat the column density from that ion as an upper limit in our ionization modeling\footnote{This upper limit can be differentiated from a non-detection upper limit as the error bars of the former are not the standard non-detection upper limit errors $+0.18,-0.30$ dex.}. 

For most of the metal ions in our survey, saturation is not a major issue owing to the lower column \hi\ densities than those of DLAs and even SLLSs (there are only 7 SLLSs in our sample and no DLA), lower metallicities on average than $z<1$, and to the  high spectral resolution of the Keck data. To determine if there is saturation, we proceed in a similar way that we did for the contamination assessment. Where there is no obvious sign of contamination and the absorption from the metal ion reaches zero flux, the absorption is automatically marked as a lower limit. The comparison of the AOD profiles for doublet ions or ions with multiple transitions can readily help determine whether there is saturation if the profiles do not match at the peak optical depth and if the apparent column density of the stronger transitions is systematically smaller than that of the weaker transitions. 

For atomic or ionic transitions that have $\Delta \log (\lambda f) \simeq 0.3$ and the difference in apparent column densities of the weak and strong transitions is $\Delta (\log N_a) \equiv \log N^{\rm weak}_a - \log N^{\rm strong}_a \le 0.13$ dex \citepalias[see][]{wotta16}, we are able to correct for the mild saturation using the method described in \citet{savage91} and \citetalias{wotta16}. In this case, we report the apparent column densities for each transition and then the adopted column density, corrected for saturation. In cases where we were not able to correct for saturation, we report the column density as a lower limit determined by the AOD method.\footnote{Contamination and saturation can be {\it a priori}\ both present at the same time and deceitful. However, it is unlikely to occur frequently over the same pixels, and therefore each can be disentangled reliably in most cases.}

For multiple transitions where there are no saturation or contamination issues, we give velocities and column densities as the weighted average. All the column densities, velocity integration ranges, averages velocities, and redshifts for the metal ions and \hi\ (see \S\ref{s-estimate-hi}) are summarized in Table~\ref{t-results}. We also provide all the results in a machine readable format (see Appendix).

\subsection{Determination of the \hi\ Properties}\label{s-estimate-hi}

To derive the metallicity, we can rely in all the cases on having column densities for several metal ions (detections and/or non-detections). However, to have information on the amount of hydrogen atoms, we rely solely on \hi. The advantage with \hi\ is that we can access, in most cases, to the entire Lyman series from 1215 \AA\ all the way down to 913 \AA\ (depending on wavelength coverage, redshift of the absorber, presence of a SLLS/DLA along the sightline, and strength of the absorption), spanning a difference in $f\lambda$ of over 2,700. However, the entire Lyman series is in the LYAF (and \lyb\ forest), and therefore many transitions can be and are often heavily contaminated, requiring a close look at all the \hi\ absorption features to assess the level of contamination. Therefore our first task was to systematically look simultaneously at several \hi\ profiles (from 1215 to 915 \AA) to assess if there is enough information (i.e., portion of the profiles of \hi\ transitions that are uncontaminated) to derive \nhi. If it was judged that the contamination could be dealt with, we proceeded by estimating the apparent column densities from the AOD profiles (and equivalent widths) and creating a mask of the \hi\ profiles that are contaminated (i.e., flagging the pixels that are clearly contaminated), a mask that is then used in the PF (see below). 

Owing to the LYAF contamination, each transition was manually handled, and we often also reassessed the continuum that can be more uncertain than in the redder part by modeling it with low-order Legendre polynomials. As mentioned above, we treated as much as possible an absorber as a single \hi\ component. For the example shown in Fig.~\ref{f-spectra-ex}, there are two components separated by 65 \km, which correspond to two absorbers at $\mzabs =2.55125$ and 2.55216. We therefore integrated the AOD profiles over each component (for the absorber at $\mzabs =2.55125$, that corresponds to the velocity interval of about $[-30,40]$ \km). For absorbers with $\mlnhi\ga 18$, this was not always possible especially if there is not coverage of the weaker Lyman series transitions (for these absorbers, the damping wings of \lya\ and \lyb\ can be used to determine \nhi, see below). However, these absorbers are rare in our sample, $\la 5\%$ (see below). For each transition, the AOD results are checked for contamination and/or saturation. For the transitions where there is no sign of contamination and/or saturation, we use a weighted-average of the integrated apparent column densities, which provides the \nhi\ estimates from the AOD method. 

The equivalent widths, estimated along with the apparent column densities and average velocities, are used for the COG analysis, which employs a $\chi^2$ minimization and $\chi^2$ error derivation approach outlined by \citet{sembach92}. A single component COG is assumed. Our procedure solves for $\log N$ and $b$ independently in estimating the errors. We started with all the \hi\ transitions for which we estimated the equivalent widths and did not appear {\it a priori} to be  contaminated based on the AOD analysis. We checked the results from the COG model and then removed only the transitions clearly departing from the model. We ran our COG program again with the new set of transitions to obtain the final estimate of \nhi\ with the COG method.  

\begin{figure}[tbp]
\epsscale{1.}
\plotone{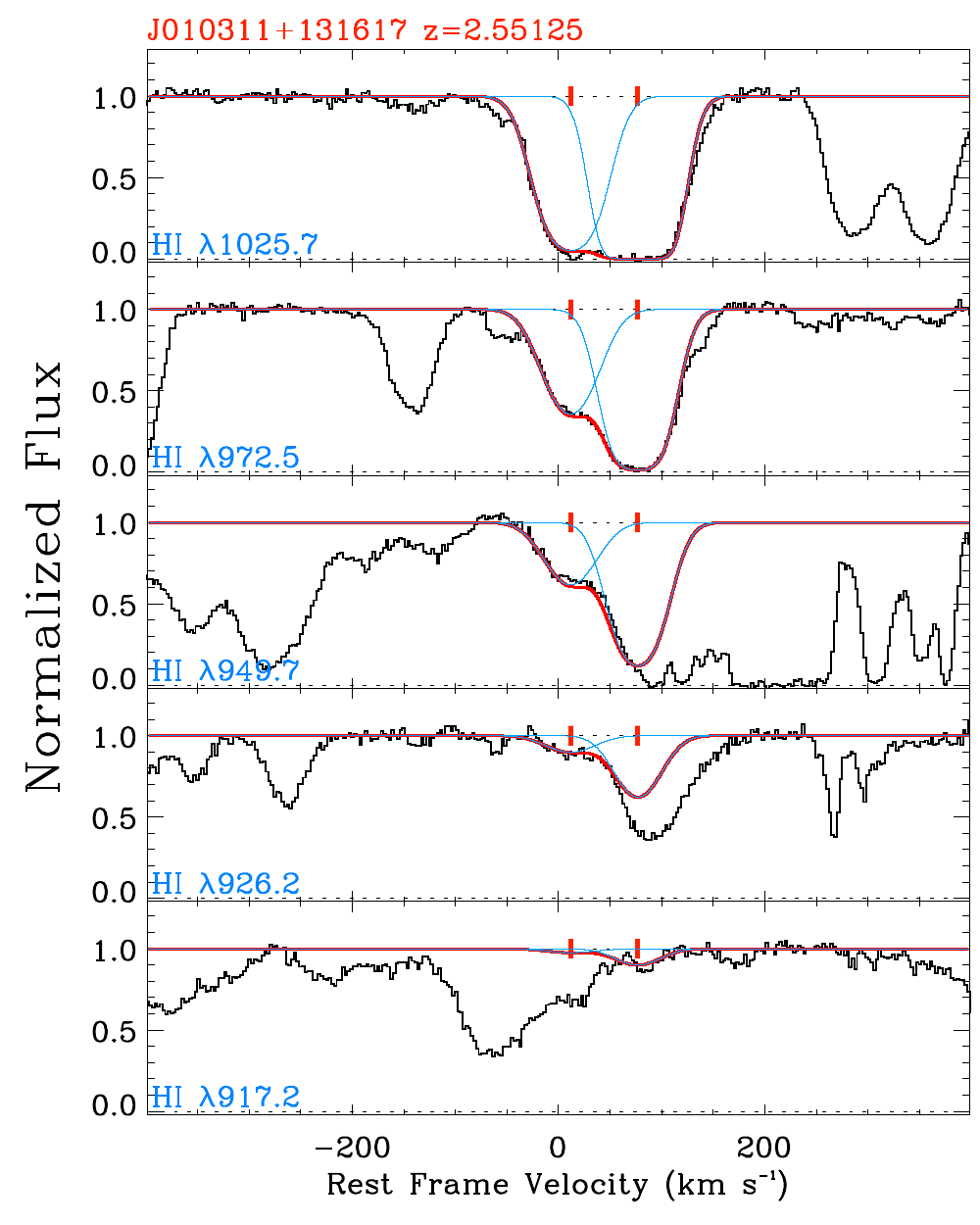}
\caption{Example of the same absorber shown in Fig.~\ref{f-spectra-ex} where we show the Voigt profile fit (in red the composite profile fit, and in blue each individual component) to the individual transitions of \hi\ (black spectra). Despite some contamination in each transition, there is enough uncontaminated regions in each profile to estimate robustly the \hi\ column density in each component. 
\label{f-fit-ex}}
\end{figure}

Using the normalized and masked \hi\ spectra, we also Voigt-profile fit the \hi\ transitions using an in-house PF program described in \citet{lehner11b,lehner14,lehner18}, which was updated and refined from \citet{fitzpatrick97}. A major difference from the previous methods is that the model profiles were convolved with the HIRES instrumental line-spread function, assumed to be Gaussian. The three parameters for each component $i$ (typically $i=1$, but in some cases $i=2$ or 3 or in rare cases more)---column density ($N_i$), Doppler parameter ($b_i$), and central velocity ($v_i$)---are input as initial guesses and were subsequently varied to minimize $\chi^2$ of the fit to all the fitted \hi\ transitions. As for the COG, we started with all the \hi\ transitions that did not appear {\it a priori} contaminated and iterated, removing any transitions that {\it a posteriori} appear contaminated (or sometimes revisiting the masked pixels). In Fig.~\ref{f-fit-ex}, we show an example of profile fitting that consists of a two-component fit corresponding to the absorbers at $\mzabs =2.55125$ and 2.55216. As it can be seen in each panel of Fig.~\ref{f-fit-ex}, there is some evidence of contamination especially in the component at 65 \km\ ($\mzabs =2.55216$), but even in that component there is enough information to derive robustly \nhi\ when using the  different \hi\ transitions simultaneously.

In general, there is a good agreement between these different methods to estimate \nhi\ for absorbers with  $14.5 \la \mlnhi \la 17.2$. The three methods explore different parameters and different transitions, allowing us to robustly estimate \nhi. When multiple approaches are valid, we simply average the values from the different methods and propagate the errors accordingly. For systems with $\mlnhi \ga 17.2$,  the PF results are most of the times solely adopted as a consequence of the need to either fit the \lya\ damping wings for absorbers with $\mlnhi \ga 19$ or  combination of weaker and stronger \hi\ transitions in order to correctly model the absorption that depends more strongly on both $N$ and and $b$ in this regime.  The eight systems with $17.5 \le \mlnhi \le 18.65$ require a more hands-on approach owing to having mostly saturated absorption in all the \hi\ transitions and only weak damping features in \lya. These were all analyzed as part of \citet{lehner14} and \citetalias{lehner16}, and we refer the reader to these papers for a full description, noting, very conservative errors were adopted with a minimum error on the \nhi\ for any LLS of $\sigma_{\log N}=0.15$ using this methodology. However, the metal lines associated with these absorbers were all reanalyzed as part of this work.

The adopted \nhi\ results are summarized in Table~\ref{t-results} and we also report in Table~\ref{t-fit} the PF results, i.e., the velocities, Doppler parameters, and column densities, and errors associated with these quantities. In that table, the listed redshift corresponds to the adopted redshift (see above) and the third column indicates the component number for each absorber.

\section{Empirical Properties of the KODIAQ-Z Survey}\label{s-immediate-results}

Before describing the ionization modeling of the absorbers, it is helpful to summarize the basic properties of the KODIAQ-Z survey, in particular in terms of \nhi\ and redshift distributions as we will compare our sample of absorbers to other surveys (such as surveys of stronger \hi\ absorbers at similar redshift or absorbers with similar \nhi\ but at lower redshift). As we show below the  \nhi\ and Doppler parameter distribution  also motivate the use of photoionization in the ionization modeling to determine the metallicity of the absorbers. 

\subsection{\hi\ Column Density and Redshift Distributions}\label{s-res-nhi-z-dist}

\begin{figure}[tbp]
\epsscale{1.1}
\plotone{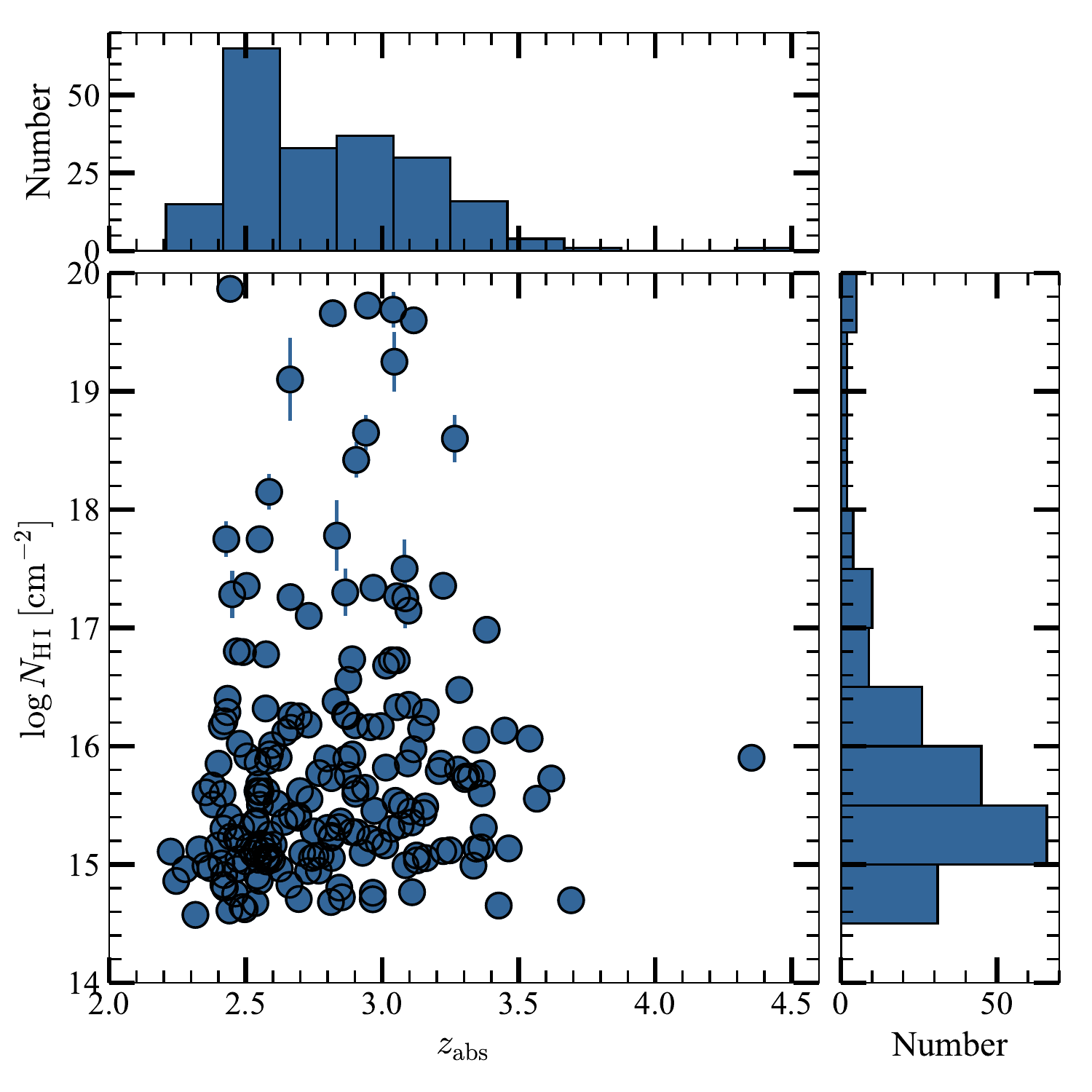}
\caption{Distribution of the redshifts and \hi\ column densities for the absorbers in the KODIAQ-Z sample. The top and right panels show the redshift and \lnhi\ distributions. 
\label{f-nhi-vs-z}}
\end{figure}

In Fig.~\ref{f-nhi-vs-z}, we show the distributions of the absorber redshifts and \hi\ column densities as well as the scatter plot between these two quantities. There are \ssz\ absorbers in KODIAQ-Z and their full \nhi\ and $z$ intervals are $14.43 \le \mlnhi \le 19.87$ and $2.23 \le \mzabs \le 4.35$. However, 99\% of the absorbers have $14.57 \le \mlnhi \le 19.87$ and $2.23 \le \mzabs \le 3.57$ with means and standard deviations $\langle \mlnhi \rangle = 15.8 \pm 1.1$ (the median being $\mlnhi = 15.5$) and $\langle \mzabs \rangle = 2.79 \pm 0.31$ (the median being $\mzabs = 2.75$), respectively. There is an evident peak in the redshift distribution at $2.4\la \mzabs \la 2.6$ where there are about 65 absorbers; otherwise between $\mzabs \simeq 2.6$ and $3.35$, the $z$-distribution is relatively flat with about 30--35 absorbers in each 0.2 redshift bin. This peak in the redshift distribution corresponds to redshifts where the optimal conditions are assembled to derive \nhi\ (access to the entire Lyman series transitions and smaller contamination from the LYAF): as $z$ decreases below 2.4, the number of \hi\ transitions diminishes rapidly; as $z$ increases above 3.6, the level of contamination from the LYAF increases (and indeed at $3.5 \la \mzabs \la 4.4$, only for 5 absorbers were we able to robustly derive \nhi).  As alluded to before, we use two other main surveys at $z>2$ for comparison with KODIAQ-Z: the HD-LLS survey \citep{prochaska15,fumagalli16} and the R12-DLA survey \citep{rafelski12}. Based on the redshift distribution of KODIAQ-Z, we limit these surveys to the redshift range $2.2 \le z \la 3.6$ when comparing the absorbers as a function of \nhi\ over the same redshift interval. 

Owing to the selection of the absorbers, the \nhi\ distribution should follow to some extent  the column density distribution of \hi\ \citep{prochaska14}, which is why the \nhi\ distribution is not flat. From approximately $10^{15}$ to $10^{16.5}$ \cmm, we see in Fig.~\ref{f-nhi-vs-z} the expected drop in the number of absorbers as \nhi\ increases. At $\mlnhi <15$, our sample is not complete as there is a sharp drop in the number of absorbers. This is because as \nhi\ decreases, we rely more and more on high S/N spectra in order to be sensitive to low metallicities, significantly reducing the sample. For absorbers with  $16.5 \la \mlnhi \la 17.5$, we have to depend more on weak Lyman series transitions to estimate \nhi, limiting the number of absorbers especially as $z$ increases  beyond $z\ga 3.2$. As described in \S\ref{s-estimate-hi}, absorbers with  $17.5 < \mlnhi \la 18.65$ are rare because they are essentially on the flat part of the COG, dramatically restricting the number of absorbers where we can estimate \nhi. And indeed although  absorbers with $18.7 \la \mlnhi \la 20$ should be less frequent than the previous category, there are more of these strong LLSs/SLLSs in our sample owing to that we can use the damping of \lya\ and \lyb\ to derive \nhi\ (see also \S\ref{s-estimate-hi}). 

\subsection{\hi\ Column Density and Doppler Parameter Distributions}\label{s-res-nhi-b-dist}

\begin{figure}[tbp]
\epsscale{1.1}
\plotone{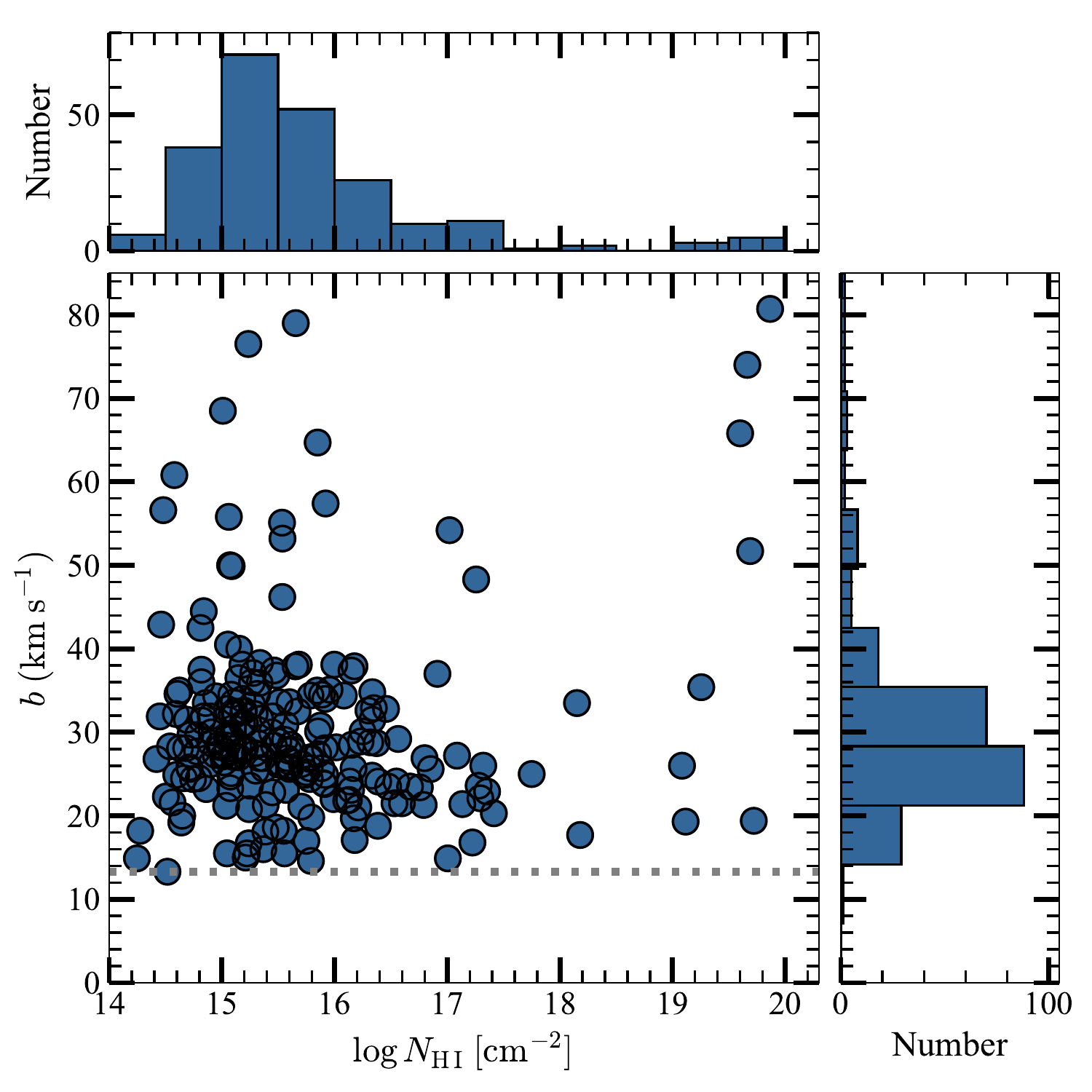}
\caption{Distribution of the Doppler parameters ($b$) and \hi\ column densities in the KODIAQ-Z sample obtained from the profile fitting of individual components. The right and top panels show the $b$ and \lnhi\ distributions. The dotted line shows the lowest fitted $b$-value (corresponding $T\la  10^4$ K); 90\% of the components have $13.3 \le b \le 40$ \km. Note that for the Keck HIRES spectra, the instrumental $b$-value is about 4--5 \km, well below the minimum $b$-value, i.e.,  narrower \hi\ absorbers than $b<12 $ \km\ could have been detected if present.
\label{f-b-vs-nhi}}
\end{figure}

In Fig.~\ref{f-b-vs-nhi}, we compare the Doppler parameter, $b$, and \nhi\ derived from the profile fitting in the individual components of \hi. There is no strong trend between $b$ and \nhi, although  broad components with $b>40$ \km\ are more frequent at $\mlnhi < 16$ ($b = 40$ \km\ implies $T<10^5$ K, which is the threshold $b$-value between broad and narrow \hi\ absorbers as defined by \citealt{lehner07}).\footnote{The Doppler parameter for \hi\ is related to the gas temperature via  $b_{\rm H\,I} = (2kT/m_{\rm H} + b^2_{\rm nt})^{0.5}  $, where $k$ is the Boltzmann constant, $T$ the temperature of the gas, $m_{\rm H}$ is the mass of hydrogen, and $ b_{\rm nt}$ is the (unknown) non-thermal component to the broadening.} We, however, note that the majority of these broad components are found in more complex velocity profiles (2 or more components) and/or lower SNRs spectra, where a broad component reduces the $\chi^2$.

Considering the entire sample, the mean and standard deviation of $b$ is $\langle b \rangle = 30 \pm 11$ \km\ and median is 28 \km. About 90\% of the components have $13.3 \le b \le 40$ \km\ where  $\langle b \rangle = 27 \pm 6$ \km\ and the median being 27 \km. That latter value corresponds to a temperature of the gas of $T<4 \times 10^4$ K, consistent with the gas being primarily photoionized. At $z<1$ for the CCC absorbers, the results are remarkably similar since 91\% of the profile-fitted absorbers have also $b<40$ \km\ and a mean of $28 \pm 8$ \km\ (\citetalias{lehner18} and Fig.~7 therein). This contrasts from the weaker LYAF absorbers with $\mlnhi \la 14$ where the median and mean $b$-values are systematically larger by 15\%--30\% at $z \la 0.5$ than at $1.5\la z \la 3.6$, implying a larger fraction of the low-$z$ IGM is hotter and/or more kinematically disturbed than the high-$z$ IGM \citep{lehner07}.

\section{Determining the Metallicity of the Absorbers}\label{s-ion-model}
\subsection{Motivations for Photoionization Modeling}\label{s-motivation}
Several empirical studies at both low and high redshifts have shown that SLFSs, pLLSs, and LLSs have properties that are consistent with the gas being predominantly photoionized (\citealt{prochaska99,prochaska17}; \citetalias{lehner13,lehner16,lehner18,wotta19}; \citealt{crighton13,fumagalli16,cooper21,zahedy21}). In \S\ref{s-res-nhi-b-dist}, we show the KODIAQ-Z absorbers have temperatures on average $T\la 4\times 10^4$ K, consistent with the gas being photoionized (the same result was found for the low redshift absorbers with similar \nhi, \citetalias{lehner18}). As illustrated in the velocity profiles provided in the appendix, when detected metal ions such as the suite of carbon ions (\cii, \ciii, \civ) or silicon ions (\siii, \siiii, \siiv) align well with \hi\ (although we emphasize again that sometimes metal ions owing to their smaller $b$ can have more than 1 component within the breadth of the \hi\ profiles even though the \hi\ absorption can be fitted with a single component---that also implies that the thermal broadening is a large contributor to the broadening). 

In contrast to the low redshift, for the ionizing radiation field and typical lower metallicities at $z \sim 2$--3 (about 0.1\%  solar or ${\rm [X/H]} = -2$, see below, \citetalias{lehner16,fumagalli16}), even strong transitions like \cii\ $\lambda$1334 and \siii\ $\lambda$1260 are often not detected, so we have to rely more on intermediate (\siiii, \ciii, \siiv) and high (\civ) ions to determine the metallicity. Because the current survey probes lower \nhi\ absorbers than our previous surveys at high redshift (\citealt{lehner14}; \citetalias{lehner16}), we find that the \ovi\ absorption can often be narrow and align with the absorption seen in \hi\ (and other lower metal-ions if present). This contrasts from the very broad and strong \ovi\ absorption frequently found in strong LLSs, SLLSs, and DLAs \citep{lehner14}. When \ovi\ is narrow, it is used to constrain the photoionization models, and we find {\it a posteriori}\ that \ovi\ is consistent with being produced by photoionization in many of these cases. 
 
 \begin{figure}[tbp]
\epsscale{1.1}
\plotone{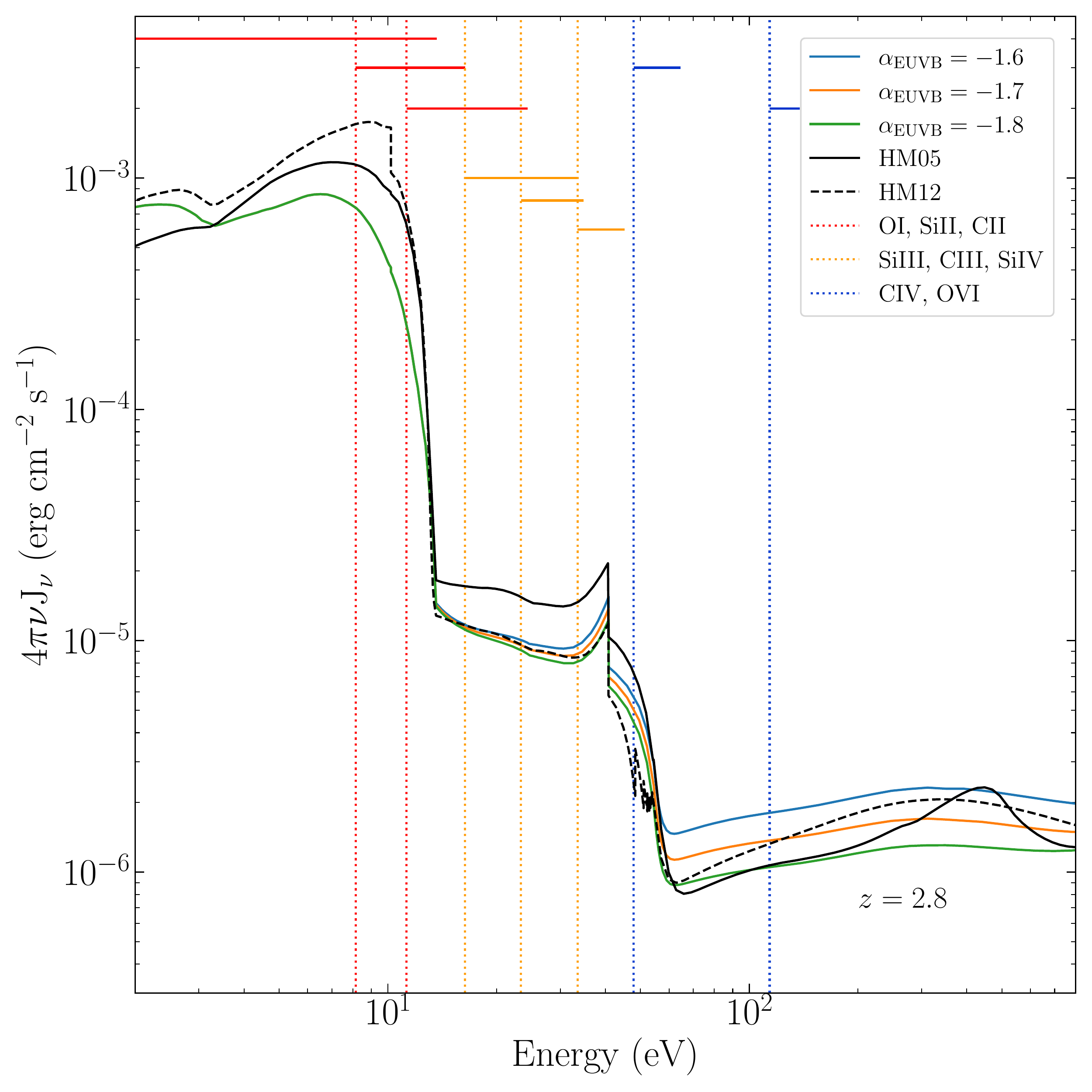}
\caption{Comparison of the several spectral energy distributions for the ionizing background radiation field. The HM05 is adopted in this work, although we systematically compare it with the results from HM12. We also show the recent EUVB models from \citet{khaire18} for comparison. The key ions used to constrain the metallicities are shown where the dotted vertical bars show their minimum ionization energy while their range of ionization energy is shown with the horizontal bar.  
\label{f-euvb}}
\end{figure}

\subsection{Photoionization Modeling}\label{s-photo-model}
To determine the metallicity for each absorber, we follow the same methodology and make similar assumptions described and discussed in \citetalias{wotta19} and \citetalias{lehner19}. As laid out in the previous section, owing to the higher redshift absorbers and the change in the EUVB, we include high ions in the photoionization modeling.  We model the photoionization using Cloudy \citep[version C13.02, see][]{ferland13}, assuming a uniform slab geometry in thermal and ionization equilibrium. In all cases the slab is illuminated with a Haardt--Madau EUVB radiation field from quasars and galaxies (HM05 and HM12, see \citealt{haardt96,haardt01,haardt12}). We adopt HM05 \citep[][as implemented in Cloudy]{haardt01} as the fiducial radiation field for KODIAQ-Z to allow for a direct comparison with the CCC results at lower redshift. However, we also use the HM12 EUVB to explore systematics associated with uncertainties in the radiation field. In Fig.~\ref{f-euvb}, we show the HM05 and HM12 EUVB at the average redshift of our survey ($z = 2.8$) along with the ionizing energy ranges for key ions used in KODIAQ-Z to constrain the metallicity of the absorbers. We also show in this figure the recent \citet{khaire18} EUVB models for their 3 favored slopes of the EUVB. While there are some differences between these EUVBs, these are not as large as at low $z$ \citep{gibson21}. And indeed we show {\it a posteriori}\ that the systematic errors arising from using a different EUVB are not as large as observed at $z\la 1$. We note that for stronger absorbers ($\mlnhi \ga 18$) at $z\sim 3$, \citet{fumagalli16}  show that local ionizing sources did not affect much the metallicity estimates, and therefore we do not expect that local sources of (galactic) ionizing radiation would drastically change the metallicities in KODIAQ-Z. We adopt the same grid of Cloudy models summarized in Table~3 in \citetalias{wotta19}. 

The main variables for the photoionization models are the ionization parameter---$U \equiv n_\gamma/n_{\rm H}={\rm H}$, the ionizing photon density/total hydrogen number density (neutral + ionized)---and the metallicity, $\xh$. We assume solar {\it relative}| heavy element abundances from \citet{asplund09} but allow for possible variation between between C and $\alpha$-elements (where $\alpha$ can be O or Si) since this ratio can be sensitive to nucleosynthesis effects (see, e.g., \citealt{cescutti09,mattsson10} for more detail). \ca\ is therefore not necessarily solar in the gas probed by absorbers studied here (see \citealt{aguirre04}, and also \citealt{lehner13}; \citetalias{wotta19} at $z<1$). 

To derive the metallicities, we compare the column densities for each absorber with a grid of photoionization models in a Bayesian context. We make use of the publicly available codes described in \citet[see also \citealt{prochaska17}]{fumagalli16} and \citetalias{wotta19}.\footnote{The code presented in \citet{fumagalli16} and \citetalias{wotta19} has been incorporated into the PyIGM Python package and is available at \url{https://github.com/pyigm/pyigm} \citep{prochaska17a}.} We perform the MCMC analysis on an absorber-by-absorber basis. For each absorber we apply Gaussian priors on \nhi\ and redshift\footnote{For the metal ions, the lower (upper) limits, the likelihoods are based on a rescaled Q-function (cumulative distribution function).}. For all the absorbers, we first assume a flat prior on $U$. After examination of the model outputs (in particular convergence of walkers, comparison of predicted and observed column densities, and ``corner plots'' showing the posterior PDFs), we can assess which absorbers can be reliably modeled with this flat prior on $U$ and which may require a Gaussian prior on \logU\ to help the models converge (see below). Similarly, during this first iteration, we also determine which absorbers can be modeled with a flat prior on $\ca$ (i.e., for which we could estimate $\ca$), which absorbers require $\ca = 0$ (e.g., those with no detections for all the ions or no carbon ions), and which absorbers may require a Gaussian prior on $\ca$ (absorbers with poorly constraints on carbon ions and/or $\alpha$-elements). 

\begin{figure}[tbp]
\epsscale{1.2}
\plotone{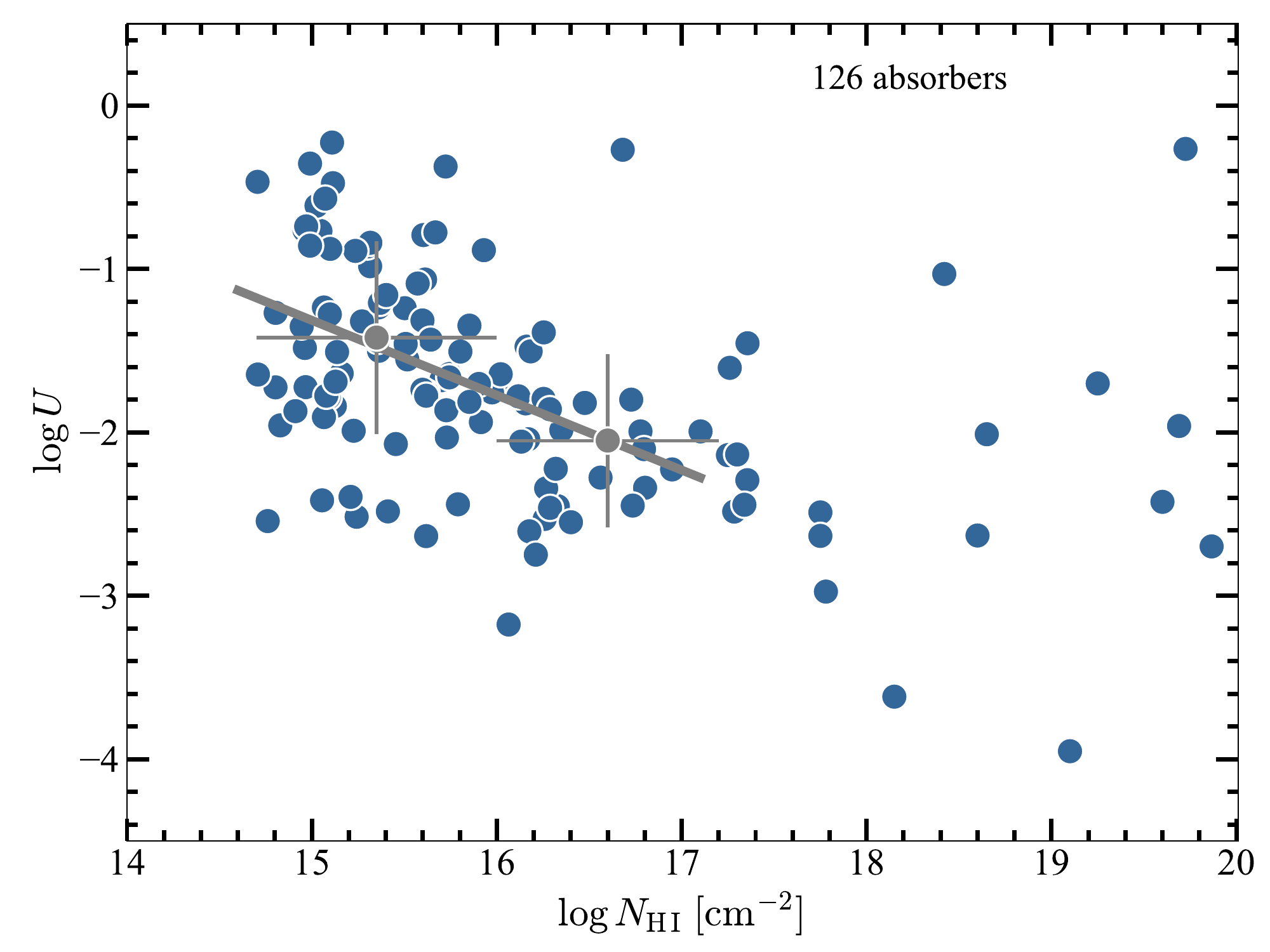}
\caption{The \logU\ median values as a function of \nhi\ for 126  absorbers in the KODIAQ-Z sample. The solid line shows a linear fit to the data with $14.6 \le \mlnhi \le 17.2$ (with a slope $-0.46 \pm 0.10$ and intercept $+5.56$). The crosses show the mean values of the median values in each interval \nhi\ interval indicated by the horizontal bar.
\label{f-logU-vs-h1}}
\end{figure}

When the constraints from the metal ions are not  optimal, we adopt a Gaussian prior on \logU\ \citepalias{wotta19}. These concern only SLFSs and pLLSs; the 16 LLSs and 7 SLLSs have enough metal-ion constraints for the ionization models to converge. Fifty eight percent  of the SLFSs and pLLSs do not require a  Gaussian prior on \logU. In Fig.~\ref{f-logU-vs-h1}, we show the \logU\ median values as a function of \nhi\ for the absorbers that can be modeled assuming a flat prior on \logU. As observed at low $z$, \citepalias{lehner19,wotta19}, a strong anti-correlation is observed between the ionization parameter and \nhi, with a slope steeper by factor 2.3 compared to the low redshift absorbers. To characterize the distribution of \logU\ appropriate for use as a Bayesian prior, we split the \nhi\ range into two bins, $14.4 \le \mlnhi<16$ and $16.2 \le \mlnhi<17.2$. In the former, 52\% (75/143) can be modeled with a flat \logU\ prior, while for the latter that number is 81\% (30/37). In both cases, we find each PDF is  well-fit with a normal distribution with $\langle \logU \rangle = -1.43 \pm 0.59$ and $-2.06 \pm 0.53$. We adopt these results for our Gaussian prior on \logU\ in the  $14.4 \le \mlnhi<16$ and $16.2 \le \mlnhi<17.2$ intervals, respectively. 

For the Gaussian prior on \ca, we similarly use absorbers with relatively good constraints from carbon and $\alpha$-element ions to determine a mean and standard deviation of $\left\langle \ca\right\rangle = -0.31\pm0.42$.  We apply these for the Gaussian prior on the \ca\ ratio for absorbers with poorly-constrained \ca.  We note that for a few (13) absorbers, \ca\ is clustered around $-1 \la \ca \la -0.9$. Not including those in the mean would of course increase the mean \ca\ value, but it would not change the posterior PDF of \ca\ for these absorbers. We discuss in \S\S\ref{s-abund-unc}, \ref{s-ca-met} in more detail the \ca\ ratio. 

We summarize the results of our photoionization modeling in Table~\ref{t-met-sum} for all the absorbers in the KODIAQ-Z sample. For each absorber, we list the sightline name, the absorber redshift ($z_{\rm abs}$), \lnhi, metallicity, ionization parameter ($\log U$), and \ca\ ratio (when estimated). Each quantity is reported with the 68\% CI and median values, except in the cases where we derive an upper or lower limit, in which case we report the 80\% CI and median. We emphasize that these values are attempts to summarize the posterior PDFs provided by the MCMC sampling of the Cloudy models. While those PDFs can be well-behaved and nearly Gaussian, some are not as well behaved (see corner plots in the appendix). 

For the HD-LLS absorbers used to compare with KODIAQ-Z, we apply the same overall methodology and same HM05 EUVB in the photoionization models in order to reduce any systematic errors between the two surveys. We, however, note that the absorption in absorber in the HD-LLS and KODIAQ-Z is not treated in the same way. In the HD-LLS survey, a  velocity interval of $\pm 500$ \km\ defines an absorber, which is driven by the methods to derive \nhi\ that use either the break at the Lyman limit or the damped wings from \lya. For the metal lines, they sum the absorption components identified within that interval even though it is rarely larger than 200 \km. As detailed in \S\ref{s-estimate-col}, we define an absorber using the Lyman series and  as much as possible we consider the absorption component by component, so in a window of $\pm 200$--500 \km, KODIAQ-Z can have several absorbers. However, as \nhi\ increases and overlaps with the HD-LLS survey, i.e., for $\mlnhi \ga 18$, we similarly lose our ability to derive the column densities in individual components, especially when we have information only from \lya\ and \lyb. In these cases, we rely on the low ions to define the integration range in the intermediate and low ions. In our approach, we make the physically motivated assumption that the optical depths of the low ions follow that of the stronger \hi\ absorption while the higher ions follow better weaker \hi\ absorbers that may not contribute much to the total \hi\ column---but the extra absorption in the higher metal-ions could dominate their column density (\S\ref{s-estimate-col}). However, {\it a posteriori}\ we will show that this different treatment does not lead to large different results.

\subsection{Uncertainties on the Metallicities}\label{s-uncertainties}
\begin{figure}[tbp]
\epsscale{1.1}
\plotone{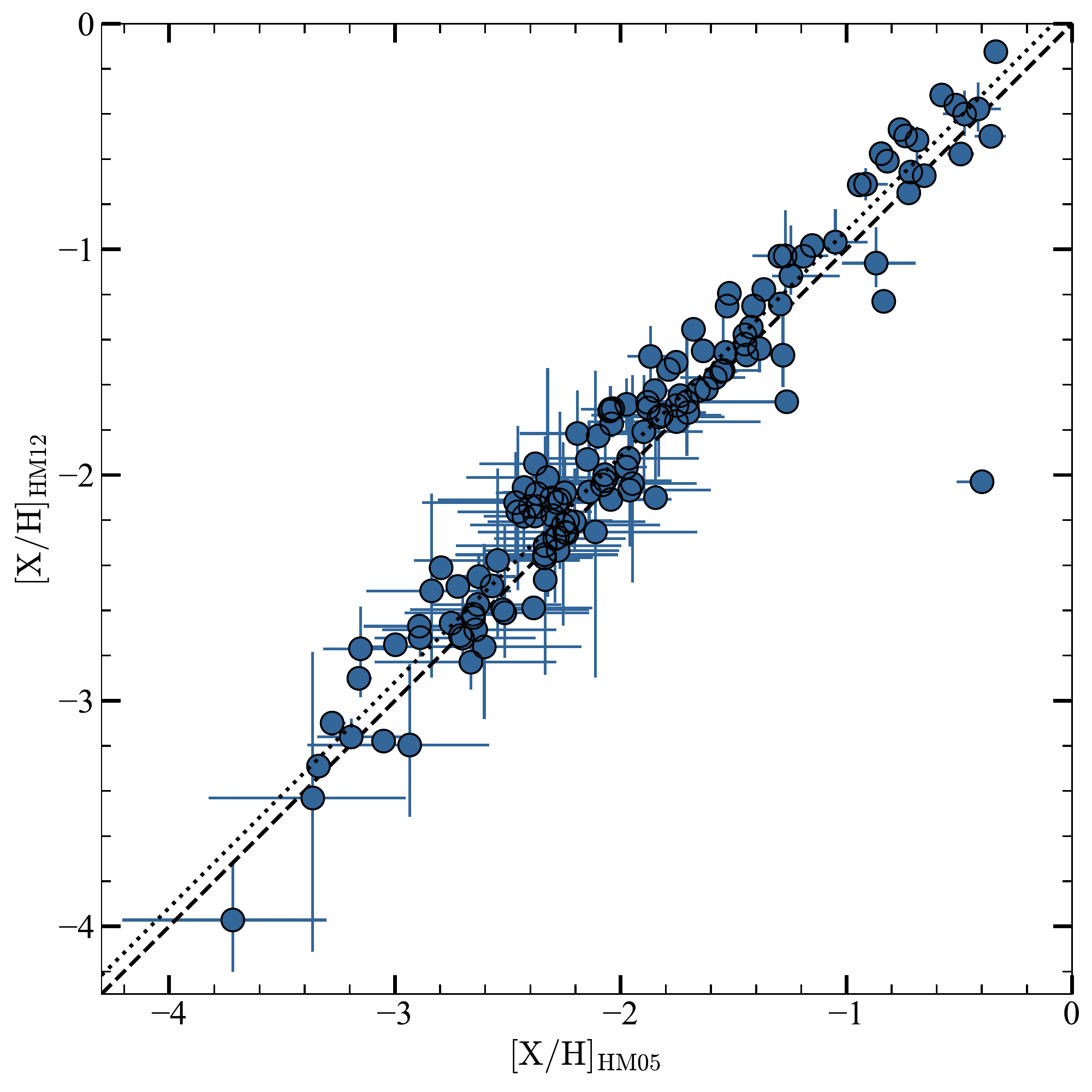}
\caption{Comparison of the median metallicities (with 68\% CI) of  KODIAQ-Z derived using the HM12 and HM05  EUVBs. The dashed line is the 1:1 relationship. The dotted line shows the mean of the differences between the metallicities derived using HM12 and HM05. 
\label{f-hm12vshm05}}
\end{figure}

All the errors for the output quantities (e.g., metallicity) reported in this work are statistical errors from the Bayesian MCMC ionization modeling. There are additional systematic and other uncertainties that we discuss here. 

\subsection{Ionizing Background}\label{s-euvb-unc}
The first one is caused by the uncertainty in the ionizing EUVB as alluded to previously. At low redshift, \citetalias{wotta19} shows that on average the metallicity is systematically shifted on average by about $+0.4$ dex for the SLFSs/pLLS and $+0.2$ dex for the LLSs when the ionizing EUVB changes from HM05 to HM12. We have done a similar experiment with KODIAQ-Z and HD-LLS absorbers showing that the impact of the EUVB is overall smaller at high $z$ than observed at low $z$. In Fig.~\ref{f-hm12vshm05}, we show the comparison between the metallicities using HM12 and HM05 for the KODIAQ-Z absorbers where we did not apply a Gaussian prior on $\log U$ (that essentially removes all the limits, although including those would not change the results). The immediate conclusion from this figure is that the systematic differences between the metallicities derived using HM05 and HM12 are quite small and much smaller than observed at low $z$. On average considering all the absorbers, we find  $\langle \xh_{\rm HM12} - \xh_{\rm HM05}\rangle =  +0.10 \pm 0.17$. The  difference is also not as systematic as low $z$ since there are several cases where $\xh_{\rm HM12}\le \xh_{\rm HM05}$. 

\begin{figure}[tbp]
\epsscale{1.1}
\plotone{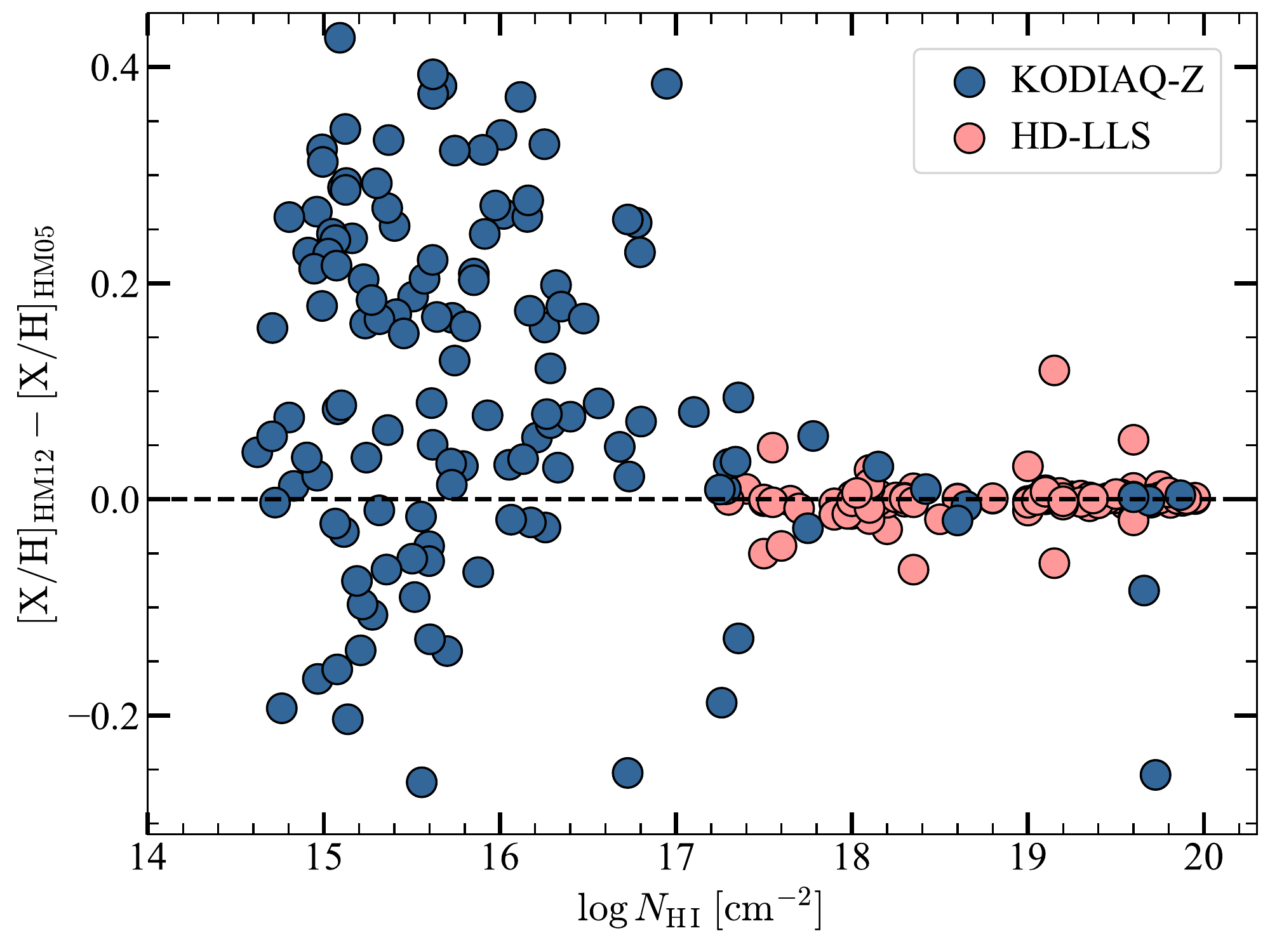}
\caption{Difference between the median metallicities for the KODIAQ-Z and HD-LLS absorbers derived using the HM12 and HM05 EUVBs against the \hi\ column densities of the absorbers. 
\label{f-hm12-hm05vsnh1}}
\end{figure}

As illustrated in Fig.~\ref{f-hm12-hm05vsnh1} where the differences of the median metallicities between HM12 and HM05 vs.\ \nhi\ are shown for both the KODIAQ-Z and HD-LLS absorbers, that difference decreases with increasing \nhi\ as observed for the low redshift absorbers. In fact, for $\mlnhi \ga 17.2$, $\langle \xh_{\rm HM12} - \xh_{\rm HM05}\rangle  \simeq 0.0 \pm 0.1$, i.e., the HM12 and HM05 EUVBs essentially give the same metallicities and any small differences are essentially random. For the SLFSs and pLLSs, $\langle \xh_{\rm HM12} - \xh_{\rm HM05}\rangle \simeq  +0.12 \pm 0.15$. These results are not entirely surprising when considering the different EUVBs in Fig.~\ref{f-euvb} with their difference being smaller than at low $z$ over similar ionizing energies \citep{gibson21}. Based on Fig.~\ref{f-euvb}, the metallicities would not change much if we had used the more recent EUVBs from \citet{khaire18}. For comparison with CCC, we adopt hereafter the results from the HM05 EUVB.

\subsection{Multiple Ionization Phases}\label{s-multi-unc}
Throughout we also assume a single ionization phase. In cases where a higher ion (\ovi\ in our survey) is clearly under produced by the photoionizing model, this ion is not included in the final model. In that latter case, this would be evidence for an absorber probing multiple gas-phases. As shown by \citet{lehner14}, this can especially happen in the strong LLSs where \ovi\ can be very strong and can rarely be modeled with a single gas-phase ionization model with the lower ions. In these cases, the velocity profiles of the \ovi\ profiles are also very different from those of the lower ions (including \civ), strongly hinting at multiple gas-phases. However, as we explore lower \nhi\ absorbers in this present survey, we find in many cases that a single ionization gas-phase can match well the observed metal-ion column densities, including those of \ovi. In these cases and in contrast to stronger \hi\ absorbers, the absorption is also often quite simple (typically a single component) and the velocity profiles of the different metal ions align well with each other (including \ovi\ when present) and with the \hi\ velocity profiles. This, of course, does not necessarily mean that a more complex ionization structure might not be present, but a  single gas-phase model is often sufficient to reproduce the observed column densities, especially in the SLFS and pLLS regimes. We adopt here therefore the simplest approach and assumption, a methodology also employed in CCC. Nevertheless, we note that at least at low redshift,  the effects on the metallicities between single versus multiple gas-phase modelings typically lead to small changes in metallicities of the order of $<0.1$--0.2 dex \citep{howk09,haislmaier21}. Therefore, we do not expect that considering a more complex ionization structures for the absorbers in our sample would drastically change the metallicities. 

\begin{figure}
\epsscale{1.2}
\plotone{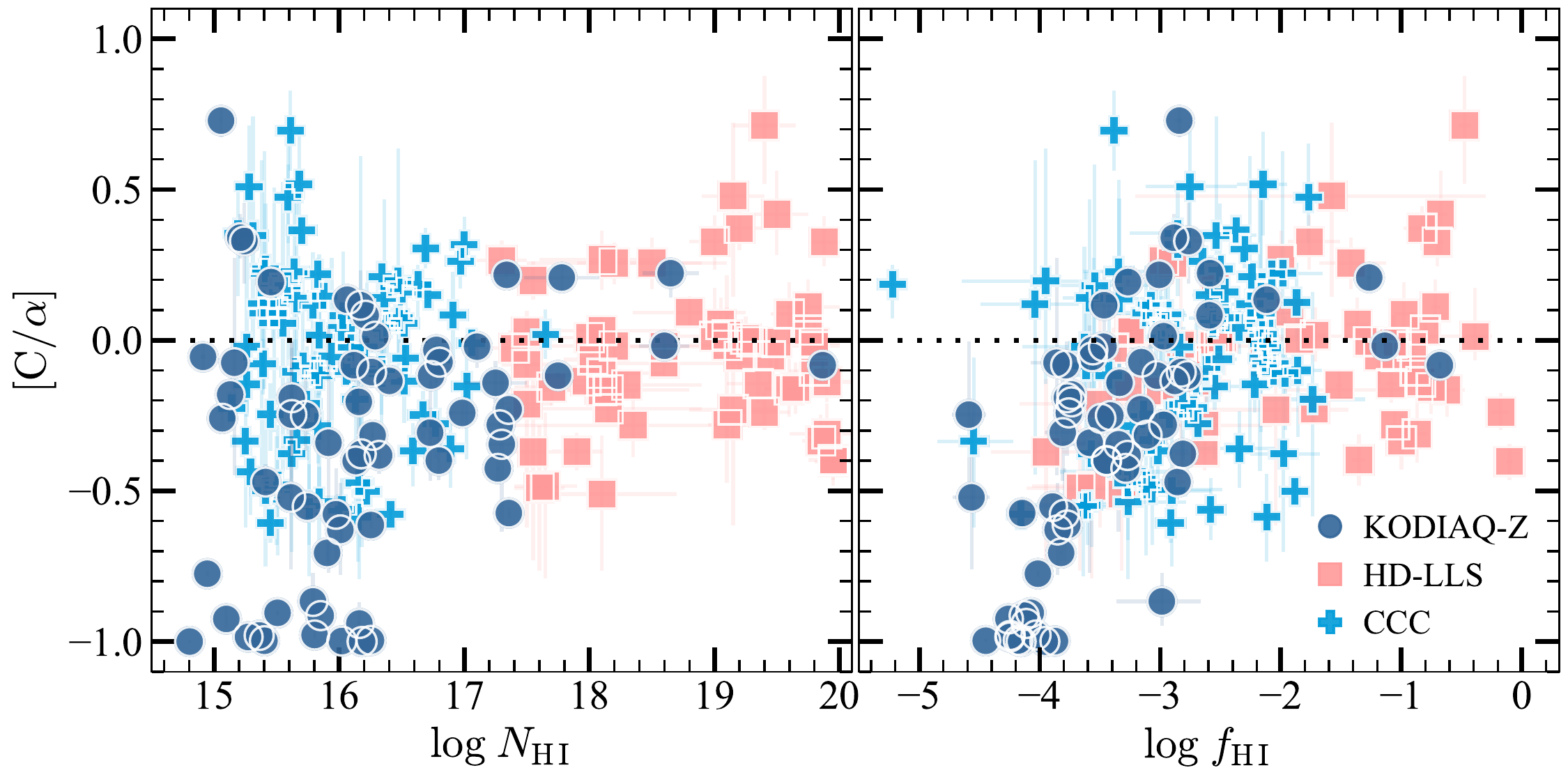}
\caption{The ratio of \ca\ against the \hi\ column density ({\it left}) and neutral fraction ({\it right}) from the KODIAQ-Z ($2.2 \la z \la 3.6$), HD-LLS ($2.2 \la z \la 3.6$), and CCC ($z\la 1$) surveys. The median values of the \ca\ posterior PDFs are adopted as the central values with 68\% CI. 
\label{f-calpha-nhi}}
\end{figure}

\subsection{Non-solar Relative Abundances}\label{s-abund-unc}

Other uncertainties may arise from dust or non-solar relative abundances. Since carbon, silicon, and oxygen ions are typically used to constrain the ionization modeling at high redshift, we explicitly take account for possible non-solar \ca\ by letting this ratio vary in the range $-1 \le \ca \le +1 $.  In Fig.~\ref{f-calpha-nhi}, we show \ca\ as a function of \nhi\ (left panel) and the neutral fraction $f_{\rm H\,I} \equiv \mnhi/(\mnhi+\mnhii)$ (right panel) for the KODIAQ-Z and HD-LLS absorbers. We also show in this figure the results from CCC at $z\la 1$ \citepalias{lehner19}. For \ca-\nhi, a Spearman rank-order test shows no strong evidence for a correlation between \ca\ and \nhi\ with a correlation coefficient $r_{\rm S} = 0.14$ and a $p$-value\,$=0.04$. On the other hand, the same statistical test shows a positive monotonic correlation between \ca\ and $f_{\rm H\,I}$ with a correlation coefficient $r_{\rm S} = 0.44$ and a $p$-value\,$ \ll 0.05\%$. However, removing the 13 KODIAQ-Z absorbers with low $\ca \la -0.9$ eliminates any evidence of correlation  between \ca\ and $f_{\rm H\,I}$ with a correlation coefficient $r_{\rm S} =-0.05$ and a $p$-value\,$=0.88$. 

So far the CCC, HD-LLS, and KODIAQ-Z were treated as a single sample. Considering the samples individually would not change the conclusions for the lack of correlation between \ca\ and \nhi. However, while neither CCC nor HD-LLS shows any correlation between  \ca\ and $f_{\rm H\,I}$, for KODIAQ-Z, the Spearman rank-order test shows a strong positive monotonic correlation between \ca\ and $f_{\rm H\,I}$ with a correlation coefficient $r_{\rm S} = 0.72$ and a $p$-value\,$ \ll 0.05\%$. In that case, removing the 13 absorbers with low $\ca \la -0.9$ still yields $r_{\rm S} = 0.55$ and a $p$-value\,$ \ll 0.05\%$. Therefore in the KODIAQ-Z sample, \ca\ is more affected by ionization correction than in the other samples, most likely because often only \civ\ is detected (the limit on \cii\ is often not very constraining and \ciii\ is often contaminated) and larger ionization correction are applied (for CCC, \cii\ and \ciii\ are used to constrain \ca, while for HD-LLS the ionization corrections are typically smaller). Based on CCC and HD-LLS where we always find $\ca \ge -0.6$, \ca-values with $\ca <  -0.6$ are most likely predominantly caused by ionization rather than nucleosynthesis effects. Excluding the 13 absorbers with low $\ca \la -0.9$, we find that $\langle \ca \rangle = -0.20 \pm 0.31$ in the KODIAQ-Z sample compared to  $\langle \ca \rangle = -0.05 \pm 0.30$ in the CCC survey and  $\langle \ca \rangle = -0.05 \pm 0.24$ in the HD-LLS survey. The similarity between CCC and HD-LLS is quite remarkable since different $z$, \nhi, and \xh\ are probed. The $1\sigma$ dispersion is about the same in the three surveys, but on average the KODIAQ-Z sample has a systematic effect of $-0.15$ dex on \ca\ most likely caused by  ionization. In conclusion, the variable C/$\alpha$ is important in that we can marginalize over that as a ``nuisance" variable in order to get a better sense of the uncertainties in the metallicity, but it may not provide deep insights on the nucleosynthesis effects that may affect C/$\alpha$ (see also \S\ref{s-ca-met}).

Finally, we note that for stronger LLSs and SLLSs as those in the HD-LLS survey \citep{fumagalli16}, dust depletion is found to be a small effect for strong LLSs and even SLLSs at high $z$, and hence not very likely to have any appreciable effect on the metallicity estimates, especially considering the typical species---carbon, silicon, oxygen---used to constrain the ionization models are known not to be strongly depleted onto dust (e.g., \citealt{savage96,jenkins09,jenkins17}).

\section{Metallicity of the IGM/CGM Absorbers at $2.2\la \lowercase{z} \la 3.6$}\label{s-metallicity}

\subsection{Metallicity Distributions}\label{s-met-pdf}
\begin{figure}[tbp]
\epsscale{1.15}
\plotone{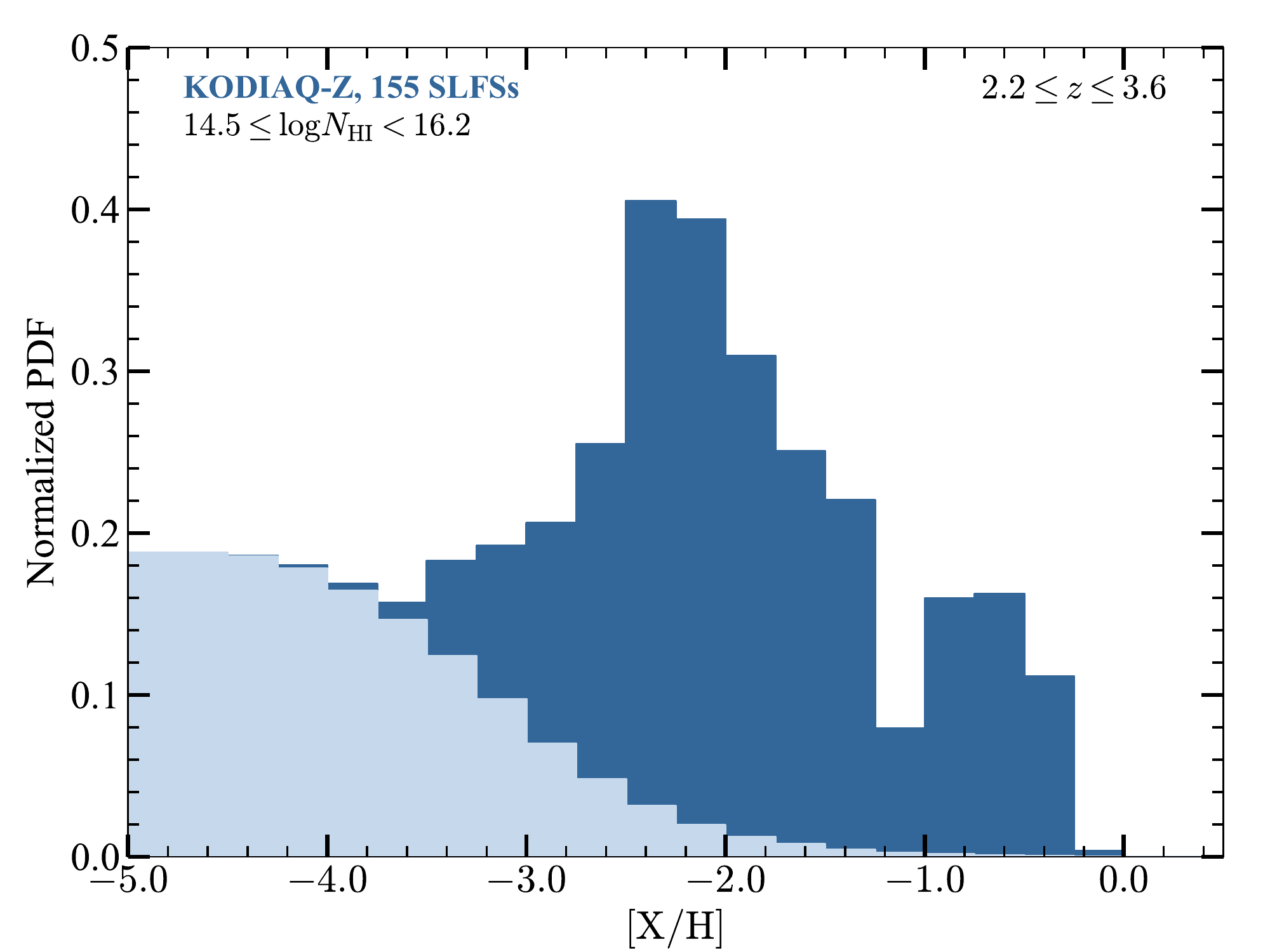}
\caption{Posterior metallicity PDF of the SLFSs in the KODIAQ-Z sample. The light-colored regions indicate the contribution from the upper limits. \label{f-pdf-slfs}}
\end{figure}

\begin{figure}[tbp]
\epsscale{1.15}
\plotone{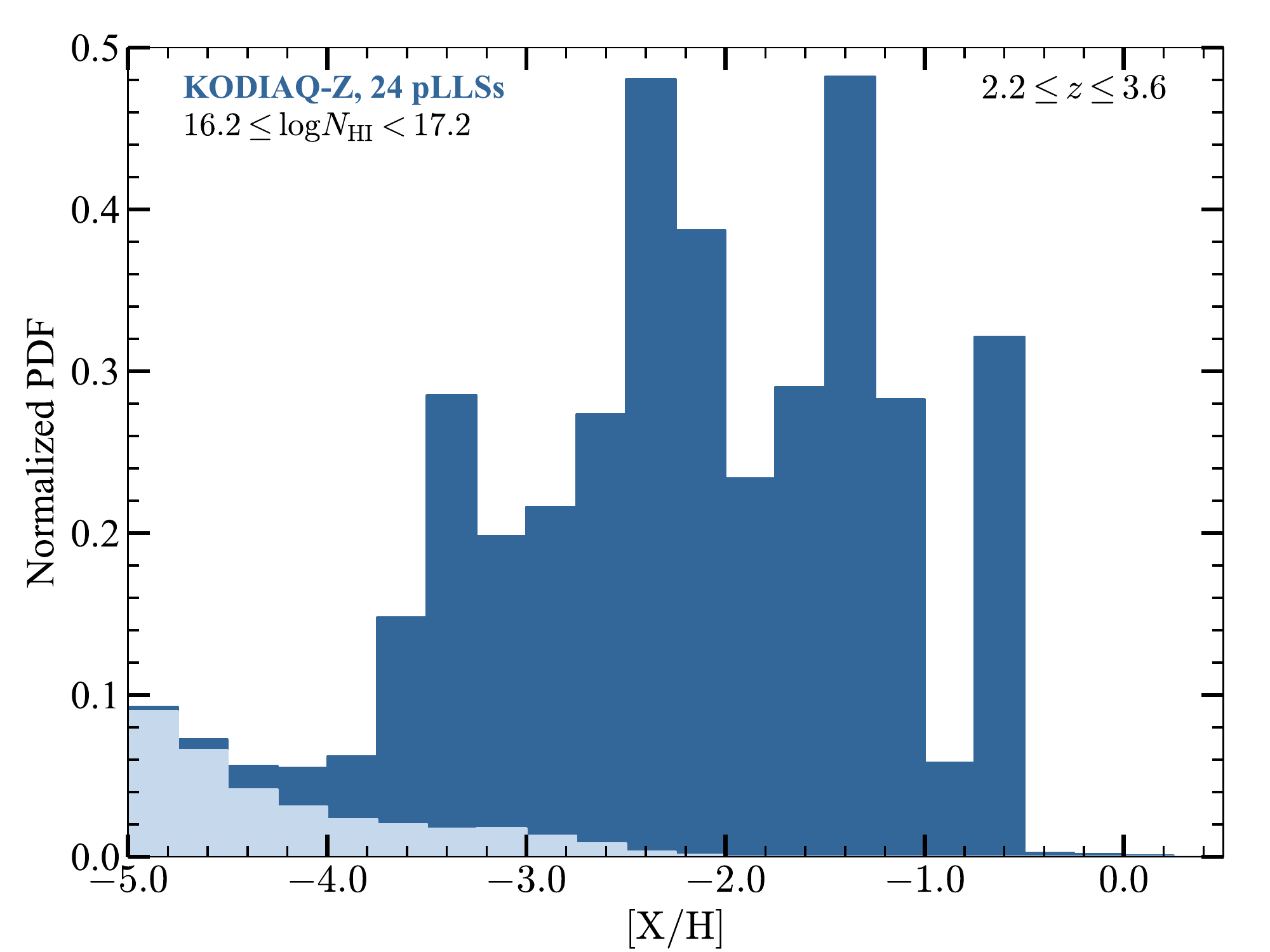}
\caption{Posterior metallicity PDFs of the pLLSs in the KODIAQ-Z sample. The light-colored regions indicate the contribution from the upper limits. \label{f-pdf-plls}}
\end{figure}
\begin{figure}[tbp]
\epsscale{1.15}
\plotone{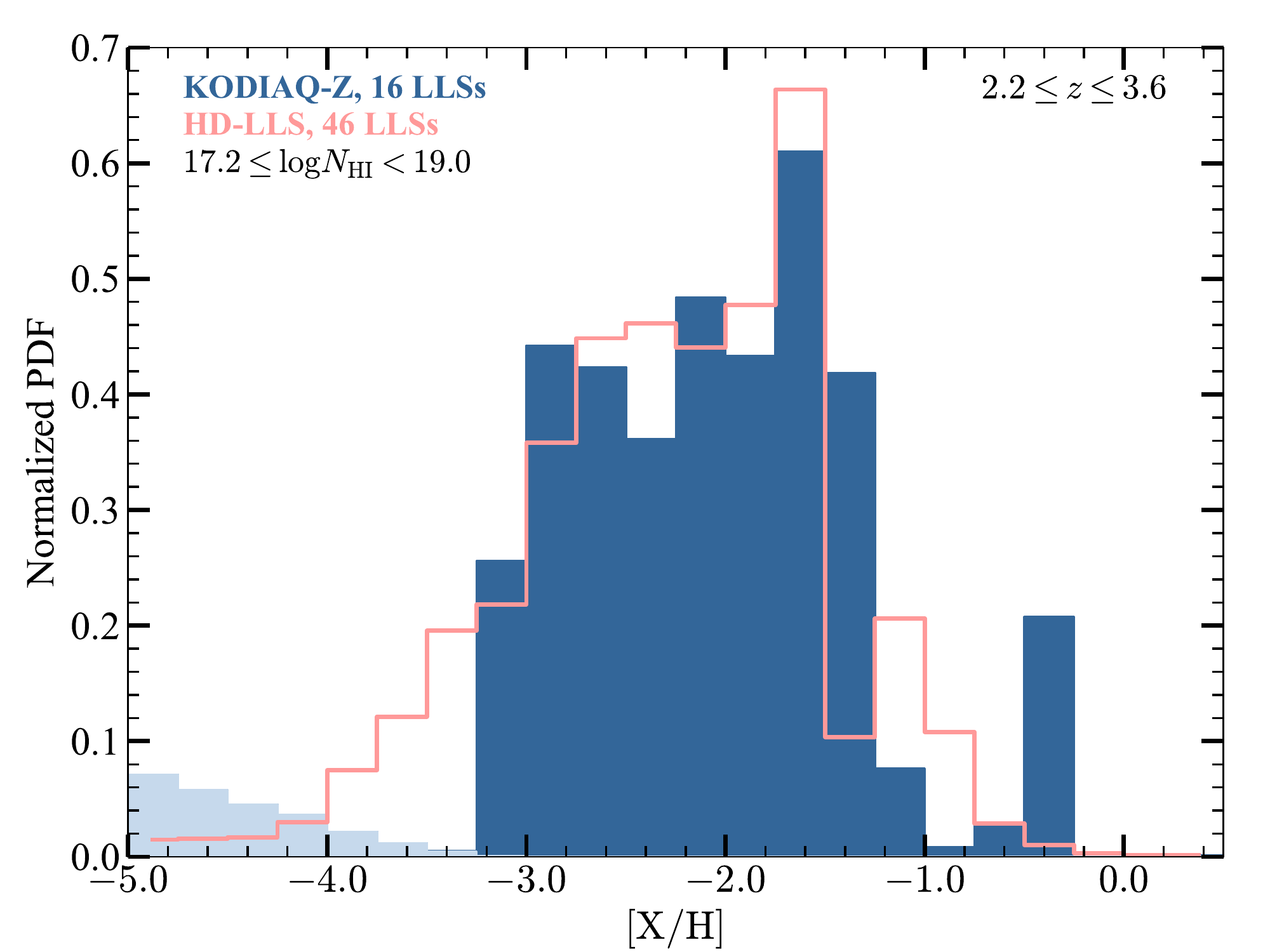}
\caption{Comparison of the posterior metallicity PDFs of the LLSs in the KODIAQ-Z and HD-LLS sample. The light-colored regions indicate the contribution from the upper limits. \label{f-pdf-lls}}
\end{figure}

In Figs.~\ref{f-pdf-slfs}, \ref{f-pdf-plls}, and \ref{f-pdf-lls}, we show the posterior probability distribution functions (PDFs) of the metallicities of the SLFSs, pLLSs, and LLSs for KODIAQ-Z. In Fig.~\ref{f-pdf-lls}, we also overplot the LLSs from HD-LLS survey, showing similar PDFs between the two surveys, and hence implying we can combine both samples to improve the statistics for the LLSs. These posterior PDFs are constructed by combining the normalized metallicity PDFs of all of the absorbers within a given \hi\ column density regime. In Table~\ref{t-met-sum}, we summarize the medians, means, standard deviations, and interquartile ranges (IQRs) for the various absorbers as well as a combination of these. We also include in this table the results for the HD-LLS, KODIAQ-Z+HD-LLS, and R12-DLA samples. 

\begin{figure}[tbp]
\epsscale{1.15}
\plotone{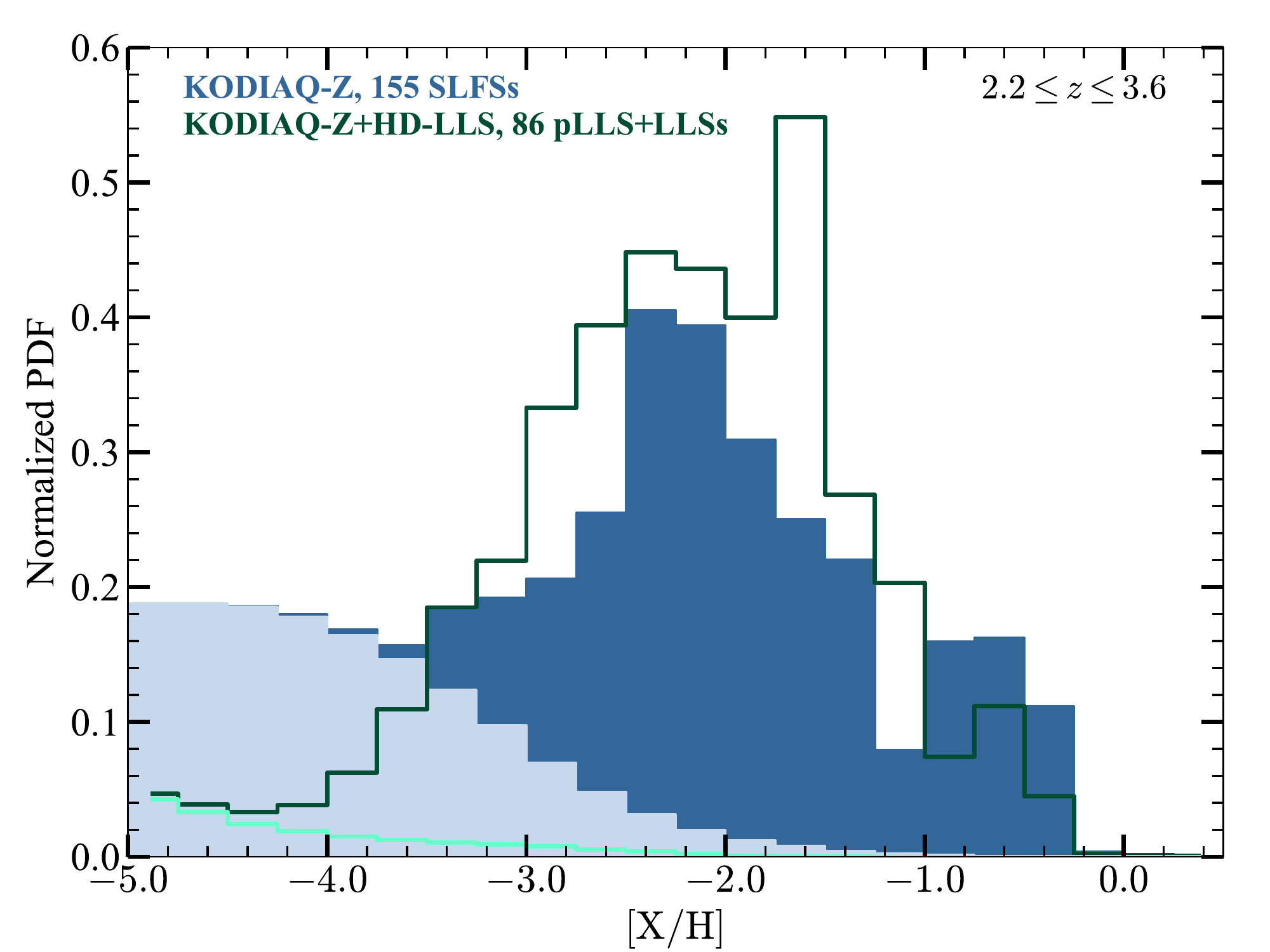}
\caption{Comparison of the posterior metallicity PDFs of the SLFSs (KODIAQ-Z) and pLLSs+LLSs (KODIAD-Z+HD-LLS). The light-colored regions indicate the contribution from the upper limits. \label{f-pdf-SLFS-LLS}}
\end{figure}

Considering first the SLFSs, the  metallicity PDF is not unimodal. It is negatively skewed with a long and prominent tail extending well below $\xh <-3$. Its main peak is around the median value of $-2.4$ dex. The metallicity PDF of the SLFSs dips around $-1.1$ dex with a second weaker peak around $-0.6$ dex. The IQR is nearly 2 dex from $-3.6$ to $-1.8$ dex. As \nhi\ increases, the overall median metallicity increases and the IQR decreases. For pLLSs+LLSs, the median metallicity is $-2.2$ and IQR only $\sim$1 dex from $-2.8$ to $-1.7$ dex. These features are distinctly illustrated in Fig.~\ref{f-pdf-SLFS-LLS} where we compare the PDFs of the SLFSs and pLLSs+LLSs: there are more frequently pLLSs+LLSs in the metallicity range $-3.2 \la \xh \la -1.2$ than SLFSs, but SLFSs become more frequent at extremely low metallicity $\xh \la -3.5$. Compared to the SLFS and pLLS PDFs, the LLS PDF is not as negatively skewed with about the same values for the mean and median.  

\begin{figure}[tbp]
\epsscale{1.15}
\plotone{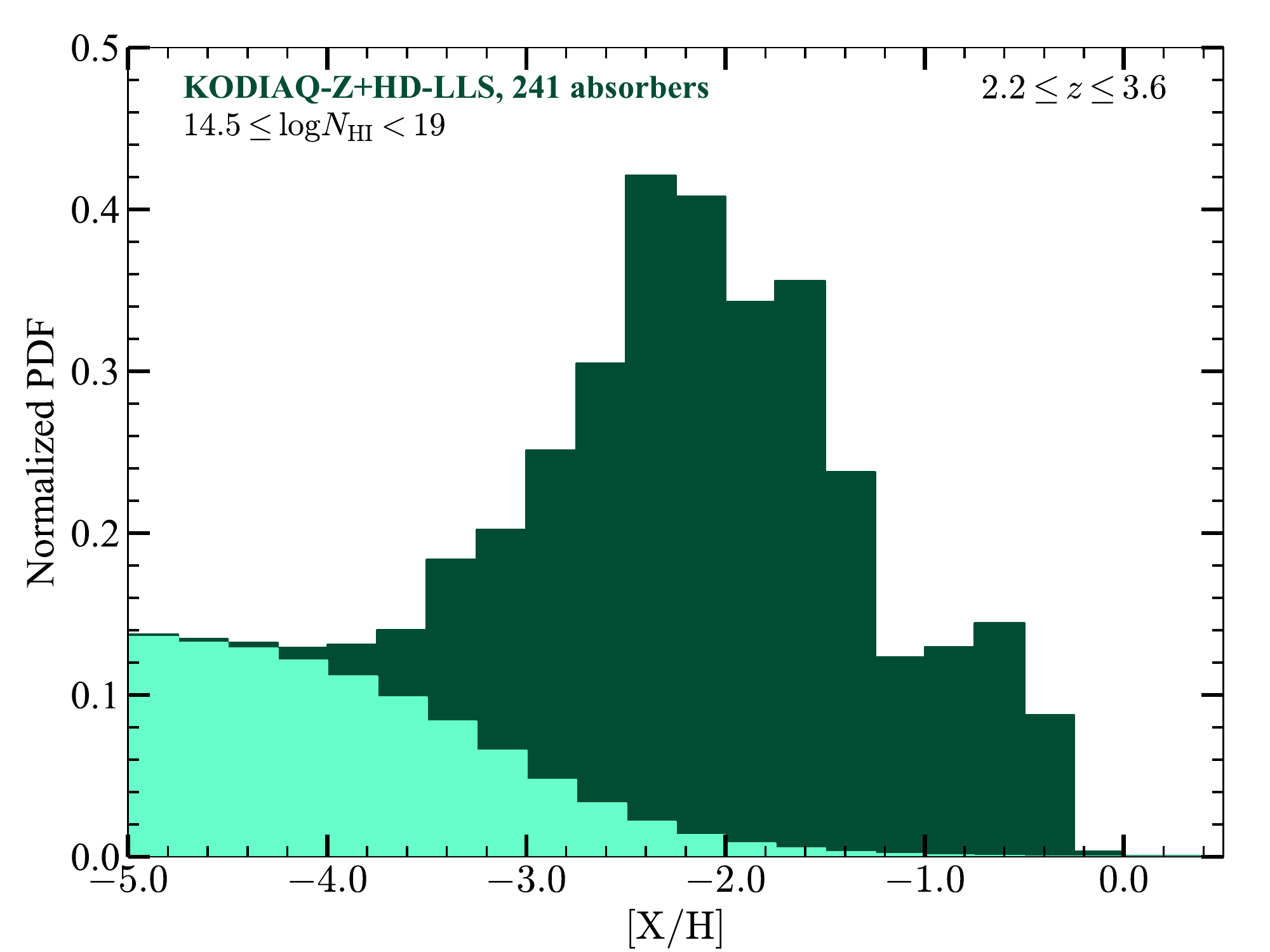}
\caption{Posterior metallicity PDF of the absorbers with $14.5 \la \mlnhi <19$ in the combined sample of KODIAQ-Z and HD-LLS. The light-colored regions indicate the contribution from the upper limits. \label{f-mdf-absorber}}
\end{figure}

For completeness and if one ignores the \hi\ column density dependence of the metallicity PDFs between the SLFSs, pLLSs, and LLSs, we show in Fig.~\ref{f-mdf-absorber} the resulting metallicity PDF of the absorbers with  $14.5<\mlnhi < 19$ at $2.2\la  z\la 3.6$. Since the SLFSs are about 2/3 of the sample, the metallicity PDF of the $14.5<\mlnhi < 19$ absorbers is not too different to that of the SLFSs with a similarly skewed distribution toward low metallicities, but also a second weaker peak around $-0.6$ dex. 

\subsection{Metallicity vs. \nhi}\label{s-met-nhi}
As shown in the previous section, there is already evidence for an overall increase of the metallicities with \nhi\ in the range $14.5 \le \mlnhi \le 19$. We now explore  further how the metallicity varies with \nhi\ considering the higher \nhi\ absorbers (the SLLSs and DLAs). From Table~\ref{t-met-sum}, the trend described above with increasing median/mean metallicity  with increasing \nhi\ continues to apply in the SLLS and DLA regimes. The IQR and standard deviation for the SLLSs are about similar to that of pLLSs+LLSs, but those for the DLAs are substantially smaller (the IQR is 0.65 dex smaller than that of the SLFSs). The negative skewness is also less pronounced for the SLLSs and absent for the DLAs. In fact, the DLA metallicity PDF, contrary to the PDFs of the other weaker absorbers, is fully consistent with a Gaussian distribution according to the Kolmogorov–Smirnov (KS) test with a $p$-value $\ll 0.1\%$ (see also \citealt{rafelski12}). 

\begin{figure}[tbp]
\epsscale{1.15}
\plotone{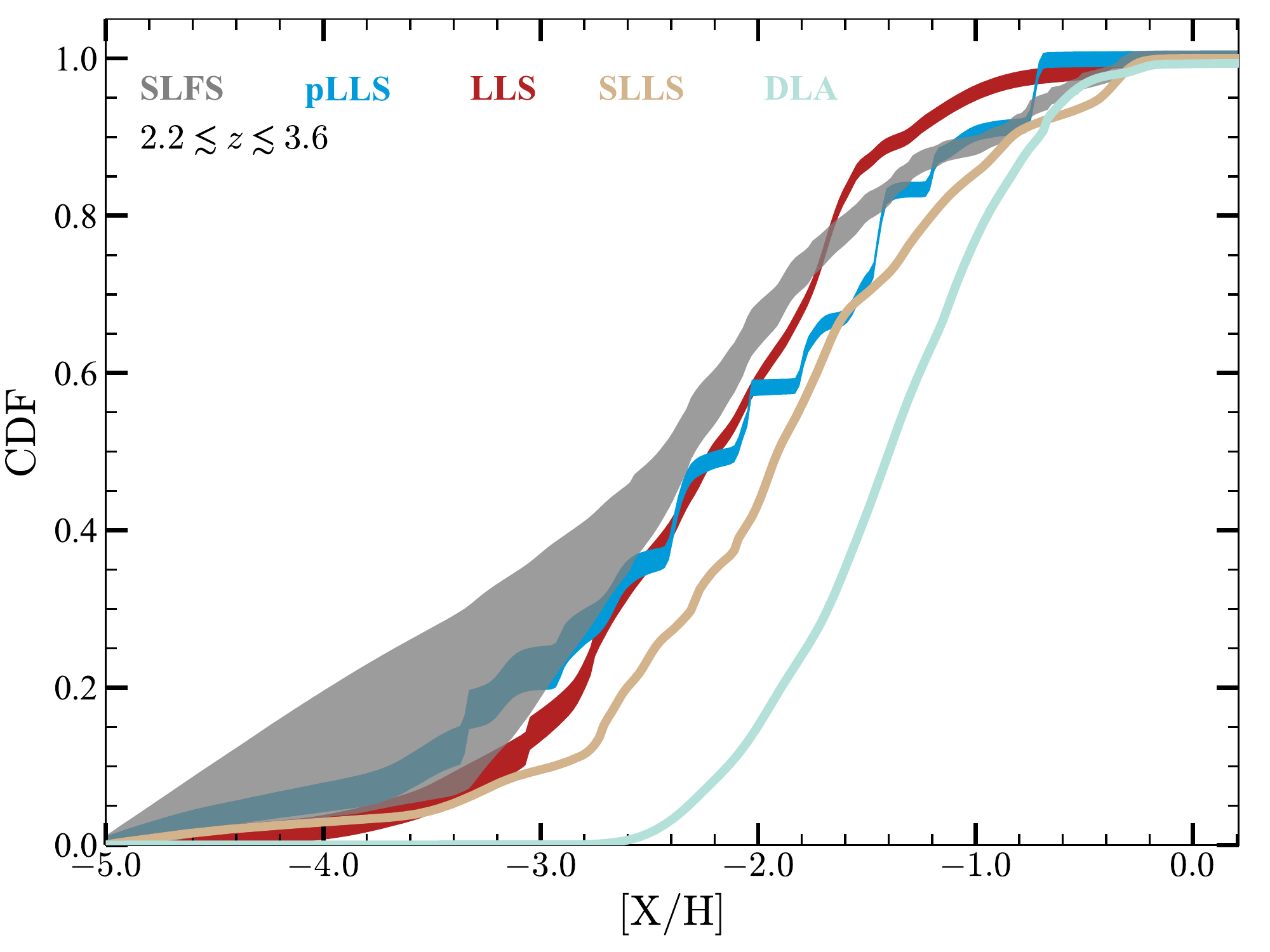}
\caption{Cumulative probabilities of the SLFS, pLLS, LLS, SLLS, and DLA metallicity PDFs. The SLFSs and pLLSs are from KODIAQ-Z, the LLSs and SLLSs from KODIAQ-Z and HD-LLS, and the DLAs from R12-DLA. \label{f-cdf}}
\end{figure}

These conclusions are strengthened considering  the cumulative distribution functions (CDFs) of the metallicity PDFs (see \citetalias{wotta19} for the description of the estimation of the CDFs including upper and lower limits). We show in Fig.~\ref{f-cdf} the metallicity CDFs for the SLFSs, pLLSs, LLSs, SLLSs, and DLAs. With only 24 pLLSs, this regime remains the least sampled. Nevertheless, the CDF confirms the overall evolution of the metallicities as \nhi\ increases: very metal poor (VMP) absorbers with $\xh \le -2.4$ are more likely to be found in the SLFS, pLLS, LLS regimes than in the SLLS or DLA regime ($\xh_{\rm VMP} = -2.4$ is the $2\sigma$ lower bound of the DLA metallicities). Notably the median metallicity of the SLFS is right at that threshold of $\xh_{\rm VMP} = -2.4$, implying that 50\% of these absorbers are VMP while only $<5\%$ of the DLAs may be VMP by definition.    

\begin{figure}[tbp]
\epsscale{1.1}
\plotone{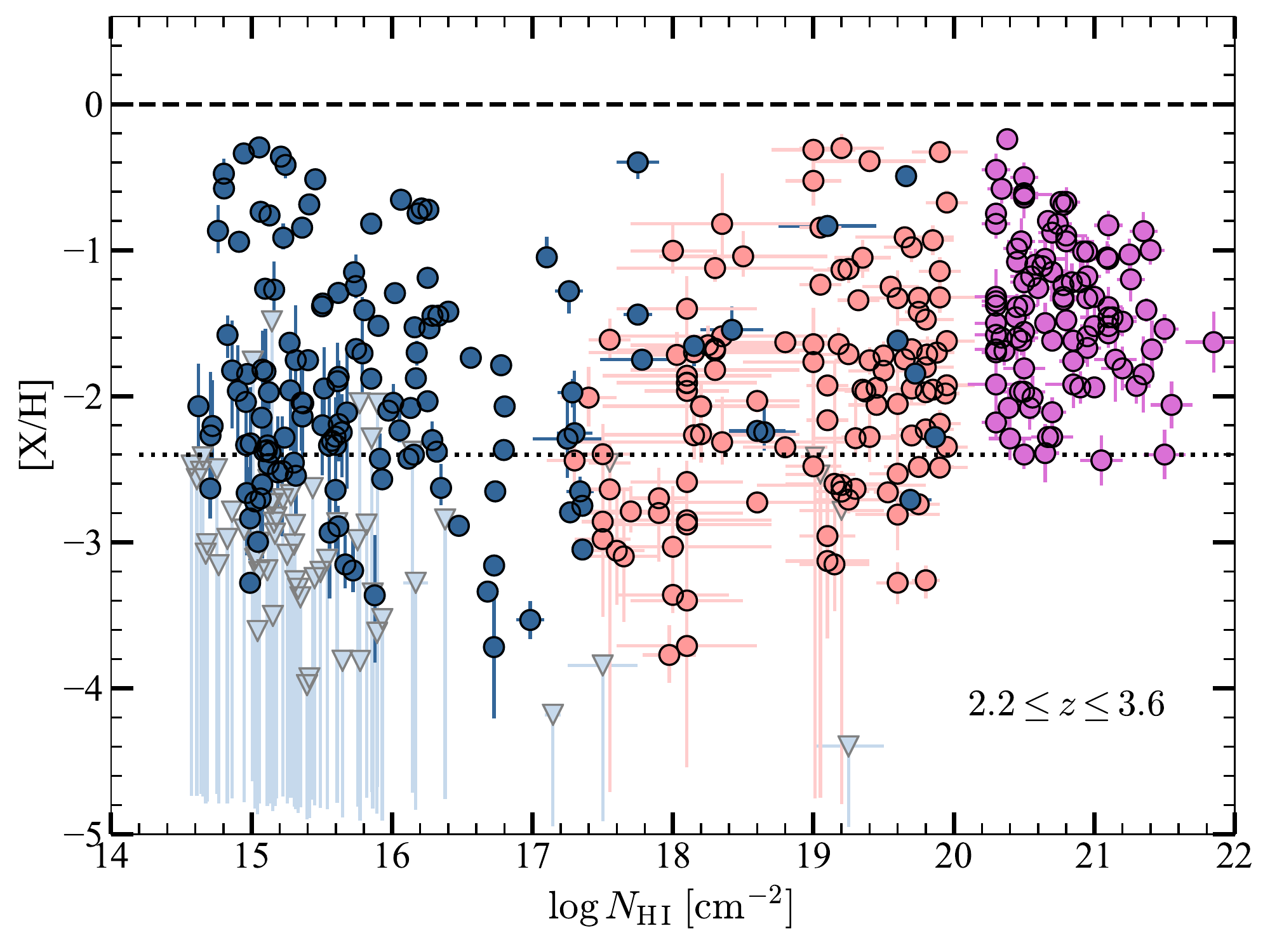}
\caption{Metallicities of the SLFSs ($14.5<\mlnhi < 16.2$), pLLSs ($16.2\le \mlnhi < 17.2$), LLSs ($17.2\le \mlnhi < 19$), SLLSs ($19\le \mlnhi < 20.3$), and DLAs ($ \mlnhi \ge 20.3$) at $2.2. \la z\la 3.6$ as a function of \nhi. The blue data from the KODIAQ-Z survey (this paper). The orange data are from the HD-LLS survey (\citealt{fumagalli16}) but we re-estimated the metallicities of these absorbers following the same EUVB and methodology as in KODIAQ-Z.  The purple data (DLAs) are from \citet{rafelski12}.   
\label{f-met_vs_nh1}}
\end{figure}

Thus far, we have separately analyzed the absorbers in different \nhi\ categories. However, we can also simply plot the metallicities of the absorbers as a function of \nhi, which is shown in Fig.~\ref{f-met_vs_nh1}. This has the obvious advantage that one does not have to make an {\it a priori}\ differentiation between the different absorbers in classes of  \hi\ column densities. For the SLFSs, pLLSs, LLSs, and SLLSs with well-constrained metallicities (i.e., not including the lower and upper limits), the central values represent the median of the posterior PDFs, and the error bars represent the 68\% CI. For the upper and lower limits, the down and upward triangles give the $90^{\rm th}$ and $10^{\rm th}$ quantiles while the vertical bar gives the 80\% CI. For the DLAs, the best estimates with their $1\sigma$ error bars are shown. The horizontal dashed line at $[{\rm X/H}]=0$ represents solar metallicity. The horizontal dotted line at $\xh =-2.4$ represents the VMP limit as defined above. Note that the lack of data between $\mlnhi = 20$ and 20.3 is purely artificial, resulting from our grid of photoionization models stopping at  $\mlnhi = 20$. 

Fig.~\ref{f-met_vs_nh1} further demonstrates the striking metallicity changes as a function of \nhi. Absorbers with  $\xh \ga -2.4$ are observed at any \nhi, but absorbers with $\xh < -2.4$ are observed only at $\mlnhi \la 19.8$. While the VMP definition implies this is VMP gas for DLAs, for low \nhi\ absorbers the VMP definition is the median value for SLFSs and  not even $1\sigma$ below the median of pLLSs and LLSs. As already pointed out above, the frequency of VMP absorbers also increases as \nhi\ decreases, which is quantitatively summarized in Table~\ref{t-frac}. We therefore also define extremely metal-poor absorbers as absorbers with  $\xh <-3$, which are mostly observed at some level in absorbers with  $\mlnhi \la 18$, see Table~\ref{t-frac}.\footnote{We emphasize that VMP DLAs with $\xh < -2.4$ exist at $z\sim 2$--4, but are found only in targeted searches \citep[e.g.,][]{cooke11a,cooke16}---they are uncommon and they do not show in sample as small as presented here since they are only  $<5\%$ of the population.}  Extremely metal-poor absorbers have metallicities that are so low that only Pop III stars may have contaminated them if they have some metals (see, e.g., \citealt{frebel07,crighton15}).

We finally note that the metallicity dip around $-1.1$ dex  observed especially in the SLFS metallicity PDF is also evident in the non-binned distribution shown in Fig.~\ref{f-met_vs_nh1}. In that figure, there is a lack of data points at $-1.1 \la \xh \la -1$ for \hi\ column density absorbers with $14.5 \la \mlnhi \la 16.4$. This is reminiscent of the metallicity dip at a similar value of $\xh \simeq -1$ observed in the pLLS+LLS PDF at $0.45 \la z \la 1$ (\citetalias{lehner19}, and see \S\ref{s-met-z} for the cosmic evolution of the metallicity). 

\subsection{Cosmic Evolution of the Metallicity}\label{s-met-z}
\begin{figure}[tbp]
\epsscale{1.15}
\plotone{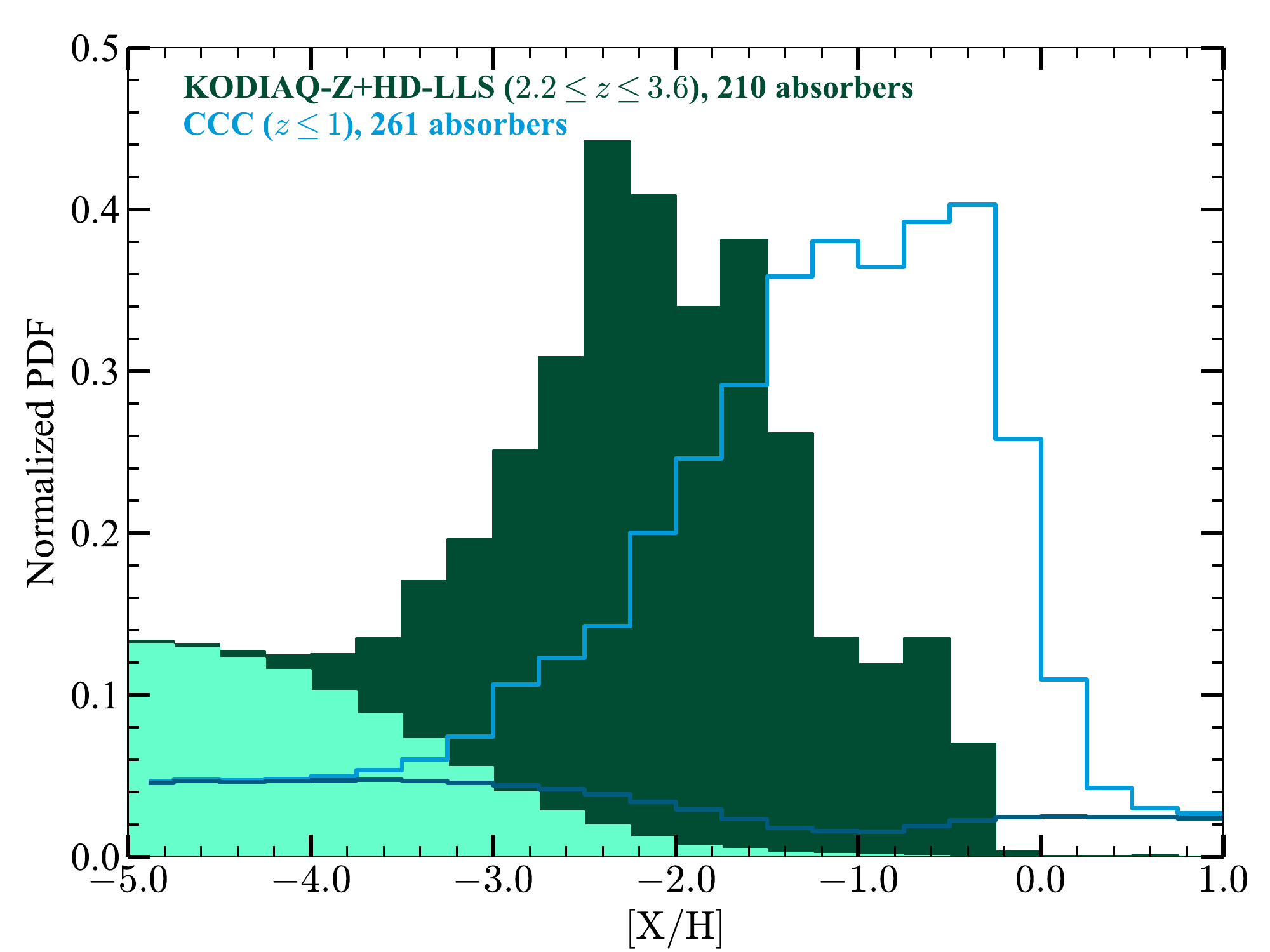}
\caption{Comparison of the posterior metallicity PDFs of the absorbers with $15 \le \mlnhi <19$ at high and low redshifts. The light-colored regions indicate the contribution from the upper limits. For CCC, The light colored histogram indicates the contributions from  upper and lower limits. 
\label{f-mdf-koa-ccc}}
\end{figure}

In \citet{lehner16}, for the first time, we studied the redshift evolution of the metallicity absorbers with $16.2 \la \mlnhi <19$, but with a much smaller sample at both low and high $z$. The samples of absorbers at all surveyed $z$ have increased at least by a factor 8 and provide far more robust results. Furthermore, we also now sample  at low and high $z$ the \nhi\ regimes below $\mlnhi \la 16.2$. In Fig.~\ref{f-mdf-koa-ccc}, we show the metallicity PDFs at low ($z<1$) and high ($2.2\le z\le 3.6 $) redshift for \hi-selected absorbers with $15 \la \mnhi <19$. There is some overlap between the two distributions between $-3 \la \xh \la -1.4$, but evidently there are more absorbers with $\xh \la -2$ at high $z$ while many more with $\xh >-1.2$ at $z<1$. In fact the latter value corresponds to the median metallicity at $z<1$ for $16.2 \la \mlnhi <19$ absorbers, and only  10\% of the high redshift absorbers have metallicities $\xh \ge -1.2$. The medians and means at $z<1$ are about $+1$ dex higher than at $2.2\le z\le 3.6 $. Both distributions are negatively skewed with similar  IQRs$\,\simeq 1.5$ at low and high $z$. 

\begin{figure}[tbp]
\epsscale{1.1}
\plotone{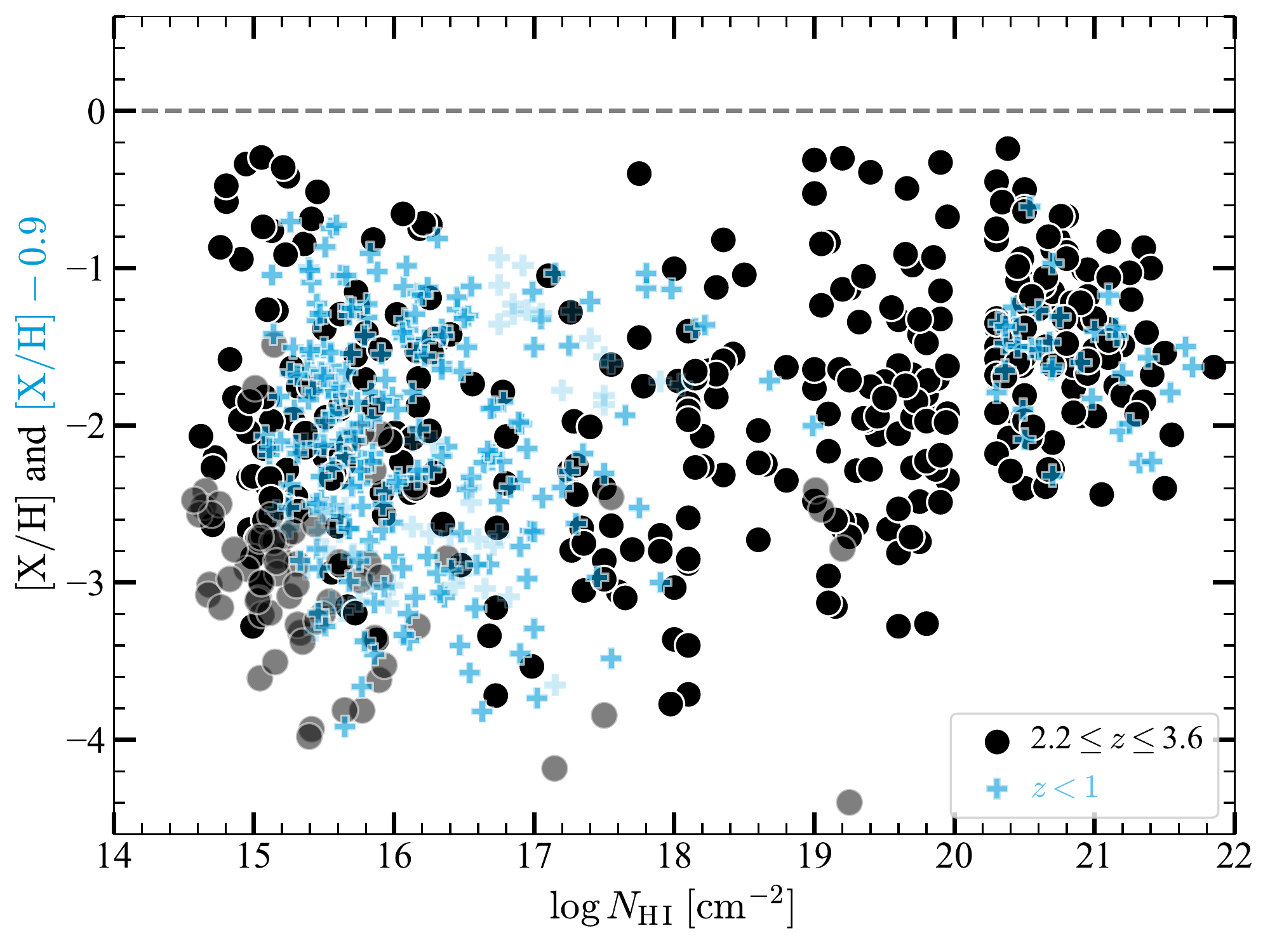}
\caption{The metallicity as a function of \nhi\ at low $z$ (CCC) and high $z$ (KODIAQ-Z, HD-LLS, R12-DLA). {\it For the  $z<1$ absorbers, we subtract the metallicity of each absorber by $0.9$ dex} (see text for more details). For the black circles, lighter symbols are upper limits on the metallicity. For the blue crosses, lighter symbols are upper limits if $\xh \le -1$ and lower limits if $\xh > -1$. For clarity, we did not plot the error bars (which are available in Fig.~\ref{f-met_vs_nh1} and Fig.~9 of \citetalias{lehner19} for the low redshift sample). Median metallicities are plotted, but that we treat upper and lower limits as Fig.~\ref{f-met_vs_nh1}, i.e.,  the lower and upper values represent the 10th and 90th percentiles, respectively.
\label{f-met-nhi-z}}
\end{figure}

In Fig.~\ref{f-met-nhi-z}, we compare the metallicities of the absorbers (following the same methodology used to produce Fig.~\ref{f-met_vs_nh1}) at $z<1$ and $2.2\le z\le 3.6 $ over the \hi\ column density range $14.5 \la \mnhi <22$. For the low redshift redshift absorbers, we shifted the metallicity by $-0.9$ dex, which is about the average difference between the mean/median metallicity of the absorbers at low and high $z$ ($-0.8$ dex for DLAs, and $-1$ dex for lower \nhi\ absorbers). The notable feature from this figure is that despite different metallicity behaviors at low and high \hi\ column densities, the offset metallicities and the overall dispersion at low and high $z$ show striking overlap at any given \nhi\ (where there is a good sampling in both datasets). A closer look reveals a small excess of higher metallicity absorbers in the SLFS, SLLS, and DLA regimes at high $z$ (i.e., a slightly larger dispersion of the metallicity distribution). However, overall the major change between $2.2\le z\le 3.6 $ and $z<1$ absorbers is an increase of their metallicity by a factor $\sim$8.

\begin{figure}[tbp]
\epsscale{1.1}
\plotone{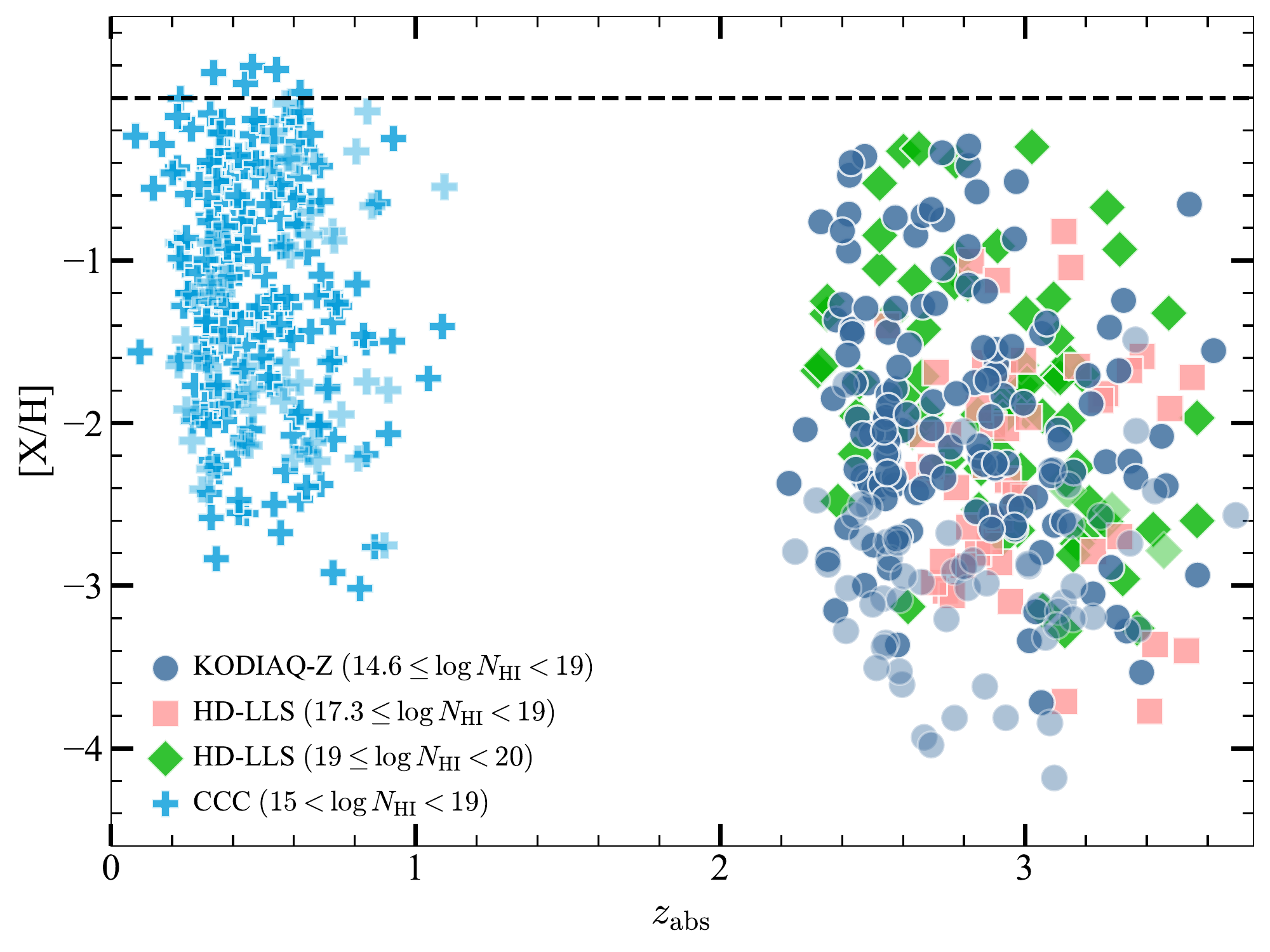}
\caption{The metallicity of different absorbers as a function of the redshift. For HD-LLS and KODIAQ-Z surveys, lighter symbols are upper limits on the metallicity. For CCC, lighter symbols are upper limits if $\xh \le -1$ and lower limits if $\xh > -1$. Median metallicities are plotted for the CCC, KODIAQ-Z, and HD-LLS surveys, but  we treat upper and lower limits as Fig.~\ref{f-met_vs_nh1}, i.e.,  the lower and upper values represent the 10th and 90th percentiles, respectively. 
\label{f-met-vs-z}}
\end{figure}

In Fig.~\ref{f-met-vs-z}, we now show the metallicities as a function of the redshift for the absorbers included in CCC, HD-LLS, and KODIAQ-Z (for clarity we do not show the R12-DLA sample in this figure). For the HD-LLS sample, we differentiate the LLSs and SLLSs.  This figure strengthens the previous conclusions, showing the overall increase in metallicities as $z$ decreases. At $z<1$, there is a substantial fraction of $\mlnhi <19$ absorbers with $-0.15 \la \xh \la +0.15$, while these are mostly absent at higher redshift.  On the other hand, at $z<1$ there is no evidence of population of absorbers with $\xh \la -3$, while there is at $2.2\le z\le 3.6 $. As already noted by \citet{lehner16} with a much smaller sample, while the VMP threshold increases by 1 dex over $2.2\le z\le 3.6 $ to $z<1$ (the 5\% quantile of DLA metallicities from $-2.4$ to $-1.4$ dex), a similar fraction of about 50\% of $\mlnhi \la 18$ VMP absorbers at low and high $z$ lie in the respective VMP regime. 

\begin{figure}[tbp]
\epsscale{1.1}
\plotone{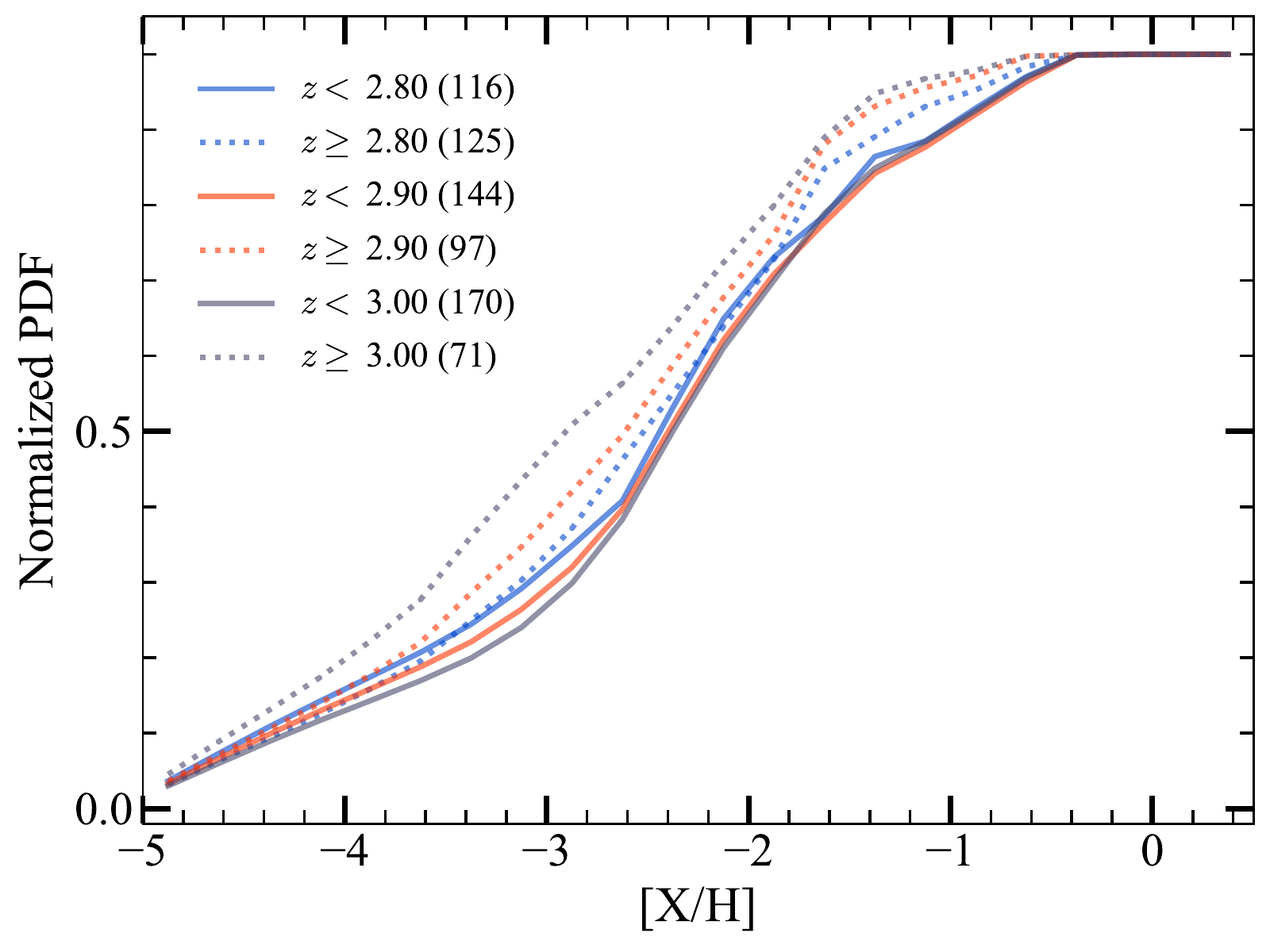}
\caption{Comparison of the metallicity CDFs  of the absorbers in KODIAQ-Z and HD-LLS with $14.5 \le \mlnhi < 19$ above and below a given redshift threshold, $z_{\rm th} = 2.8$, 2.9, 3.0. The numbers between parentheses indicate the number of absorbers in each redshift interval. 
\label{f-pdfall-vs-z}}
\end{figure}

In Fig.~\ref{f-met-vs-z}, there is also some evidence that the metallicity distribution somewhat changes around $z\sim 2.85$. Considering the KODIAQ-Z and HD-LLS sample with $\mlnhi < 19$ absorbers at $z > 2.85$, the fraction of absorbers with $\xh \ge -1$ is very small ($4/112\simeq 3.6\%$) while at  $z \le 2.85$ it is a factor $\sim$4 larger ($17/130\simeq 13.1\%$). To explore that potential of the change of the metallicity with $z$, in Fig.~\ref{f-pdfall-vs-z}, we show the comparison between the metallicity CDFs of the absorbers with $14.5 \le \mlnhi < 19$ from KODIAQ-Z and HD-LLS above and below several redshift thresholds, $z_{\rm th} = 2.8$, 2.9, 3.0. The large number of absorbers with $\xh \ga -1$ at $z \la 2.9$ compared to higher $z$ is further demonstrated in this figure. However, this figure also shows that at $z\ga 2.9$, and in particular at $z\ge 3.0$, there is a much larger fraction of metal-poor absorbers with $\xh -2.8$ than at lower redshifts, demonstrating that within the redshift interval $2.2 \la z \la 3.6$, there is a change in the metallicity PDFs with an overall increase of the metallicity at $2.2\la z<2.8$ compared to $2.8\la z \la 3.6$.

\subsection{Metallicity Variation over Small Velocity Scales}\label{s-paired}
Owing to the treatment of the absorbers where we separate individual \hi\ components as much as possible (see \S\ref{s-estimate-col}), our  KODIAQ-Z survey provides also information on the metallicity variation over small redshift/velocity separation between absorbers. Following \citetalias{lehner19}, we define here paired absorbers as absorbers that are closely separated in redshift space so that $\Delta v \equiv |(z^2_{\rm abs} - z^1_{\rm abs})/(1+z^1_{\rm abs})\, c| < 500$ \km. We have a sample of 37 such paired absorbers.

\begin{figure}[tbp]
\epsscale{1.1}
\plotone{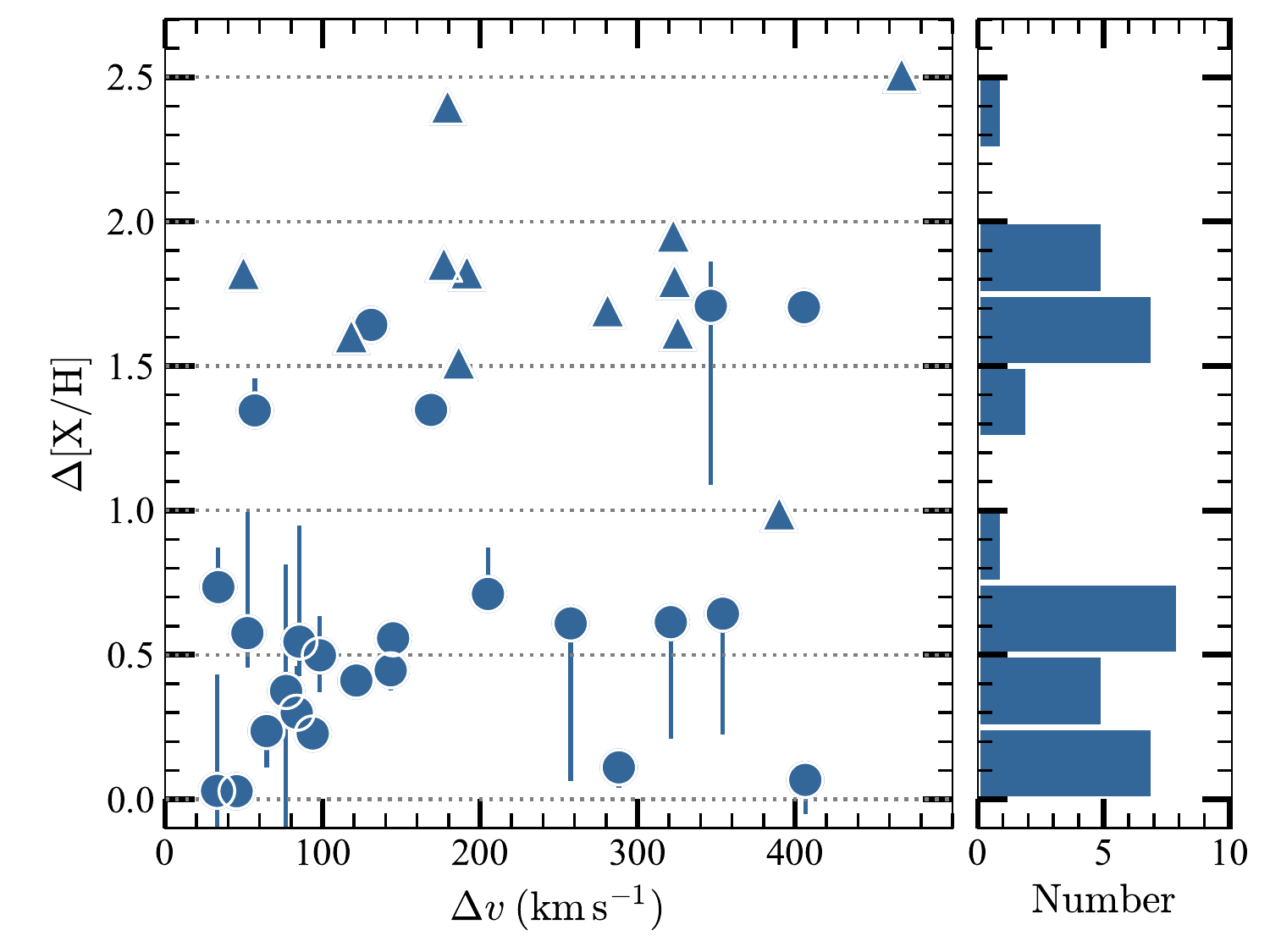}
\caption{{\it Left}: scatter plot of the difference between the higher and lower median metallicities for the closely-redshift separated absorbers (a.k.a. paired-absorbers) as a function of their absolute velocity difference. For the lower limits (blue triangles), we use the lower bound of the 80\% CI to be the most conservative. {\it Right}: distribution of the metallicity differences between paired-absorbers.
\label{f-met-dv}}
\end{figure}

In Fig.~\ref{f-met-dv}, we show in the left panel the absolute metallicity difference ($\Delta \xh$) between the higher and lower median metallicities of the paired absorbers against their absolute velocity separation and in the right panel the histogram distribution of $\Delta \xh$. As observed at low $z$ \citepalias{lehner19}, we find no obvious relation between $\Delta v$ and the level of the metallicity variation. At any velocity separation, there is a large scatter from about 0.2 dex to $>2$ dex. Using the survival method to account for the lower limits \citep{feigelson85}, we estimate a mean $\langle \Delta\xh \rangle = 1.20 \pm 0.16$ (error on the mean) and the IQR  in $\Delta \xh$ is $[0.38,1.81]$. There appear to be two main clusters of data separated at $\Delta \xh \simeq 1 $ with about 50\% of the sample in each bin and with a mean $\langle \Delta\xh \rangle = 0.39 \pm 0.05$ and $2.15 \pm 0.13$  below and above this limit, respectively. Therefore for the about half of the paired absorbers, there is evidence for  metallicity variations over $\Delta v \la 500$ \km\ of a factor 2--3 and for the other half of a factor $>140$.

Owing to the large metallicity variation over $\Delta v \la 500$ \km, {\it a posteriori}\ it is justified to treat these paired absorbers as individual absorbers. The other direct consequence is, when possible, components in absorbers should be treated individually to derive the true metallicity of the gas rather than a column-density weighted average metallicity between very different types of gaseous components, a consequence first inferred by \citet{prochter10} at similar redshifts for one LLS and then in the CCC survey at $z<1$ for 30 paired-absorbers (\citetalias{wotta19,lehner19}; see also \citealt{kacprzak19,zahedy21}). We discuss the implication of these findings in \S\ref{s-disc-var} in more detail.

\subsection{The \ca\ ratio as a Function of the Metallicity}\label{s-ca-met}

\begin{figure}[tbp]
\epsscale{1.1}
\plotone{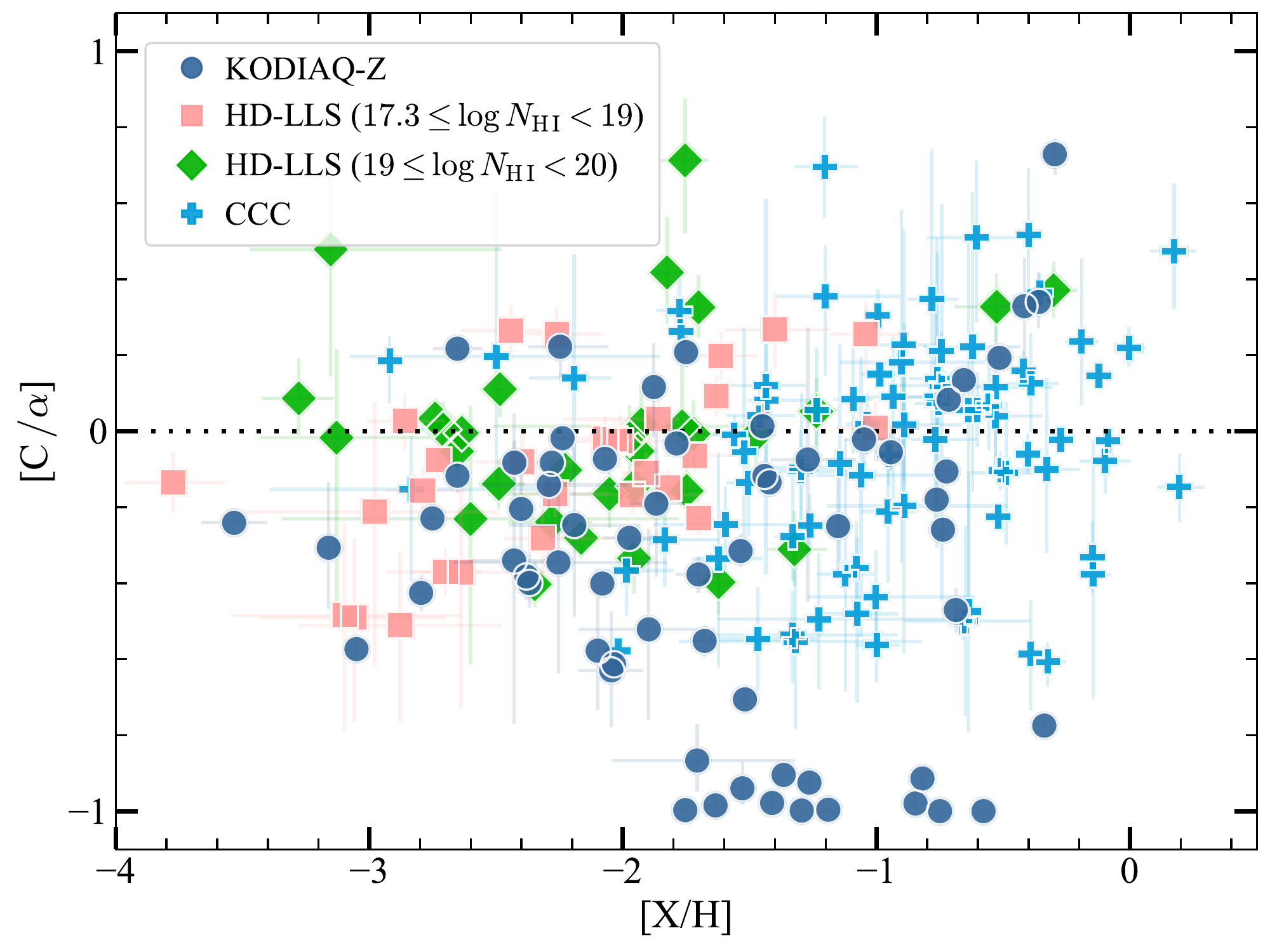}
\caption{The  \ca\ ratio against the metallicity of the absorbers from the KODIAQ-Z ($2.2 \la z \la 3.6$), HD-LLS ($2.2 \la z \la 3.6$), and CCC ($z\la 1$) surveys.  The median values of the \ca\ and \xh\ posterior PDFs are adopted with 68\% CI. Note that \ca-values with $\ca < -0.6$ are most likely predominantly caused by ionization corrections rather than nucleosynthesis effects (see text for more details).  
\label{f-calpha-met}}
\end{figure}

In the photoionization modeling of the absorbers to determine their metallicities, the \ca\ ratio is allowed to vary in the range $-1 \la \ca \la +1$ to take into account possible nucleosynthesis effects that likely arise from the time-lag in the production between $\alpha$-elements and carbon (e.g., \citealt{cescutti09,mattsson10}). The general trend between \ca\ and \ah\ seen in stars \citep[e.g.,][]{akerman04,fabbian09} and SLLSs/DLAs \citep[e.g.,][]{pettini08,penprase10,cooke11a} is a decrease of \ca\ from about $\ca \simeq 0$ (up to $+0.2$ dex) at $\ah \ga 0$ to $\ca \simeq -0.6$ at $\ah \simeq -0.5$ where \ca\ plateaus around that metallicity value up to a metallicity of $\ah \simeq -2$ where an upturn is observed in \ca\ with an increase of \ca\ to $\ca \ga +0.2$ at $\ah \le -2.8$. 

As discussed in \citetalias{lehner19} at low redshift and \citet{lehner16} at high redshift, the SLFSs, pLLSs, and LLSs do not quite follow this overall trend. In fact, little trend is observed between \ca\ and \ah\ for these absorbers with  a large scatter in \ca\ at any metallicity. While no obvious trend is observed between \ca\ and \ah\ for these absorbers, \citetalias{lehner19} note three characteristics:  1) a lower floor level for the distribution at about  $\ca \simeq -0.6$,  similar to that observed for the stars, \hii\ regions, and SLLSs/DLAs; 2) an upper ceiling of the distribution at  $\ca \simeq +0.5$, but with a concentration of data around $0< \ca \sim +0.3$, similar again to the highest values observed in  stars, \hii\ regions, and SLLSs/DLAs; and 3) a tentative upturn in \ca\ at $\xh \la -2$, which would be similar to the one observed in stars and SLLSs/DLAs. 

With KODIAQ-Z and our re-modeling of the HD-LLS absorbers, we can now revisit these conclusions with a much larger sample at high redshift, especially in the low metallicity regime. In Fig.~\ref{f-calpha-met}, we show \ca\ as a function of \xh\ for all the absorbers in KODIAQ-Z, HD-LLS, and CCC where the relative abundance and metallicity are simultaneously constrained. In that figure, we use different symbols to differentiate LLSs from SLLSs in the HD-LLS survey (there is only one SLLS in KODIAQ-Z with information on \ca, see Fig.~\ref{f-calpha-nhi}, which we did not highlight in Fig.~\ref{f-calpha-met}). Excluding the 13 absorbers with $\ca \la -0.9$ (because those are more largely affected by ionization, see \ref{s-uncertainties} and below), the addition of the KODIAQ-Z and HD-LLS data confirm the first two characteristics described above and summarized in \citetalias{lehner19}. However, at  $\xh \la -2$, there is no clear upturn in \ca\ for the absorbers with $\mlnhi <19$, although the scatter in the \ca\ distribution becomes smaller. On the other hand, considering the SLLSs from HD-LLS (diamond symbols in Fig.~\ref{f-calpha-met}), the upturn in \ca\ is apparent when the metallicity becomes smaller than $\xh \la -2.2$. 

Since the derived ratio of \ca\ is controlled by both ionization corrections and nucleosynthesis effects, it is plausible that the ionization corrections are washing away some of the trends that would appear if only nucleosynthesis effects were at play (as for example for the DLAs). In \S\ref{s-abund-unc}, we show that \ca\ in the KODIAQ-Z sample is more affected by ionization than in the two other samples in view of the correlation between  \ca\ and $f_{\rm H\,I}$. However, no such correlation is observed in CCC and HD-LLS, and the overall dispersion is  similar in these three surveys. The mean values for \ca\ in CCC and HD-LLS are also the same (see \S\ref{s-abund-unc}), despite widely different ionization correction levels. Therefore the \ca\ ratio with $\ca>-0.6$ determined from the ionization modeling does not seem to have a strong dependence with the level of the ionization correction. Nevertheless, the ionization correction essentially adds noise in the $\ca$ distribution at the level of $\pm 0.3$ dex at any \nhi\ and metallicity, which is large enough to hide subtle changes of \ca\ with \xh\ at the expected level of 0.2--0.5 dex. Hence, the  scatter observed in Fig.~\ref{f-calpha-met} must be largely caused by ionization while the range $-0.6 \la \ca \la +0.5$ where most of the absorbers lie may be controlled by nucleosynthesis effects based on its similarity observed in DLAs, stars, and \hii\ regions. 

\section{Physical Properties of the SLFS\lowercase{s}, pLLS\lowercase{s}, LLS\lowercase{s}, and SLLS\lowercase{s} at $2.2 \la \lowercase{z} \la 3.6$}\label{s-phys-cond}

The photoionization models discussed in \S\ref{s-ion-model} also predict the physical properties of the absorbers. In particular, these include the neutral fraction of the gas ($f_{\rm HI}$), the density of the gas (\nnh), the total  H column density ($\mnh  \equiv \mnhi + \mnhii$), the length-scale of the cloud ($l \equiv \mnnh/\mnh$), and the temperature of the gas ($T$). These can in turn be used to constrain the cosmological baryon and metal budgets for the absorbers (\citealt{fumagalli16}; \citetalias{lehner19}). A caveat to keep in mind and discussed at length in \citet{fumagalli16} is that there is a degeneracy between the density and intensity of the radiation field. This dependency produces much less robust inference on the density and sizes of absorbers than for the metallicity, which depends only on the shape of the radiation field. We also remind the reader that only photoionization by the EUVB has been used in this work, and it is quite plausible that more than one ionization process is at play, which would also affect more the physical quantities discussed in this section.  

\subsection{Densities, H Column Densities, Temperatures, and Length Scales}\label{s-phys}

\begin{figure*}[tbp]
\epsscale{0.9}
\plotone{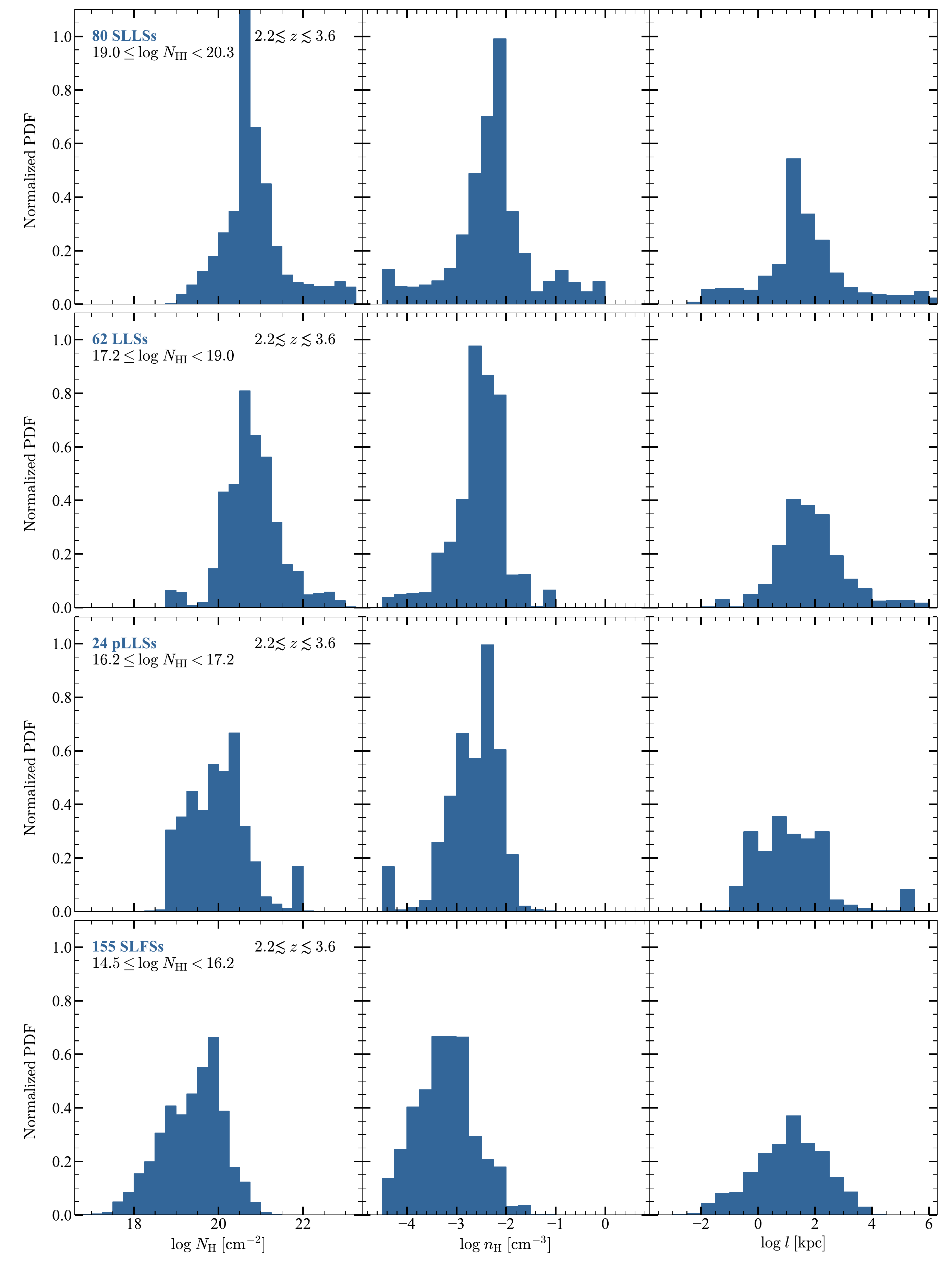}
\caption{Posterior PDFs of the total H column density ({\it left}), H density ({\it middle}), and path-length ({\it right}) for SLLSs, LLSs, pLLSs, and SLFSs for the entire sample from the KODIAQ-Z survey and HD-LLSs. \label{f-pdf-phys}}
\end{figure*}

We summarize the physical quantities (\nnh, \nh, $l$, and $T$) derived from our modeling for the KODIAQ-Z sample in Table~\ref{t-physical_properties}. Values of $\mlnnh$ with a trailing colon represents absorbers for which we adopted a Gaussian prior on \logU, and therefore on  $\mlnnh$.  In Table~\ref{t-phys-stat}, we summarize the statistical properties of the physical properties derived for the entire sample (median, mean, and dispersion of \nhi, \nnh, \nh, $l$, and $T$) and a restricted sample where only a flat prior on $\log U$ was used. Comparing the results from Table~\ref{t-phys-stat} between the entire and restricted samples, it is apparent that the results are not statistically different. We emphasize that most of the pLLSs, LLSs, SLLSs did not require a Gaussian prior on  $\log U$. However, for about 45\% of the pLLSs, we had to use a Gaussian prior on  \logU, and the results are essentially the same in both sample. Hereafter, we therefore consider the results from the entire sample. In Fig.~\ref{f-pdf-phys}, we show the posterior PDFs of \nh, and \nnh, and $l$ for the SLLSs and LLSs (using results from KODIAQ-Z and HD-LLS) in the top two rows and the pLLSs and SLFSs (from KODIAQ-Z) in the bottom two rows. These PDFs contrast remarkably from the metallicity PDFs with much narrower distributions. 

Temperatures of ionized gas predicted by Cloudy are based on the balance of atomic heating (photoionization) and cooling (recombination and collisional effects). As such, temperatures are dependent on the radiation field shape and intensity, density, and metallicity of the models. The distribution of predicted temperatures has overall small dispersion, and there is a clear trend of decreasing temperatures as \nhi\ increases of about 0.1 dex decreases between each absorber category from the SLFSs, pLLSs, LLSs, to SLLSs (see Table~\ref{t-phys-stat}). For the SLFSs, the average temperature is around $3\times 10^4$ K, while it is a factor 2 less on average for the SLLS. In \S\ref{s-res-nhi-b-dist}, we show that for about 90\% of the \hi\ components $\langle b\rangle = 27 \pm 6$ \km. This places an upper limit on the temperature from the observations of $T<4\times 10^4$ K, consistent with the ionization modeling. It also implies that for the SLFSs, on average the thermal broadening component is somewhat larger than the non-thermal component, while for the other absorbers, non-thermal motions take over the thermal ones. Compared to the \citetalias{lehner19}, the predicted temperatures of the $z<1$ SLFSs, pLLSs, and LLSs are about a factor 1.4 times smaller than at $2.2 \la z \la 3.6$.

Considering the SLFSs, pLLSs, and LLSs, the total H column densities decrease with decreasing \nhi, but their distributions overlap within $1\sigma$ dispersion (that are large for these quantities, see Table~\ref{t-phys-stat} and Fig.~\ref{f-pdf-phys}). The differences between the mean \lnhi\ for SLFSs compared with the pLLSs and the pLLSs to the LLSs are a factor 4 and 28, respectively. In contrast, the differences in total H are significantly smaller with differences of a factor  $\sim 4$ and 6.5 between the SLFSs and pLLSs and pLLSs and LLSs, respectively. Thus, despite their lower \nhi\ values, the contribution of pLLSs and SLFSs  to the total mass and baryon budgets are higher compared with the LLSs than their \hi\ columns would suggest, especially since they are more numerous (e.g., \citealt{prochaska14} and see \S\ref{s-budget}). The \nh\ column densities for SLLSs are similar to those for LLSs, \nh\ is about the same even though \nhi\ is on average a factor 34 smaller for the LLSs than for SLLSs. Given the higher incidence of LLSs, they make a larger contribution to the cosmic baryon budget than the SLLSs. Compared to \citetalias{lehner19}, \nh\ at the $z<1$ SLFSs, pLLSs, and LLSs are about a factor $\sim10$--15  smaller than at $2.2 \la z \la 3.6$.

The estimated hydrogen densities increase as \nhi\ increases. The SLLSs has the largest dispersion on \nnh, which is caused by a long tail at low and high \nnh\ values (see Fig.~\ref{f-pdf-phys}). While there is an increase in \nnh\ with increasing \nhi, the bulk of the hydrogen density for the pLLSs, LLS, and LLSs are within about $\mlnnh \simeq -3.4$ and $-1.8$, while for SLFSs it is between $\mlnnh \simeq -4.2$ and $-2.5$. At $z\sim 2.8$ (the mean redshift of KODIAQ-Z), the mean cosmic density is about $\bar{n}_{\rm H} \simeq 10^{-5}$ \cmmm\ \citep{schaye01b}. The pLLSs, LLSs, and SLLSs therefore probe over-densities of $\delta \equiv \mnnh / \bar{n}_{\rm H} -1  \sim 40$--1600 while for the lower column SLFSs $\delta \sim 5$--320. A large part of the observed pLLSs, LLSs, and SLLSs  largely probe virialized (mini)halos ($\delta >100$,  \citealt{schaye01b}). For SLFSs, there is more a mixture of systems probing over-densities below and above 60, at the transition between the IGM and virialized halos \citep{schaye01b}. Compared to the \citetalias{lehner19}, \nnh\ at the $z<1$ SLFSs, pLLSs, LLSs, SLLSs are about a factor $\sim 1.5$--2.5   smaller than at $2.2 \la z \la 3.6$. However, owing to the redshift dependence of the mean cosmic density that is $\propto ((1+z)/4)^3$ \citep{schaye01b}, the density contrast at $z<1$ for the SLFSs and other stronger \hi\ absorbers is always $\delta >100$, even for the lower densities probed by SLFSs (although we note that at $z<1$, only $\mlnhi >15$ SLFSs could be probed). 

The length-scales, which are directly related to \nnh\ and \nh, have the largest dispersion among all these physical quantities. For the SLFSs and pLLSs, the interquartile range (IQR) of length-scale of the absorbers is  about $4 \la l \la 150$ kpc. For the LLSs, IQR for $l$ is $18 \la l \la 288$ kpc and for the SLLSs it is $8 \la l \la 80$ kpc. In contrast at $z<1$, the IQR of length-scale of these absorbers is $0.3 \la l \la 3.2$ kpc \citepalias{lehner19}. At $2.2 \la z\la 3.6$,  the SLFSs, pLLSs, LLSs, and even the SLLSs may probe a variety of structures from gas embedded in virialized halos to large-scale structures. Recent MUSE galaxy surveys show strong hints that at least LLSs may probe a wide variety of environments at $z>2$, including gaseous filaments, CGM of galaxies, or  intra-group gas \citep[e.g.,][]{fumagalli16a,fumagalli17,lofthouse20}.

\subsection{Neutral Fractions}\label{s-neut-frac}
\begin{figure}[tbp]
\epsscale{1.1}
\plotone{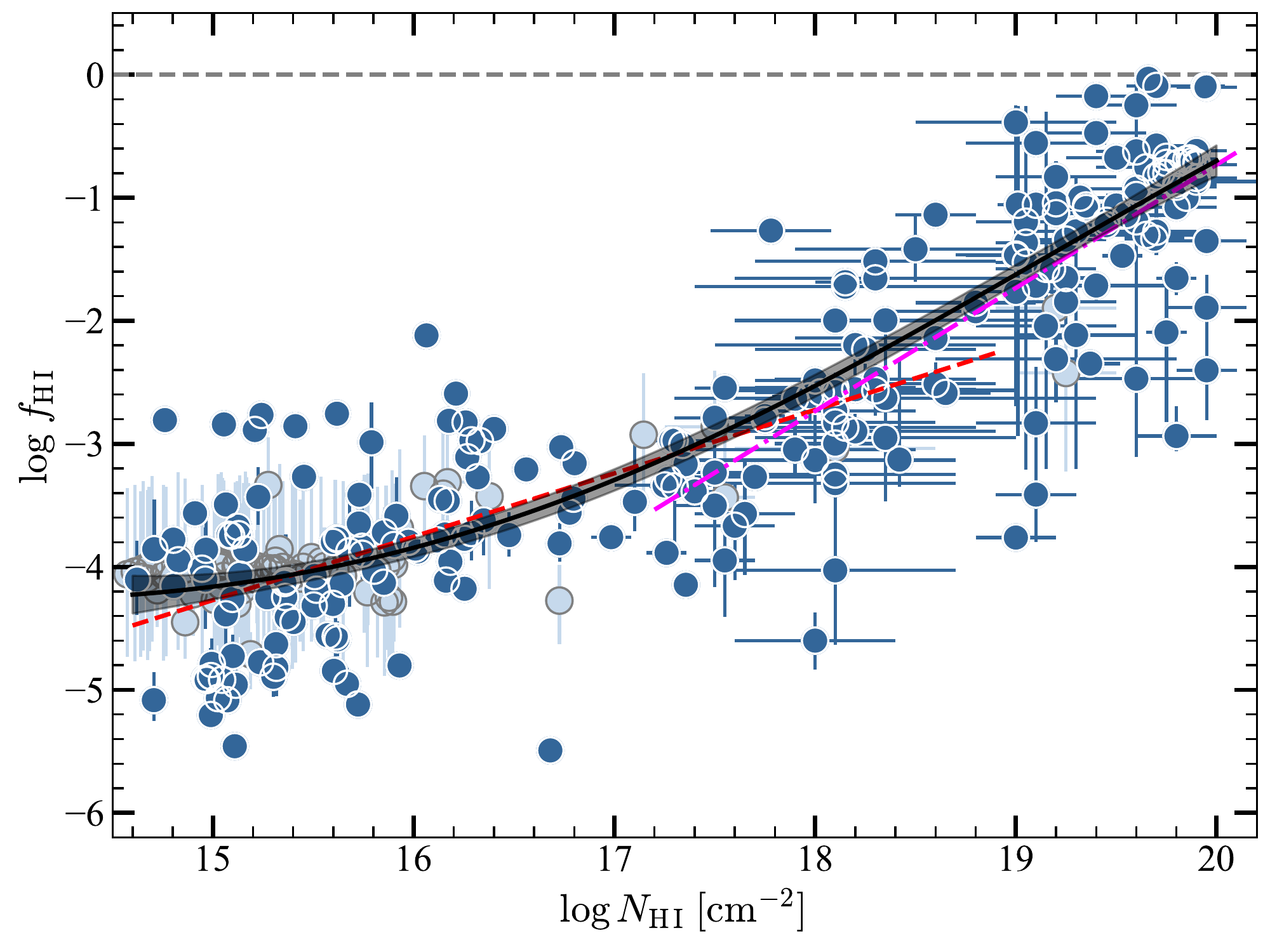}
\caption{The neutral fraction of \hi\ versus \nhi\ using the KODIAQ-Z and HD-LLS surveys. The median values of the posterior PDFs are adopted with 68\% CI. Light blue data were estimated using a Gaussian prior on $\log U$, while the other data have a flat prior on $\log U$. The black curve is a GP-derived model to the entire \nhi\ range shown in the figure (the dark area corresponds to the dispersion derived from the GP model). The dashed red line is a linear fit to the data with $14.6 \le \mlnhi < 19$ (i.e., the SLFSs, pLLSs, and LLSs). The dash-dotted magenta line is  a linear fit to the data with $17.2 < \mlnhi \le 20$ (i.e., the LLSs and SLLSs). For all the models, we consider only data with a flat prior on $\log U$ (although the results are similar if the entire dataset was considered instead).
\label{f-fnhi-nhi}}
\end{figure}

The behaviour of the neutral fraction with \nhi\ can be used  to  constrain the cosmic baryon and metal budgets (see \S\ref{s-budget}). In Fig.~\ref{f-fnhi-nhi}, we show the neutral fraction of the absorbers estimated from the Cloudy models as a function of \nhi\ from $\mlnhi\ \simeq 14.6$ to 20 from KODIAQ-Z and HD-LLS (see \S\ref{s-photo-model}, and see also the recent work for the SLLSs from \citealt{berg21}). There is a clear correlation between $f_{\rm H\,I}$ and \nhi. This is confirmed by the Spearman rank-order test that shows a strong positive monotonic correlation between $f_{\rm H\,I}$ and \nhi\ with a correlation coefficient $r_{\rm S} = 0.9$ and a $p$-value\,$ \ll 0.05\%$. However, the slope of the correlation is shallower between the SLFS, pLLS, LLS regimes compared to the LLS, SLLS regimes. The dashed red line in Fig.~\ref{f-fnhi-nhi} is a linear fit to the data with $14.6 \le \mlnhi < 19$, yielding $\log f_{\rm H\,I} = (0.51 \pm 0.04) \mlnhi -12.00$. The slope is quite similar to that found at $z<1$ for absorbers with $15.1 \le \mlnhi < 19$ \citepalias{lehner19}. The dash-dotted magenta line  is  a linear fit to the data with $17.2 < \mlnhi \le 20$, $\log f_{\rm H\,I} = (1.00 \pm 0.08) \mlnhi -20.76$, which is consistent with the linear fit derived in the HD-LLS survey for the LLSs+SLLSs \citep{fumagalli16}.

To model this overall trend between $f_{\rm H\,I}$ and \nhi, we use a Gaussian process (GP) model, providing a generic supervised learning method designed to solve a regression. We refer the reader to Appendix~F in \citet{lehner20} for more details about this method, but in short, the advantage of this method is that the prediction interpolates the observations in a nonparametric way and is probabilistic so that empirical confidence intervals can be computed. We use the Python {\sc Gaussian Process Regression} package within {\sc scikit-learn} \citep{pedregosa11,buitinck13} to model the data with a squared-exponential kernel (the use of a more complex kernel like a Matern kernel would yield similar results when using similar bounds).  The overall scatter in the data is typically much larger than the errors on single data point, and we use the $1\sigma$ dispersion derived for SLFSs, pLLSs, LLSs, and SLLSs to inform the GP model of the intrinsic scatter in the data  (this can be understood as a prior factor to smooth out the scatter of the data). The black curve in Fig.~\ref{f-fnhi-nhi} is the GP model with the dark area showing the standard deviation determined by the GP model. There is a large overlap with the two linear fits, but the GP fit has the advantage of providing a global model to the data without {\it a priori}\ selecting manually the inflection point in the curve. The black curve is well-fitted by a third order polynomial, $\log f_{\rm H\,I} = -0.0104 x^3 + 0.6268 x^2 - 11.62 x + 64.23$ (where $x = \mlnhi$). 

\section{Cosmological Budgets}\label{s-budget}

\subsection{Cosmological Baryon Budget}\label{s-budget-baryon} Observations of the LYAF show that about 80\%--90\% of the baryons at $2<z<4$ are found in the cool photoionized phase of IGM  \citep[e.g.,][]{rauch97,weinberg97,kim01}. In \citet{lehner14}, we show that the photo- and collisionally-ionized gas found in $\tau_{\rm LL}>1$ systems is very likely the second largest contributor to the baryon budget at $z\sim 2$--3. However, this result could be affected by small-number statistics as \citeauthor{lehner14} sample included only 7 LLSs and  8 SLLSs. Our earlier study also did not consider the baryon contributions from the SLFSs and pLLSs. Here we review the impact of the cool, photoionized gas in these absorbers on the  cosmic baryon budget at $2.2 \la z \la 3.6$. 

The  mass gas density relative to the critical density, $\Omega_{\rm g}$,  can be estimated via (e.g., \citealt{tytler87,lehner07,omeara07,prochaska10}; \citetalias{lehner19}):
\begin{equation}
\Omega_{\rm g} =  \frac{\mu_{\rm H} m_{\rm H} H_0}{\rho_c\, c } \int \frac{\mnhi}{f_{\rm H\,I}} \mathcal{F}(\mnhi) d \mnhi\,,  \\ 
\end{equation}
where $m_{\rm H} = 1.673\times 10^{-24}$ g is the  mass of hydrogen, $\mu_{\rm H} = 1.3 $ corrects for the presence of helium, $c = 2.9979\times 10^{10}$ cm\,s$^{-1}$ is the speed of light, $\rho_c = 3 H^2_0/(8\pi G) = 8.62 \times 10^{-30}$ g\,cm$^{-3}$ is  the $z=0$ critical density, $H_0 = 2.20\times 10^{-18}$ s$^{-1}$ is the Hubble constant,\footnote{We adopt here the Planck 2015 results (\citealt{planck16}).} and $\mathcal{F}(N_{\rm H I})$ is the column density distribution function. For $\mathcal{F}(N_{\rm H I})$, we adopt the spline function model from \citet{prochaska14} at $z\approx 2.5$ (see their Table 2 and Fig.~7). For the neutral fraction, $f_{\rm H\,I}$, we adopt the GP model from \S\ref{s-neut-frac}. Integrating the above equation over the different \nhi\ regimes of the SLFSs, pLLSs, LLSs, and SLLSs, we find that $\Omega_{\rm g} \simeq (2.7, 1.1,1.4,1.1)\times 10^{-3}$ for the SLFSs, pLLSs, LLSs, and SLLSs, respectively. The contributions of $\Omega_{\rm g}$ to the total baryon density are $\Omega_{\rm g}/\Omega_{\rm b} \simeq 5.5\%,2.2\%,2.8\%, 2.3\% $ for the SLFSs, pLLSs, LLSs, and SLLSs, respectively ($\Omega_{\rm b} = 0.0486$, \citealt{planck16}). The values for the LLSs and SLLSs are consistent with the ranges in \citet{lehner14} (their Table~6, column with \siiv).\footnote{Although we remind the reader that here the SLLSs include only absorbers with $19\le \mlnhi \le 20$ owing to the grid of ionization models stopping at $\mlnhi =20$}.  

Importantly these estimates do not include a contribution from the highly ionized, hotter gas that is not photoionized. \citet{lehner14} found that the \ovi\ collisionally ionized gas in LLSs and SLLSs could contribute to 0.4--10\% to baryonic budget.  As noted in \S\ref{s-motivation}, for the pLLSs and SLFSs, we often find that the \ovi\ absorption is narrow and aligns with the absorption seen in \hi\ (and other lower metal-ions if they are detected), i.e., the properties of the \ovi\ are quite different as \nhi\ decreases to $\mlnhi <17$ compared to stronger \hi\ absorbers. The observed column densities for these narrow \ovi\ components are also well-reproduced by single-phase photoionization models as those described in \S\ref{s-ion-model} where the source of the ionizing photons is EUVB. Providing a full description of the \ovi\ associated with the KODIAQ-Z absorbers is beyond the scope of this paper, but based on a visual inspection of the profiles and our ionization modeling, very broad, strong \ovi\ absorbers associated with SLFSs and pLLSs are rare, whereas they are common in LLSs, SLLSs, and even DLAs. Therefore the contribution of broad, strong \ovi\ associated with SLFSs and pLLSs to the cosmic baryon budget is very likely small. 

The photoionized gas associated with pLLSs, LLSs, and SLLSs combined therefore contribute $\Omega_{\rm g}/\Omega_{\rm b} \simeq 7\%$ and the SLFSs contribute to another 5.5\% to the total baryon budget.\footnote{We separate the SLFSs from the other absorbers because, e.g., \citet{kim01} included those in the LYAF contribute where they estimate that 90\% of all baryons reside in the LYAF at $1.5 <z<4$.} In contrast, the neutral gas that is mostly found in DLAs yield only $\Omega_{\rm g}/\Omega_{\rm b} \simeq 2\%$ at $z \sim 2$--5.  \citep[e.g.,][]{prochaska05,wolfe05,lehner14,peroux20}. At $z<1$, \citetalias{lehner19} find fractional contributions to baryonic budget for the SLFSs, pLLSs, and LLSs that are about 10 times smaller. At $z<1$, the SLFSs, pLLSs, and LLSs are probing much larger overdensities, and at these redshifts only $\sim 5\%$ of the baryons are thought to be in the CGM of galaxies \citep{shull12}.

\subsection{Cosmological Metal Budget}\label{s-budget-metal}
The metal density of the absorbers can be estimated with the baryon density via $\Omega_{\rm m} = \Omega_{\rm g} Z$, where $ Z =10^{\xh} Z_\sun $ is the metallicity of the gas in mass units and $Z_\sun = 0.0142$ is  the bulk solar metallicity in mass units from \citet{asplund09}. For the metallicity, we use the linear mean metallicities in about 0.5 dex bin of \nhi\ from $\mlnhi =14.5$ to $\mlnhi =20$. We note that the linear mean metallicities for SLFSs, pLLSs, and LLSs are $0.035, 0.032, 0.028 Z_\sun$, respectively. We find  $\Omega_{\rm m}  = (1.4, 4.7, 5.9)\times 10^{-7}$ for the SLFSs, pLLSs, and LLSs, respectively. The latter value is  consistent with that derived by \citet{fumagalli16}, $\Omega_{\rm m}  = 5.1\times 10^{-7}$. It is not surprising that our value is somewhat larger since KODIAQ-Z added a few LLSs on the high metallicity side, including one with $\xh\simeq -0.4$ (see, e.g, Fig.~\ref{f-met_vs_nh1})  and since the estimate of $\Omega_{\rm m}$ that depends on the mean linear metallicity is much more sensitive to the positive tail end of the logarithmic metallicity PDF. Since we do not have SLLSs with $\mlnhi >20$ in our sample, we use the result from the HD-LLS survey where \citet{fumagalli16} derived $\Omega_{\rm m}  = 1.6\times 10^{-6}$ for the SLLSs. For  reasons (sample variance, using the arithmetic mean) discussed in detail in \citeauthor{fumagalli16}, these values are uncertain by a factor $\sim 3$ (see also discussion in \citealt{berg21}). We note that the comoving metal-mass density can be estimated via $\rho_{\rm m} =\Omega_{\rm m}/ \rho_c$ (e.g., \citealt{peroux20}). Therefore, for the combined photoionized SLFSs, pLLSs, and LLSs, we find $\rho_{\rm m} = 1.5\times 10^5$ M$_\odot$\,cMpc$^{-3}$, while combined with the SLLSs, this increases to $3.5\times 10^5$ M$_\odot$\,cMpc$^{-3}$.

The total expected cosmic metal density from the output of stars  at $z\sim 2.8$ is about $\Omega^{\rm exp}_{\rm m} \simeq 9\times 10^{-6}$ or expressed in terms of cosmic metal mass density $\rho^{\rm exp}_{\rm m} \simeq 10^6$ M$_\odot$\,cMpc$^{-3}$  \citep{peroux20}. \citet{peroux20} also recently estimated for the DLAs, i.e., for the neutral gas,  that $\Omega_{\rm m} \simeq 3.5\times 10^{-6}$  and for the dust $\Omega_{\rm m} \simeq 1\times 10^{-6}$ at $2.2 \la z \la 3.6$, , i.e., about 40\% of the metals are in the neutral gas and about 10\% are in form of dust  at $2.2 \la z \la 3.6$.  The SLLSs account for 17\%, while the SLFSs, pLLSs, and LLSs  account for 1.5\%, 5.1\%, 6.4\%, and all combined they account for about 30\% of the metals  at $2.2 \la z \la 3.6$. We note that the recent estimates from \citet{peroux20} for the SLLSs and LLSs are about within $1\sigma$ errors from those derived here and \citet{fumagalli16}. As for the baryons, another contribution is from the hot \ovi\ gas associated with LLSs, SLLSs, and DLAs that can contribute another $\ga 7\%$ \citep{lehner14}. 

While most of the baryons at  $2.2 \la z \la 3.6$ are in the LYAF, this is not the case for the metals since the  DLAs, dust, SLLSs, LLSs, pLLSs, and SLFSs contribute to about 90\% of the cosmic metal budget. Direct estimates of the metal budgets in the LYAF are uncertain owing to ionization correction and the fact that some of the metals are not solely associated with $\mlnhi \ga 14.5$ absorbers. Using the recent results on $\Omega_{\rm C\,IV}$ from \citet{dodorico13,dodorico10} (but see also, e.g., \citealt{schaye03,simcoe04,songaila05}) and correcting their estimate from \civ\ to carbon and carbon to the total metal content yields  $\Omega_{\rm m} \ga 3.4\times 10^{-7 }$ at $z\sim 2.8$ for the ``IGM" or about $\ga 4\%$ of the total cosmic metal budget (although we note this may include contributions from higher \hi\ column density absorbers than LYAF absorbers). Overall the ionized universe  probed by absorbers $\mlnhi <20$ contains about 40\% of the metals at $2.2 \la z \la 3.6$. Combining these results with those from \citet{peroux20}, the majority of metals are in the neutral and ionized gas at $z\sim 2.8$.

At $z<1$,  $\Omega_{\rm m}  = (4.6, 6.5, 25.9)\times 10^{-7}$ for the SLFSs, pLLSs, and LLSs, respectively, \citepalias{lehner19}.\footnote{In \citetalias{lehner19}, we inadvertently used the mean logarithmic metallicity, instead of the linear mean, which does not change the value for the LLSs, but does for the SLFSs and pLLSs, which are, respectively, $0.177 Z_\sun$ and $0.197 Z_\sun$. This increases $\Omega_{\rm m}$  by a factor 5 for these absorbers. As the LLSs dominate the metal budget, the effect on the total $\Omega_{\rm m}$ for the SLFSs, pLLSs, and LLSs  is only an increase by a factor 1.3, i.e., $\Omega_{\rm m} \simeq 3.7\times 10^{-6}$ for these absorbers at $z<1$. } In terms of comoving mass density,  the combined SLFSs, pLLSs, and LLSs at $z<1$ have $\rho_{\rm m} = 4.7\times 10^5$ M$_\odot$\,cMpc$^{-3}$. Therefore the amount of metals in SLFSs, pLLSs, and LLSs has increased by a factor 3 from $z\simeq 2.2$--3.6 to $z\la 1$. However, the total expected cosmic metal density  at $z\sim 0.5$ is about $\Omega^{\rm exp}_{\rm m} \simeq 6\times 10^{-5}$ or cosmic metal mass density $\rho^{\rm exp}_{\rm m} \simeq 8\times 10^6$ M$_\odot$\,cMpc$^{-3}$ \citep{peroux20}, a factor 7 larger than at $z\sim 2.8$. This implies that the (photoionized) metals associated with the SLFSs, pLLSs, and LLSs account for about 6\% of the cosmic metal budget at low redshift, a factor 2 times lower than the contribution of the same absorbers at $2.2 \la z \la 3.6$. As shown by \citet{peroux20}, the majority of the metals at low redshift remain in galaxies (including stars), in their vicinity, and in the hot intra-cluster medium, demonstrating a major shift in the distribution of the metals from 2.3 Gyr to $\ga 9$ Gyr  post-Big Bang.

\section{Comparison to the FOGGIE Simulations}\label{s-disc-sim}

\subsection{Motivation and Methodology}\label{s-foggie-motivation}
In this section we compare the KODIAQ-Z metallicity measurements as a function of \hi\ column density to the simulated CGM from the FOGGIE project \citep[][and \url{https://foggie.science/}]{peeples19}. This project is comprised of cosmological zoom simulations of six Milky Way--like halos. While KODIAQ-Z selects for absorbers without knowing their galaxy associations, absorbers with the highest column densities probe overdensities that are consistent with being virialized (mini)halos of galaxies (see \S\ref{s-phys} and \S\ref{s-foggie-insight}). The FOGGIE simulations let us address whether or not the metallicity-\hi\ column density distribution seen in KODIAQ-Z could be explained by only the CGM and nearby IGM of progenitors to Milky Way-mass galaxies.

Many groups, including FOGGIE, have recently shown that better spatial resolution in the simulated CGM leads to a natural development of small clouds \citep{vandevoort18,peeples19,rhodin19,suresh19}, largely because once the cooling length of the gas is resolved the gas can cool and condense into smaller structures \citep{hummels19,zheng20}. Importantly for comparisons to observables such as metallicity, improved spatial resolution in the low-density gas also reduces the artificial mixing of metals, thereby allowing gas to remain at ``high'' and ``low'' metallicities even in the CGM; that is, while the median metallicities do not change much, the {\em scatter} about that median increases with improved resolution \citep{corlies20}. In the 
FOGGIE cosmological zoom-in simulations analyzed here the CGM resolution at $z=2$ is 91 proper pc in the cooler gas, with a coarsest resolution of 365 proper pc. (The resolutions are correspondingly a factor of $0.85$ and $0.75$ smaller at $z=2.5$ and $z=3$, owing to the spatial resolution being at a fixed comoving size). This spatial resolution profile enables  both the development of small relatively-dense and cool cloudlets {\em and}\ for the hotter, more rarefied, generally more metal-rich gas to be well-resolved. 
Our analysis uses the $z=2$, 2.5, and 3 outputs for each of the six FOGGIE halos. As presented in \citet{simons20}, these halos are selected to be ``Milky Way--like'' in that they are roughly $\sim 10^{12}$\,M$_{\odot}$ at $z=0$ and have no mergers of mass ratios more than 1:10 after $z=2$ (see also \citealt{zheng20} and \citealt{lochhaas20} for the full histories and $z=0$ properties of the Tempest, Squall, and Maelstrom halos). While the KODIAQ-Z absorbers are not selected to be associated with Milky Way--like halos, galaxy halos of this size account for most of the stellar mass in the universe and therefore most of the chemical enrichment.

We extracted \hi\ absorbers from each sightline using the Synthetic Absorption Line Surveyor Application (SALSA, \citealp{boyd20}) and its Simple Procedure for Iterative Cloud Extraction (SPICE) method. Across all six FOGGIE galaxies, we found 8825 \hi\ absorbers at $z=3$, 8207 at $z=2.5$, and 7694 at $z=2$, for a total of 24,726 \hi\ absorbers. SPICE identifies absorbers based on \hi\ number density, without the need for synthetic spectra. First, a number density cutoff is determined such that 80\% of the total column density lies above the cutoff\footnote{Adjusting this cutoff has a slight effect on the total number of absorbers but does not significantly impact the resulting distribution of properties. Using a 70\% threshold results in a 6\% drop in the number of \hi\ absorbers, and a 90\% threshold a 11\% increase.}. An interval is defined encompassing the sightline's cells that lie above this cutoff. This interval is then masked, and a new number density cutoff is found. The intervals from these two generations are then combined if they overlap along the line of sight and their average line-of-sight velocities are within 10 \kms\ of each other. These steps are repeated until the total column density of the remaining, unmasked cells falls below $\mlnhi = 12$. SALSA then calculates the total column density and column-density weighted average metallicity of each identified absorber.

While KODIAQ-Z is an \hi-selected survey, we also extracted \civ\ and \siiv\ absorbers from these same sightlines in order to assess how our sample of simulated absorbers would differ if we also required the presence of metal absorption. Carbon and silicon are not tracked by the FOGGIE simulations, so they are inferred from the metallicity field assuming solar abundances. Their ionization states are estimated using the Trident code \citep*{hummels17}, based on a table of Cloudy equilibrium models \citep{ferland13} using the \citet{haardt12} UV background with self-shielding \citep{emerick19}.  We pair our \hi\ and metal absorbers based on their position within the simulations and their velocities. To be paired, the two absorbers must be within the same sightline, within 100~pc of each other, and have line-of-sight velocities within 10 \km\ of each other. Pairing absorbers in this way is robust to minor changes in distance and velocity. Of our 24,736 \hi\ absorbers at redshifts 2, 2.5, and 3, 1189 have an associated \civ\ absorber and 358 are associated with an \siiv\ absorber.

\subsection{Comparison of the Metallicities}\label{s-foggie-com}

\begin{figure*}[tbp]
\epsscale{1.1}
\plotone{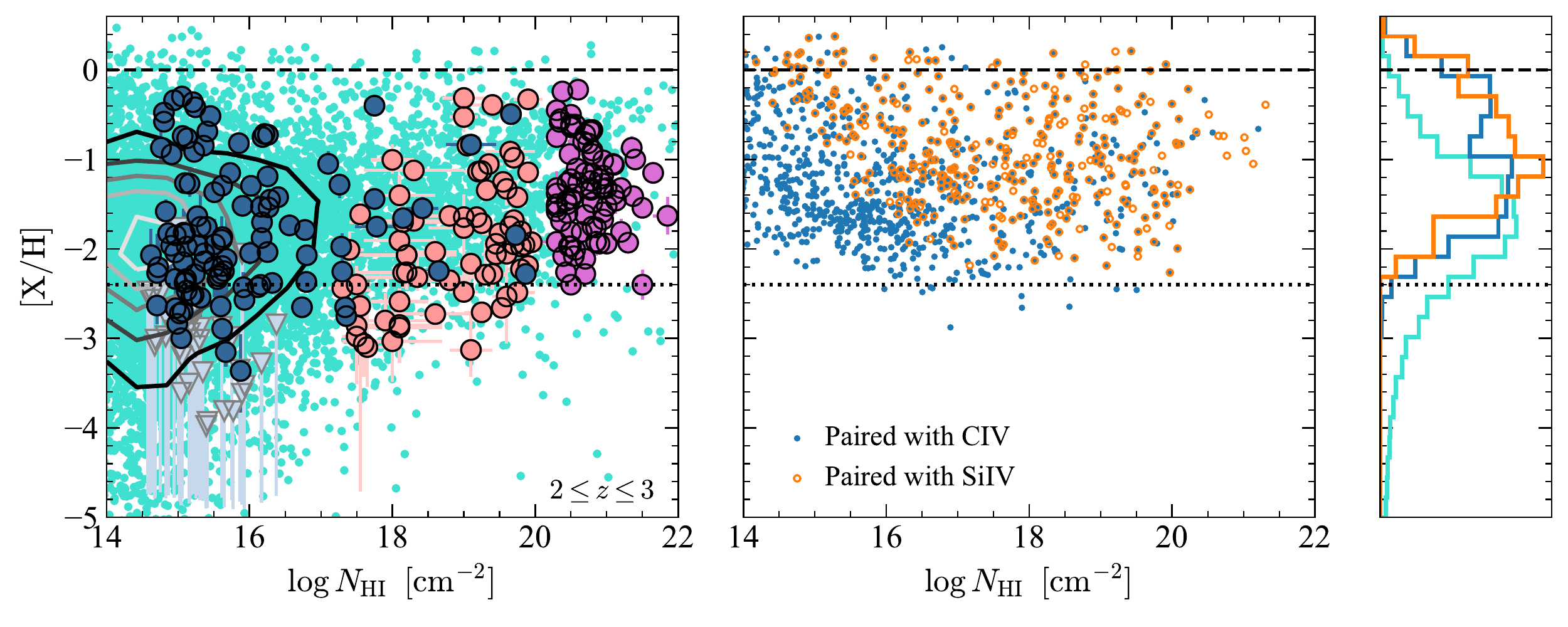}
\caption{Metallicities of the \hi\ absorbers extracted from the FOGGIE cosmological zoom-in simulations \citep{peeples19,simons20}. On the left, all \hi\ absorbers (cyan) are compared to the KODIAQ-Z, HD-LLS, R12-DLA surveys (where we limit these surveys to $2 \le z\le 3$ to  match the FOGGIE redshift range). Contours have been added where the density of cyan scatter plot points is highest. In the middle panel, we only show \hi\ absobers that have been paired with a \civ\ (blue filled circles) or \siiv\ (orange open circles) absorber, analogous to requiring detection of \civ\ or \siiv\ with $\log N_{\rm X} > 12$. Probability densities of both the full FOGGIE \hi\ absorber population (cyan) and the subset with paired metal absorbers (blue and orange) are shown on the right. The dashed and dotted lines are the solar and VMP metallicity levels, respectively.
\label{f-met_vs_nh1_sim}}
\end{figure*}

\begin{figure}[tbp]
\epsscale{1.1}
\plotone{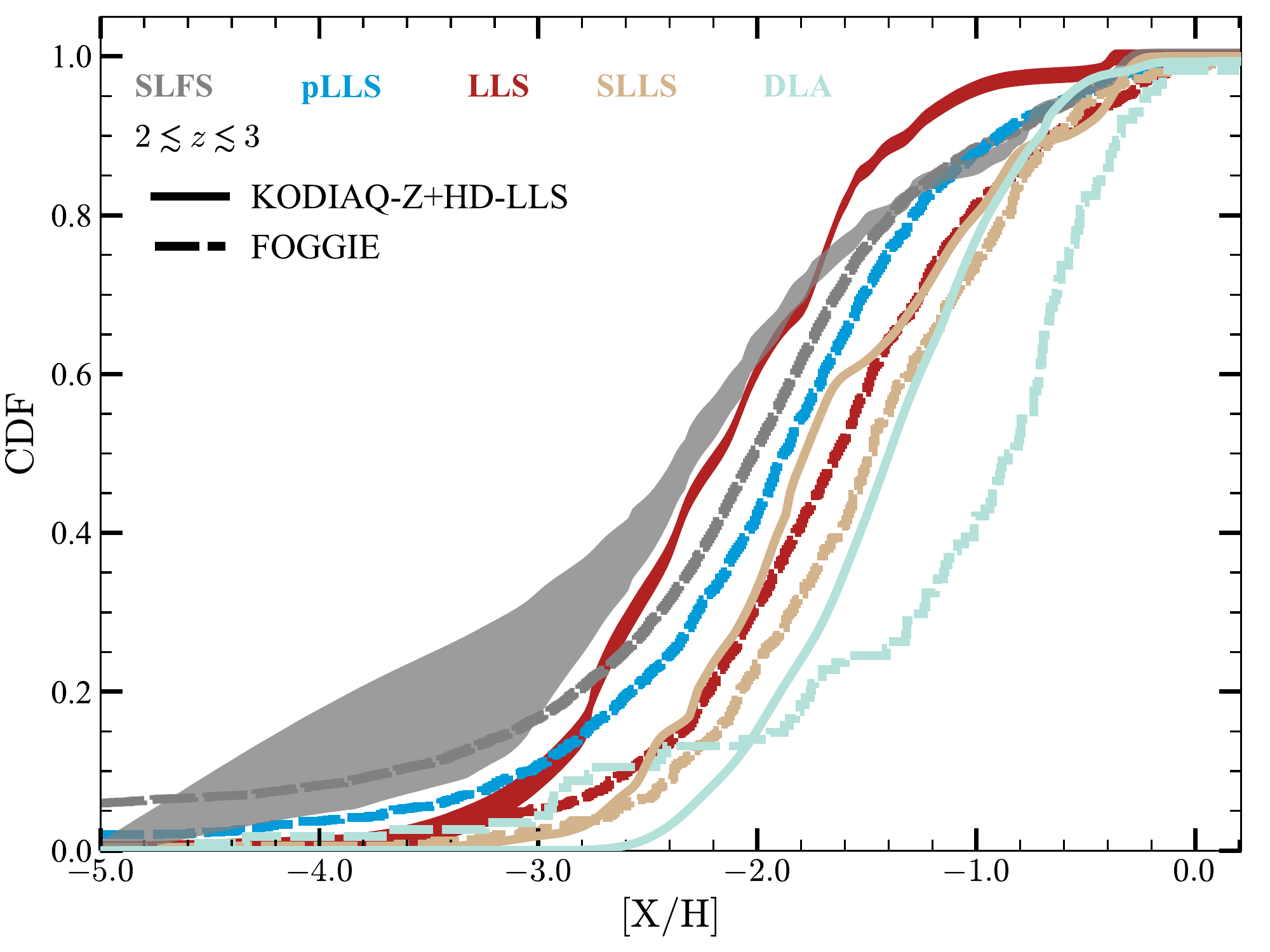}
\caption{Comparison of the metallicity CDFs for the different categories of absorbers in the KODIAQ-Z+HD-LLS+R12 observations and FOGGIE simulations at $2\le z\le 3$ (the sample of pLLSs in this redshift interval for the observation sample is too small to produce a reliable CDF and therefore is not shown here).
\label{f-met_cdf_sim}}
\end{figure}


The column density and average metallicity of our extracted FOGGIE \hi\ absorbers are shown in  Fig.~\ref{f-met_vs_nh1_sim}. The plot bounds and dotted and dashed lines are the same as in Fig.~\ref{f-met_vs_nh1}. The left panel shows all \hi\ absorbers in cyan, with the observational samples overplotted. Gray-scale contours have been added to illustrate the structure of the scatter plot where the density of points is highest. For the rightmost panel, no metallicity limit or metal line requirement has been imposed on the \hi\ absorbers. The middle panel only shows \hi\ absorbers that are close to a \civ\ or \siiv\ absorber in position and velocity space, which---due to column density limits on our selection---is analogous to requiring a \civ\ or \siiv\ detection with column densities $>10^{12}$ \cmm. This limits the metallicity of the \hi\ absorbers to $\xh \ga -2.4$, which is the limit defining the metal-poor regime for KODIAQ-Z. Imposing requirements on metal line detection can therefore limit the observed metallicity range (reinforcing the idea that a \hi-selection is required to probe the entire metallicity distribution). It should be noted that the gradual increase in the absorber metallicity as \nhi\ decreases seen in the middle panel is caused by the column density requirement of $\log N_{\rm X} > 12$ placed on the simulated \civ\ and \siiv\ absorbers. The far right-panel in this figure compares probability densities for the full, \civ-paired, and \siiv-paired \hi\ absorber populations in FOGGIE.

There is a broad agreeement between the FOGGIE \hi\ absorbers and those seen by the KODIAQ-Z, HD-LLS, and R12-DLA surveys. Both the observed and simulated absorbers follow similar trends in metallicity with \nhi. Table~\ref{t-stat-foggie} gives the median, mean, standard deviation, and IQR range on the metallicity for different classes of \hi\ absorber, which can be compared to the observed populations in Table~\ref{t-stat} (although note that the observations probe a larger redshift range). Both the mean and median metallicities increase with \nhi, with the spread tending to decrease as \nhi\ increases. This is remarkable because neither EAGLE nor FIRE zoom simulations could reproduce the similar trend seen at low redshift \citepalias{wotta19,lehner19}. In Fig.~\ref{f-met_cdf_sim}, we compare the metallicity CDFs for the various absorbers at $2 \la z\la 3$ in the FOGGIE simulations and observations, further demonstrating the similar trends of increasing metallicity with increasing \nhi.\footnote{The FOGGIE \hi\ absorbers can have metallicity as low as $\log(Z/Z_\odot) = -8$ (the FOGGIE metallicity floor). There are absorbers with $\log(Z/Z_\odot) < -5$ classified as SLFSs and pLLSs. These absorbers are not excluded from the CDFs in Fig.~\ref{f-met_cdf_sim}, so these CDFs do not start at zero. } However, this figure also shows that the metallicity of the FOGGIE absorbers is higher than that seen in the observations (even though their IQR or standard deviations are similar), especially for the LLSs and DLAs. 

The FOGGIE absorbers therefore mimic the general trend seen in the observations, where higher \hi\ column density regimes are dominated by higher metallicities. In FOGGIE, however, the metallicity at which a particular \hi\ regime is most prevalent is higher than seen in the observational data. FOGGIE's SLFSs have a higher metallicity than both the observed SLFSs and the observed LLSs. Moreover, the FOGGIE LLSs more closely match the CDF of the observed SLLSs. Similarly, the FOGGIE SLLS CDF is closer in metallicity to the observed DLAs. The FOGGIE DLAs have a substantial higher metallicity (the FOGGIE median metallicity is 0.5 dex higher or a factor 3 higher than in the R12-DLA sample). 

The metallicities used for the FOGGIE absorbers are conceptually slightly different from the metallicities derived in KODIAQ-Z and other spectral surveys, in that the metallicities reported for FOGGIE are the column density-weighted average of the gas cells identified as the absorber rather than the result of a fit to single-phase ionization models. However, \citet{marra21} show that these two approaches, when applied to the same physical system, do not lead to large discrepancies in the average metallicity of the gas. The precise metallicities of FOGGIE's absorbers likely depend strongly on its implementation of supernova feedback, and these would be expected to differ slightly (e.g., be generally lower or higher) for different implementations. Yet despite FOGGIE's overall higher metallicity, the general similarity between the observed and simulated absorbers in Figure~\ref{f-met_vs_nh1_sim} seems to be a good indication that the {\it trends}\ observed in the KODIAQ-Z, HD-LLS, and DLA surveys can in principle be explained by CGM and nearby IGM gas around typical galaxies at these redshifts.

\begin{figure}
\epsscale{1.2}
\plotone{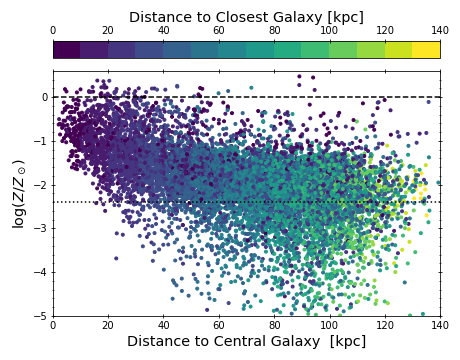}
\caption{Metallicities of the FOGGIE \hi\ absorbers versus their distance from the central galaxy, colored by the distance to their closest galaxy. Possible ``closest galaxies'' include the central and all satellites with more than 1\% of the central's stellar mass. High metallicity absorbers far from the central galaxy tend to be close to a satellite galaxy. Dashed and dotted lines are the same as in Figure \ref{f-met_vs_nh1}}
\label{f-met_vs_nh1_sim_dist}
\end{figure}

\subsection{Insights from FOGGIE Simulations}\label{s-foggie-insight}

Given some similarities in the metallicity trends with \nhi\ between our samples of observed and simulated absorbers, we can use the FOGGIE simulations to shed light on the origin(s) of these absorbers and their connection to galaxies. From the empirical results from KODIAQ-Z, the overdensities (at least for the pLLSs, LLSs, and SLLSs) imply that these absorbers should largely probe gas within virialized (mini)halos (see \S\ref{s-phys}). The Keck Baryonic Structure Survey (KBSS) also shows that at  $z \sim 2$--3 there is a strong incidence of absorbers with $\mlnhi >14.5$ with galaxies at transverse physical distance $\la 300$ kpc and velocity separation between the absorber and galaxy redshifts $\la 300$ \km\ \citep{rudie12}. This connection is not observed for lower \nhi\ absorbers, also implying some connection between $\mlnhi>14.5$ absorbers and  galaxies. On the other hand, recent MUSE IFU observations also show at least some LLSs may probe IGM filaments \citep{fumagalli16a,lofthouse20}. With the FOGGIE simulations, we can explore the origins of these absorbers without observational constraints due to, e.g., due to brightness limit of the galaxies. It is possible, however, that biases due to the selection criteria of the FOGGIE galaxies as Milky Way-like analogues at $z=0$ could bias the results in some way. Importantly, we can also explore if and how the origins may depend on the metal enrichment of the absorbers. 

Each of the central FOGGIE galaxies is surrounded by a number of satellites at $z=2$--3. These satellite galaxies likely contribute to the enrichment seen in our absorbers, and it is worth determining whether a given absorbers is more likely to be associated with such a satellite than with the central galaxy. Including only those satellites with at least 1\% of the central galaxy's stellar mass, we identify which galaxy each absorber is physically closest to. In Fig.~\ref{f-met_vs_nh1_sim_dist}, we compare the absorber metallicities to their distance from the central galaxy and the distance to their closest galaxies, be that a satellite or the central galaxy. Generally, distance to an absorber's closest galaxy increases with distance from the central galaxy, which is the behavior we expect for absorbers whose closest galaxy is the central galaxy. Metallicity also tends to decrease with increasing distance from the central galaxy. Yet, there are important exceptions to these two trends: some absorbers are far from the central galaxy yet close to a satellite, and these absorbers span the range of extracted metallicites. This includes some absorbers distant from the central galaxy that have metallicity close to and above solar. Therefore, not all high metallicity absorbers should be assumed to be associated with the central galaxy.

\begin{figure}
\epsscale{1.2}
\plotone{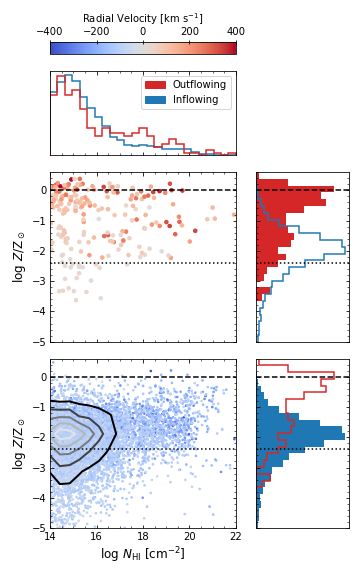}
\caption{Metallicities of the FOGGIE \hi\ absorbers in Figure \ref{f-met_vs_nh1_sim}, split into outflowing (red) and inflowing (blue) based on $|\hat{v}\cdot\hat{r}| > 0.5$ in the frame of the host central galaxy. Saturation shows the magnitude of the radial velocity. Probability densities are provided as histograms along the side and top. Both metallicity distributions are repeated in the side panels, with the filled histogram corresponding to the adjacent scatter plot. Contours have been added to show where the density of inflowing absorbers is highest.}
\label{f-met_vs_nh1_sim_split}
\end{figure}

\begin{figure*}
\epsscale{0.8}
\plotone{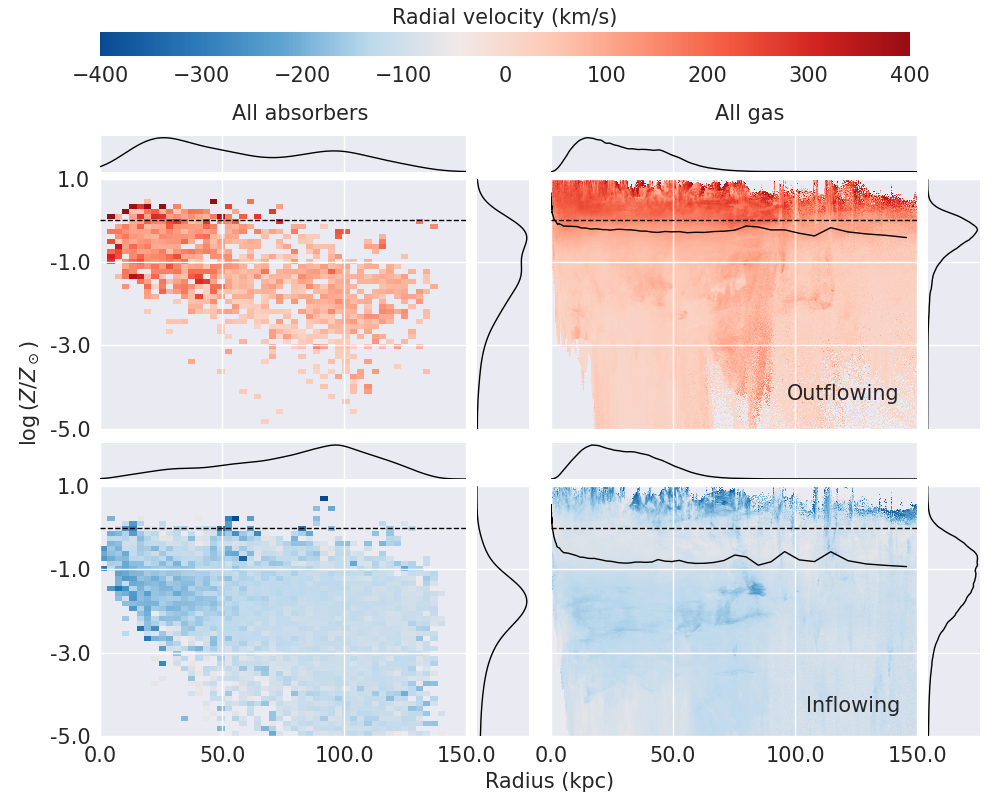}
\caption{On the left, the metallicities of outflowing (red) and inflowing (blue) FOGGIE absorbers are binned versus their distance from their host central galaxies, colored by the average radial velocity for that bin. On the right, the gas in all eighteen FOGGIE snapshots is shown in the same configuration. The mean metallicity ($\log \langle Z/Z_\odot \rangle$) is marked with a black line. The dashed line indicates the solar metallicity level. Probability densities are shown at the top and sides of each panel. }
\label{f-met_vs_radius_sim_split}
\end{figure*}

In Figs.~\ref{f-met_vs_nh1_sim_split} and \ref{f-met_vs_radius_sim_split}, we select FOGGIE absorbers that are inflowing toward or outflowing from the central galaxy. This separation is made based on the angle  between an absorber's 3D velocity vector and its position vector, both of which are calculated in the rest frame of the relevant central FOGGIE galaxy. Outflowing gas must have an angle $\theta > 60^\circ$, while inflowing absorbers must have $\theta < 120^\circ$. This is equivalent to $\hat{v} \cdot \hat{r} = \cos\theta = \pm 0.5$. This cut prevents gas that is neither strongly inflowing nor outflowing from being selected.

Fig.~\ref{f-met_vs_nh1_sim_split} shows the metallicity and \hi\ column densities of our inflowing (blue) and outflowing (red) FOGGIE absorbers. PDFs are also included for these two quantities, with both metallicity PDFs repeated in the upper and lower panels for comparison. The inflowing absorbers dominate by number, but both categories have similar \hi\ column distributions with absorbers becoming more numerous at lower \hi\ column densities. The outflowing absorbers skew towards the high metallicity end of the distribution. This is not surprising, as in FOGGIE winds are launched by the same supernova feedback that also enriches the surrounding gas. There is a plateau in the outflowing metallicity distribution around $-2 \la \log Z/Z_\odot \la -1$, which may be tracing older outflows that have slowed and mixed with the ambient CGM. These absorbers number significantly less than the inflowing absorbers at the same metallicity, however, so it would be difficult for observations to disentangle these less enriched outflows. The inflowing absorbers follow a more symmetric metallicity distribution, with a peak about 0.5~dex above the metal-poor limit for KODIAQ-Z; though, as stated above, the FOGGIE absorbers generally tend to have higher metallicities than seen in the random sightlines observed by KODIAQ-Z. The inflowing absorbers also extend to the lowest metallicities. Even though Fig.~\ref{f-met_vs_nh1_sim_dist} reveals that some absorbers may be closer to a satellite than to the central galaxy, the trends in Fig.~\ref{f-met_vs_nh1_sim_split} are unchanged when the absorbers are limited to those whose closest galaxy is the central. The biggest difference is that there are no outflowing absorbers with $ \log Z/Z_\odot < -2.4$.

Although the outflowing absorbers dominate the high end of the metallicity PDF, they are far outnumbered by the inflowing absorbers even at the highest metallicities. This is largely due to the selection on \hi\ and the underlying physics included in the FOGGIE simulations. There is very little cool gas that would form \hi\ within the outflows as these are driven only by thermal pressure. Simulations with additional non-thermal pressure sources such as magnetic fields and cosmic rays \citep{butsky18,ji20} demonstrate an increase in \hi\ at large radii within the CGM. It is therefore quite likely that the number of \hi\ absorbers classified as either inflowing or outflowing following our definition would change if these physical processes were included in the FOGGIE simulations, though it is likely that the numbers would still favor inflowing gas.

With Fig.~\ref{f-met_vs_radius_sim_split}, we look at the metallicity and velocity of the FOGGIE absorbers as a function of their distance from the central simulated galaxy. We also compare the absorbers to the underlying gas distributions. This distribution compiles together the gas in all 18 FOGGIE outputs we sampled. Probability densities are shown for the metallicity and radial distributions, and the mean metallicity of the gas is shown with a black line. Colors indicate the mean radial velocity at that metallicity and radius; we note that, unlike with the absorbers, the gas radial velocity has no selection based on the angle toward or away from the galaxy.

First, we consider outflowing material: outflowing absorbers with the highest metallicities tend to be at or within the virial radius of these central halos, which at $z=2$--3 is about 30--50 kpc, and have the highest velocities. These velocities are generally higher than what is seen in the gas at similar radii and metallicities, and appears consistent with fast-moving enriched material launched by stellar feedback. On the other hand, high metallicity gas can be found even beyond the virial radius, implying that galactic outflows become more diffuse as they travel farther away from the galaxy. The outflowing absorbers with the lowest metallicity are found well outside the virial radius, and their velocities are comparable to the gas in this regime. This supports the interpretation that these absorbers come from previously ejected gas.

Inflowing absorbers can be found at all radii, with their metallicity generally increasing the closer they are to their central galaxy, similar to the outflowing gas. The velocity of these inflowing absorbers is also higher near the galaxy. The FOGGIE zoom simulations contain several satellite and pre-merger galaxies that are producing their own enriched outflows, which explains why high metallicity absorbers can be found at large distances from the central galaxy. These absorbers do not appear to reflect the fastest inflowing gas, which has very high metallicity and can again be found at all radii.

Together, Figs.~\ref{f-met_vs_nh1_sim_split} and  \ref{f-met_vs_radius_sim_split} imply that, given a set of \hi-selected observed absorbers that are known or assumed to be associated with a galaxy system, most of these absorbers are likely to be inflowing towards a central galaxy. The precise number of inflowing and outflowing absorbers predicted by simulation will depend on the included physics, especially non-thermal pressure sources, though inflowing absorbers will likely still dominate. This dominance is true across column densities and metallicities, as satellites and nearby companions enrich gas flowing towards the central galaxy. Whether inflowing or outflowing, more enriched absorbers are more likely to be located physically close to the central galaxy. 

\section{Discussion}\label{s-disc}

\subsection{Bridging the LYAF and DLAs}\label{s-disc-bridge}
The history of metallicity in the universe provides an important constraint on models of galaxy formation and evolution. The metals in the universe represent a fossil record of star formation; characterizing how the metallicities change with \hi\ column density (and hence overdensities or densities) yields information on the transport of metals and  efficiency of that transport from the densest to the most diffuse regions of the universe. At high redshift, much of the effort has focused on the metal enrichment of the most diffuse regions probed by the LYAF (e.g., \citealt{cowie95,songaila98,ellison00,schaye00,schaye03,aguirre04,simcoe04,simcoe11b}) and the densest regions of the universe probed by DLAS (e.g., \citealt{pettini97,pettini99,prochaska99,prochaska03,rafelski12,jorgenson13}). The HD-LLS survey (\citealt{prochaska15,fumagalli16}, and see also \citealt{berg21} for the XQ-100 survey of SLLSs) has surveyed lower \hi\ column density absorbers mostly in the range $17.8 \la \mlnhi <20.3$. With KODIAQ-Z (this work, and  earlier KODIAQ surveys \citealt{lehner14,lehner16}), we bridge the gap between the DLA/SLLS regimes and the LYAF. \footnote{We  show in \S\ref{s-met-z} that similar trends of the metallicities with \nhi\ are observed at low and high redshifts in the range $15 \la \mlnhi \la 22$. However, the current UV observations at low $z$ do not have sensitivity to probe much lower metallicities than $\xh <-1$ in the LYAF at $z<1$ (this would need to await a future $\sim$6-m UV space-based telescope), and we cannot yet assess the full range of metallicities for absorbers with $\mlnhi <15$ at $z<1$. Therefore, we keep our discussion focused on the high redshift universe.}

In \S\ref{s-met-nhi}, we show that the scatter of absorber metallicities at $\xh \ge -2.4$ is quite similar for the DLAs  and the $14.6 \la \mlnhi \la 20$ regime, i.e., metal enriched gas is observed not only in the densest regions (in or near galaxies) but has also spread to the more diffuse regions down to overdensities $\delta \la 10$ (see Fig.~\ref{f-met_vs_nh1}). However, as \nhi\ decreases below $\mlnhi \la 14.4$ (diffuse IGM), most of the absorbers have metallicities $\xh \la -1.6$ (see Fig.~5 in \citealt{simcoe04}), while gas with $-1.6 \la \xh \la -0.2$ is commonly observed over the entire range $14.6 \la \mlnhi \la 22$. Therefore the diffuse IGM probed by  LYAF absorbers has not been enriched at the same level as stronger \hi\ absorbers. 

In contrast, for metallicities $\xh \la -2.4$, blind surveys of DLAs do not reveal a significant population of VMP DLAs \citep[e.g.,][]{rafelski12,jorgenson13}. Yet, VMP absorbers are regularly observed at $\mlnhi \la 20$, and their frequency increases with decreasing \nhi\ (see Table~\ref{t-frac} and Fig.~\ref{f-met_vs_nh1}). These VMP absorbers are of course also observed in the LYAF. The median metallicity of the LYAF at $z\sim 2.5$ is around $\xh =-2.8$ (\citealt{simcoe04,simcoe11b} and see also, e.g., \citealt{ellison00,schaye03,aguirre04}), which is a factor 2.5 lower than the median metallicity of the SLFSs at $\langle z\rangle \simeq 2.8$ (Table~\ref{t-met-sum}). However, owing to sensitivity issues, the lowest metallicities found for LYAF absorbers are around $\xh \sim -3.5$, while our survey reveals a population of absorbers with $\xh<-3.5$ that is nearly and completely metal-free  (see \S\ref{s-disc-pris}, and also \citealt{fumagalli11b,lehner16}). \citet{simcoe04} estimated that about 70\%--80\% of the LYAF has been enriched to $\xh \ga -3$, but 20--30\% might be chemically pristine gas at low densities, which is about the same as the fraction of SLFSs with $\xh <-3$. 

Therefore, gas in the intermediate overdensity regime between the DLAs and LYAF has metallicities that are found in both the most diffuse and densest region of the universe. Gas probed by absorbers with $14.6 \la \mlnhi \la 20$ has a much larger dispersion in metallicity than observed either in $\mlnhi \la 14.4$ or $\ge 20.3$ gas.  The simultaneous presence of abundant pristine, metal-rich, intermediate metallicity gas is unique to the \hi\ column density range $14.6 \la \mlnhi \la 20$. Below we discuss in more details these various levels of chemical enrichment and the mixing of metals in the $14.6 \la \mlnhi \la 20$ range.

\subsection{Pristine Gas}\label{s-disc-pris}
From the KODIAQ-Z survey and CCC at low redshift, we find that there is an abundance of VMP absorbers in relatively high overdensity (except for the DLAs by definition). While the metallicities have increased by about a factor 8--10 from $2.2 \la z \la 3.6$ to $z<1$,  the fractions of VMP absorbers are strikingly similar at low and high redshifts in $\mlnhi <19$ absorbers. On the other hand, owing to the overall increase of the metallicity from $2.2 \la z \la 3.6$ to $z<1$, it is not surprising that extremely metal poor gas with $\xh < -3.5$ or even $<-3.0$ is scant at $z<1$ \citepalias{lehner19}.  With KODIAQ-Z, we determine that the fractions of SLFSs and pLLSs with $\xh<-3.5$ and $<-3.0$ are about 3\%--10\% and 15\%--25\% (90\% confidence intervals, see Table~\ref{t-frac}). Combining KODIAQ-Z and HD-LLS, similar numbers are derived for the LLSs. For the SLLSs, the fractions of absorbers with these metallicities are smaller by about a factor 2. 

The pristine LLSs reported by \citet{fumagalli11b} at $z\sim 3.5$ have $\xh <-3.8$ and $<-4.2$, but at the time of this discovery it was impossible to say how common this population was. With KODIAQ-Z, we can revisit this question (see also \citealt{lehner16}). The lowest metallicity where some metals are detected is around $\xh \la -3.8$ in KODIAQ-Z (see Fig.\ref{f-met_vs_nh1}). We use that value of $\xh \simeq -3.8$ to separate pristine (no metal) from metal-enriched (even at very low levels) absorbers. Considering Fig.~\ref{f-met_vs_nh1}, we can separate the population of \hi\ absorbers into two broad categories owing to the impact that upper limits on \xh\ can have for interpreting the frequency of pristine absorbers:

1) For absorbers with $16 \le \mlnhi\le 20$, there are only 11 upper limits on \xh: 3 out 179 these absorbers have $\xh <-3.8$, implying a frequency of pristine absorbers in the range 0.7\%--4.2\% (90\% CI) in this column density interval. This number cannot increase by much more than a factor $\sim$2 including data with higher upper limits. Even if all the upper limits represent pristine gas for absorbers with $\mlnhi\ge 16$, this would increase the frequency of pristine absorbers to only 3.7\%--9.6\% (90\% CI). Hence, the fraction of pristine $\mlnhi\ge 16$ absorbers (i.e., gas with overdensities $\delta \ga 50$--100) is at least 1\% and at most 10\%  at $2.2 \la z \la 3.6$. 

2) The second regime is for absorbers with $14.6 \la \mlnhi<16$ where the upper limits on the metallicities are much more numerous. Taking strictly $\xh <-3.8$, there are 4 absorbers out of 142, implying a frequency of pristine absorbers of 1.3\%--6.1\% (90\% CI), a similar percentage to that of absorbers with higher \nhi. However, if all the $\mlnhi<16$ absorbers with upper limits such as $\xh<-3$ probe pristine absorbers, then the frequency of  pristine $\mlnhi<16$ absorbers would increase to 13.3\%--23.9\% (90\% CI). This would be similar to the possible population of metal-free LYAF absorbers at similar redshifts  (see previous section and \citealt{simcoe04}). The fraction of pristine gas in overdensities $\delta \ga 50$--100 at $2.2 \la z \la 3.6$ is therefore at the level of a few percent, while for smaller overdensities it is likely in the range 10\%--20\%. 

The frequency of mock absorbers with $\xh <-3.8$ in the FOGGIE simulations is very similar to the frequencies found in KODIAQ-Z, despite the FOGGIE absorbers on average have higher metallicities than observed. In contrast, cruder resolution simulations show a tendency to have over-enriched pLLSs and LLSs (see discussion in \citealt{fumagalli11b}). In the  EAGLE HiRes cosmological hydrodynamical simulations, \citet{rahmati18} find a fraction of pristine gas with $\xh <-3.8$ of about 10\% in the pLLS and LLS regimes. However, contrary to FOGGIE and the observations reported here or at low redshift \citepalias{wotta19,lehner19}, the EAGLE simulations show little change of the metallicity with \nhi\ (see also \citetalias{wotta19,lehner19}); the FIRE simulations also show similar results \citep{hafen17}. Consistent with the lack of \nhi\ dependence, these simulations find a significant population of $\xh<-3$ DLAs that is not observed. This highlights some issues in simulations that may be related to the implementation of  feedback physics, but is probably also related to insufficient resolution in the more diffuse gas. As discussed in \S\ref{s-foggie-com}, low-metallicity inflowing gas is able to naturally get much closer to the galaxy when the CGM structures are resolved than in the standard-resolution simulations because the metals are not forced to over-mix in the regions around galaxies (see also \citealt{peeples19,vandevoort18,hummels19,suresh19}).  As discussed in \S\ref{s-foggie-insight} and shown in Figs.~\ref{f-met_vs_nh1_sim_dist}, \ref{f-met_vs_nh1_sim_split}, \ref{f-met_vs_radius_sim_split}, the pristine  and the VMP absorbers predominantly probe inflowing cool gas (see, also \citealt{hafen17,rahmati18,suresh19}), consistent with the idea that these absorbers  are ideal candidates for the cold-mode accretion \citep[e.g.,][]{keres05,dekel09,fumagalli11a}.

\subsection{Extreme Metallicities and Metal-rich gas}\label{s-disc-rich}
The metal-enriched diffuse gas constitutes a record of the transport of metals from the densest to the most diffuse regions of the universe. In \S\ref{s-disc-bridge}, we discuss that while the LYAF shows evidence for metal enrichment, it is not enriched to levels seen in higher \hi\ column absorbers have. Metallicities in the range $-1.6 \la \xh \la -0.2$ are commonly observed over the entire range $14.6 \la \mlnhi \la 22$, but not in the LYAF. Therefore, at $2.2 \la z \la 3.6$, galaxies have not yet polluted the diffuse IGM with metals at the same level as denser regions of the universe. 

In \citet{lehner16}, we reported the discovery of one supersolar pLLS at $z\simeq 2.5$ with $\xh \simeq +0.2$.\footnote{The HIRES spectrum of this QSO was not released in KODIAQ DR2 and therefore is not part of the present KODIAQ-Z survey.} Besides its metallicity, this pLLS  is unique on several other levels including the detection of \oi, its small physical (0.35 pc), relatively high density ($\mnnh \simeq 0.2$ \cmmm), and multiphase nature, with \civ\ having a very different velocity profile compared to the low ions (see Fig.~14 in \citealt{lehner16}). We argued in that paper that its high metallicity and multiphase nature strongly suggest that it directly probes an active outflow from a proto-galaxy at $z\simeq 2.5$. At  $z\sim 1.8$, \citet{prochaska06} uncovered two supersolar SLLSs, suggesting that these metal-rich absorbers may represent a significant metal reservoir in the young universe. 

However, the HD-LLS survey does not report any supersolar SLLS (see Fig.~\ref{f-met_vs_nh1}), and the XQ-100 survey only reports one supersolar SLLS at $z\sim 2.5$ (see Fig.~2 in \citealt{berg21}; we exclude poorly-constrained upper limits on the metallicity that are above solar in that survey). Adding that supersolar pLLS to the KODIAQ-Z+HD-LLS survey would imply that only 1 out 242 absorbers has a supersolar metallicity at $2.2\la z \la 3.6$, i.e., $0.41\%^{+1.42\%}_{-0.32\%}$ (90\% CI) of the absorbers have $\xh>0$. None is found in the R12-DLA either, implying that the fraction of supersolar absorbers in the \hi\ column density range  $14.6 \la \mlnhi < 22$ at $2.2\la z \la 3.6$ drops to $0.29\%^{+1.00\%}_{-0.23\%}$. Supersolar metallicity absorbers are therefore very rare at $2.2\la z \la 3.6$ for any \nhi\ absorber. 

At low redshift, the situation is quite different, with at least 2\%--6\% of supersolar absorbers  at $z<1$ (see Fig.~\ref{f-met-vs-z}, and \citetalias{lehner19}). This number could increase by a factor $\sim$3 if a harder EUVB was used instead of the HM05 EUVB to model the absorbers (at low redshift the effect of the EUVB is more important, see \citetalias{wotta19} and \S\ref{s-ion-model}). This increase of supersolar absorbers is consistent with the overall increase of the metallicities by a factor 8--10 from $2.2\la z \la 3.6$ to $z<1$ (see Fig.~\ref{f-met-nhi-z}). 

The larger fraction of metal-enriched and even super-metallicity absorbers at low redshift indicates a much larger volume of the universe has been exposed to metal pollution than at $2.2\la z\la 3.6$.  \citet{ferrara00a} developed a model that predicts at $z<1$ essentially all absorbers should have associated metal absorption, and the spread in metallicity should be less than 1 dex. This is not case. Even at $z<1$, the enrichment of the most diffuse gas in the universe is still very  inhomogeneous in view of the nearly 3 dex spread in metallicities at $z<1$ for absorbers with $15\la \mlnhi\la 19$ (see Fig.~\ref{f-met-vs-z} and next section).

\subsection{Inhomogeneous Metal Mixing}\label{s-disc-var}
There is ample of evidence for inhomogeneous abundances in \hi\ column density $14.6\la \mlnhi \la 20$ gas where the metallicities range from pristine ($\xh <-4$) to about $\xh \simeq -0.2$ dex, a factor at 6,000 spread from the lowest to highest metallicities. Similar results are found at $z<1$, but a factor somewhat smaller with $1,000$ variation between the highest and lowest metallicities (\citetalias{lehner19} and Fig.~\ref{f-met-vs-z}). However, our results also show there is inhomogeneous metal mixing in single halos as directly evidenced by the large variation in metallicity in closely redshift-separated absorbers (see Fig.~\ref{f-met-dv}). Such large metallicity variation over small redshift separations was initially reported by \citet{prochter10} between a SLFS, a pLLS, and a LLS at $z\sim 3.5$ separated in velocities along the same QSO from about 130 to 180 \km\ where the metallicity differences were a factor 3, 40, and 158. Now with a sample of 37 paired absorbers, we find similarly large metallicity variations  between absorbers separated by less than $\Delta v<500$ \km\ along a given QSO sightline where about half of paired absorbers have metallicity variations of a factor 2--3 and for the other half of a factor $>140$. It is plausible that the smaller variations may be related to the variation within a halo, while the large variations to gas in inter-halos. 

Thus, within the overdensities of a few to several hundreds, the gas probed by absorbers with $14.6\la \mlnhi \la 20$ is chemically inhomogeneous where both metal-poor and metal-rich gas are observed. This  again appears unique to the $14.6\la \mlnhi \la 20$ column density range. While for DLAs determining the metallicity in individual components is complicated since the \hi\ velocity structure cannot be recovered over $\Delta v<500$ \km, the metallicity spread for the DLAs is much smaller than observed at $14.6\la \mlnhi \la 20$ (see \ref{s-disc-bridge}). It is also much smaller for the LYAF \citep[e.g.,][]{simcoe04}, implying overall a more chemically homogeneous gas in the LYAF and DLA regimes. Therefore, in and quite near galaxies, in the gas traced by DLAs, the metal enrichment from supernovae is quite efficient, but beyond the immediate vicinity of the galaxies, the volume filling factor for metals must be low, which  is consistent with models of metal ejection from supernovae that is not expected to be efficient and confined to small regions \citep[e.g.,][]{ferrara00a,scannapieco05}. 

\section{Summary and Concluding Remarks}\label{s-sum}
Using the KODIAQ DR2 data from the KOA \citep{omeara15}, we have built a sample of 202 \hi-selected absorbers with $14.6\le \mlnhi \le 20$ at $2.2 \le z \le 3.6$ comparable in size to our companion survey, CCC, at $z <1$. The \hi\ selection and the \hi\ column density range both ensure that no bias is introduced in the metallicity distribution of these absorbers, i.e., we are sensitive to any absorbers with $\xh \ga -3.5$. The \hi\ selection also provides a clean separation of the absorbers based on \hi\ column density, which is largely used to separate the LYAF, absorbers with $\mlnhi \la 14.5$ (IGM), from the denser regions of the universe probed by stronger \hi\ absorbers. By definition, the SLFSs all have over-densities $\delta >3$ in our sample ($z = 3.6$, the maximum redshift in our statistical sample corresponds to this $\delta$ value using the analytical expression in \citealt{schaye01b}). In contrast, a metal-selection using \civ\ or \ovi\ can include a wide range of \hi\ column density absorbers from the LYAF to the pLLS regime.

In the KODIAQ-Z survey, we have 155 SLFSs, 24 pLLSs, 16 LLSs, and also 7 SLLSs, for a total of \ssz\ absorbers. To increase the number of LLSs and SLLSs in our sample, we use the \hi-selected absorbers from the HD-LLS survey that have a \hi\  column density range $17.2\le \mlnhi \le 20$, but a majority have $17.7 \la \mlnhi \le 20$  \citep{prochaska15,fumagalli16}. We also use the results from survey from \citet{rafelski12} to study the metallicity changes with \nhi\ in the DLA regime. For both of these surveys, we restrict them survey to absorbers at $2.2 \le z \le 3.6$, unless otherwise stated. 

For all the absorbers with $\mlnhi \la 20$ in the KODIAQ-Z and HD-LLS, we derive the posterior PDFs of the metallicities and other physical quantities using a Bayesian formalism that employs a MCMC sampling of a grid of Cloudy photoionization models (where we assume photons from the HM05 EUVB provide the source of photoionization). This follows directly from the methodology used in HD-LLS \citep{fumagalli16} and CCC \citepalias{wotta19,lehner19}. For absorbers with less than ideal constraints (about half of the SLFS sample and a few pLLSs), we adopt the ``low-resolution" method developed at low redshift by \citet{wotta16} and refined in \citetalias{wotta19} to estimate the metallicities. Using SLFSs and pLLSs with reliable constraints from the metal ions,  we find that \logU\ can be reasonably well modeled by a Gaussian for the SLFSs and pLLSs, which can then be used reliably as a prior in the Bayesian MCMC modeling. We  explore the effects of changing the EUVB from HM05 and HM12 to estimate the metallicities (Figs.~\ref{f-hm12vshm05}, \ref{f-hm12-hm05vsnh1}), and we find that the changes in the metallicities between these two EUVBs are negligible for the absorbers with $\mlnhi \ga 17.2$. For the SLFSs and pLLSs, the effect is only an increase of $+0.12 \pm 0.15$. The effect from changing the EUVBs on the metallicities is therefore much smaller than at $z<1$.

Our main findings on the properties of our statistical sample of \hi\ selected absorbers with  $14.6\le \mlnhi \le 20$ at $2.2 \le z \le 3.6$ can be summarized as follows.
\begin{enumerate}[wide, labelwidth=!, labelindent=0pt]
\item From the comparison of the absorption profiles and the width of the profiles, we conclude that singly, doubly, triply ionized species (e.g., \siii, \siiii, \siiv, \cii, \civ) and \hi\ often trace the same gas in absorbers with $\mlnhi < 19$. For SLFSs and pLLSs, we also find that when \ovi\ absorption is present, it has often a similar velocity structure than lower ions, i.e., the \ovi\ absorption is commonly narrow  in these absorbers. This contrasts remarkably from the \ovi\ that is frequently strong and broad when detected in absorbers with $\mlnhi \ga 17.8$. 
\item From the profile fitting of the \hi\ transitions, we show that 90\% of the components have $13.3 \le b \le 40$ \km\ with a mean  $\langle b \rangle = 27 \pm 6$ \km. This implies a temperature of the gas of $T<4 \times 10^4$ K, consistent with the gas being primarily photoionized. 
\item We find that a single-phase photoionization model is appropriate to match the column densities of the low ions to high ions (including \ovi) for the majority of the SLFSs and pLLSs. For the LLSs and SLLSs, when \ovi\ is detected, a single-phase photoionization model cannot commonly reproduce the observed \ovi\ column density, implying that as \nhi\ increases, the multiple gas-phase nature becomes more important. 
\item In our ionization models, \ca\ is allowed to vary in the range $[-1,+1]$ to accommodate for non-solar relative abundances between carbon and $\alpha$-elements caused by nucleosynthesis effects. This approach allows us to have a better sense of the uncertainties in the metallicity caused by the variability of C/$\alpha$. Overall, we find that \ca\ is in the range $-0.6 \la \ca\ \la+0.5$ as observed in other environments (DLAs, stars, \hii\ regions), but robustly studying the variation of \ca\ with \xh\ for these absorbers is hindered by the ionization corrections that add too much noise in the \ca\ distribution, hiding any subtle changes between \ca\ with \xh\ that are expected to be at the level of a factor 2--3. 
\item The 155 \hi-selected SLFSs ($14.6 \le \mlnhi <16.2$) probe a wide range of metallicities from $\xh <-4$ to $\xh \simeq -0.2$. The metallicity posterior PDF is negatively skewed with a prominent tail extending well below $\xh < -3$. It has a main peak around the median value $\xh \simeq -2.4$ and another smaller peak around $-0.6$ dex solar (the dip around $-1.1$ is observed in both binned and unbinned data). 
\item The 24 \hi-selected pLLSs ($16.2 \le \mlnhi <17.2$) probe a range of metallicity from $\xh <-4.2$ to about $\xh \simeq -1$. The sample is too small to robustly characterize the distribution and the lack of pLLSs with $\xh >-1$ could be mainly due to small-number statistics. However, both the median ($-2.1$ dex) and IQR metallicities imply an overall increase in metallicities compared to the SLFSs despite the lack of $\xh >-1$ pLLSs. 
\item Our sample of \hi-selected LLSs ($17.2 \le \mlnhi <19$) consists of 16 absorbers that we combine with the HD-LLS survey to reach a size sample of 62 LLSs. The full range and median of the LLS metallicity PDF are quite similar to that of the pLLSs. Combining the pLLSs+LLSs, the median metallicity is $-2.2$ and IQR only 1 dex compared to $-2.42$ and 2 dex, respectively, for the SLFSs. The pLLSs+LLSs  are more frequent in the metallicity range $-3.2\la \xh \la -1.2$ but less frequent at very low metallicity $\xh \la -3.5$ than the SLFSs, showing there is a shift in the metallicity enrichment properties of these absorbers below and above $\mlnhi \simeq 16.2$ (at $2.2 \le z \le 3.6$, this corresponds to overdensities of $\delta \simeq 140$--50). 
\item Combining the 7 SLLSs from KODIAQ-Z and 73 SLLSs from HD-LLS ($19 \le \mlnhi \le 20$) and using the \citetalias{rafelski12} sample of 101 DLAs ($\mlnhi \ge 20.3$), the overall metallicity trend observed with \nhi\ continues in these regimes: the median/mean metallicity increases and IQR decreases with increasing \nhi. For the SLLSs, the median and IQR metallicities $-1.9$ and $1.1$ dex, while these are $-1.4$ and $0.65$ dex for the DLAs. DLAs have therefore rarely gas with $\xh <-2.4$, while at $\mlnhi \la 20$, VMP absorbers are not rare at $2.2 \le z \le 3.6$. 
\item The fractions of extremely metal-poor systems with $\xh<-3.5$ for the pLLSs, pLLSs, LLSs are about the same, around 3--10\% (90\% confidence interval). For the SLLSs, it is most likely somewhat smaller with a fraction in the range 0.3--5.4\%.  Yet, the most metal poor absorber in our sample is a SLLS with $\mlnhi = 19.25 \pm 0.25$ and $\xh<-4.4$ with a cosmic overdensity of several thousands. 
\item We find that (pristine) gas clouds with no metals down to limit $\xh <-3.8$ constitutes about  1\%--10\% of the $16< \mlnhi <20$ absorbers at $2.2 \le z \le 3.6$, and increases to 10\%--20\% for $14.6\le \mlnhi \le 16$ absorbers, the latter being similar to the amount of pristine gas found in the diffuse IGM. On the other hand, supersolar absorbers at any \nhi\ at $2.2 \la z \la 3.6$ are very uncommon ($<1\%$). 
\item Using paired absorbers with velocity separations of  $\Delta v \la 500$ \km\ along the same QSO sightlines, we find there is a large scatter in the metallicity from about 0.2 dex to $>2$ dex.  For about half of the paired absorbers, there is evidence for  metallicity variations over $\Delta v \la 500$ \km\ of a factor 2--3, while for the other half of a factor $>140$. It is plausible that the smaller variations correspond to the change within a single galaxy halo, while the large variations to differences between galaxy halos. Both the large metallicity range and metallicity variation between paired absorbers imply that the transport of metals from their formation sites (galaxies) into the CGM and the IGM is very inhomogeneous. 
\item The photoionized gas associated with pLLSs, LLSs, and SLLSs contribute to the cosmic baryon budget to $\Omega_{\rm g}/\Omega_{\rm b} \simeq 7\%$ and the SLFSs contribute to another 5.5\%. From the first KODIAQ survey \citep{lehner14}, the \ovi-bearing hot collisionally ionized gas associated with LLSs and SLLSs is likely to contribute to about the same level. These are the second largest contributors  to the cosmic baryon budget at high redshift behind the LYAF.  
\item We show that about 2\%, 5\%, 7\% of the metals ever produced at $2.2 \la z \la 3.6$ are in photoionized gas associated with SLFSs, pLLSs, and LLSs, respectively. The SLLSs account for another 18\%, and therefore about 30\% of the metals  at $2.2 \la z \la 3.6$ in photoionized absorbers with $14.6 \le \mlnhi < 20.3$. Combining the SLFSs, pLLSs, LLSs, and SLLSs, their comoving metal mass density of the photoionized gas probed by these absorbers is  $\rho_{\rm m} = 3.5\times 10^5$ M$_\odot$\,cMpc$^{-3}$. Another possible 5\%–20\% of the cosmic metal budget may also be in form of highly ionized metals in gas with $\mlnhi \ga 17.8$ \citep{lehner14}. 
\end{enumerate}

To study their cosmic evolution, we combine our results with the CCC survey that explores the properties of similar absorbers at $z<1$ \citepalias{lehner18,wotta19,lehner19}. The cosmic evolution of absorbers from $2.2 \la z \la 3.6$ to $z\la 1$ with $\mlnhi \ga 15$ is summarized as follows.
\begin{enumerate}[wide, labelwidth=!, labelindent=0pt]\addtocounter{enumi}{13}
\item We find that from  $2.2\le z\le 3.6 $ to $z<1$, there is an overall increase of the metallicity of the gas probed by SLFSs, pLLSs, LLSs, SLLSs, and DLAs by a factor $\sim$8. 
\item While the metallicity threshold for the VMP absorbers increases by about 1 dex from $2.2\le z\le 3.6 $ to $z<1$, a similar fraction of about 50\% of absorbers with $\mlnhi \la 18$ are VMP  at low and high $z$. 
\item Although there is plenty of {\em primitive} gas around $z\la 1$ galaxies that has largely not been polluted, the fraction of pristine SLFSs, pLLSs, LLSs with $\xh <-3$ at  $z\la 1$ is $<1$\% at the 90\% confidence level over the range $15<\mlnhi <19$. In contrast 10\%--25\% of similar absorbers at $2.2 \la z \la 3.6$ have $\xh <-3$, implying that although the transport of metals outside galaxies is still very inhomogeneous at $z<1$, all regions with $\mlnhi \ga 15$ have been polluted to some level by $z<1$. 
\item  The hydrogen column density (\nh) is a factor 10--15 smaller at $z<1$ than at $2.2 \la z \la 3.6$, and therefore the contribution to baryonic budget from the SLFSs, pLLSs, and LLSs  are about 10 times smaller at $z<1$ than at $2.2 \la z \la 3.6$.
\item The photoionized metals associated with the SLFSs, pLLSs, and LLSs take about 6\% of the cosmic metal budget at low redshift. This is about a factor 2 decrease compared the contribution of the same absorbers at $2.2 \la z \la 3.6$, i.e., at low redshift most of the metals are found in or near galaxies and in the intra-cluster medium (see \citealt{peroux20}). This is a major change in the distribution of metals in the universe from $z\sim 2.8$ to $z\sim 0.5$.    
\end{enumerate}

We set our empirical results in the context of the cosmological zoom simulations using simulations from the FOGGIE project, and the main conclusions from that study are as follows.
\begin{enumerate}[wide, labelwidth=!, labelindent=0pt]\addtocounter{enumi}{18}
\item Contrary to cruder resolution simulations (especially in the CGM), a striking feature of the FOGGIE cosmological zoom simulations is that the behavior of the metallicity as a function of \nhi\ is broadly similar to the observed empirical relationship: as \nhi\ increases the overall metallicity increases and the dispersion of the metallicity decreases. However, as other cosmological simulations, FOGGIE appears to have too many metals at any \nhi\ (e.g., supersolar metallicity gas is not uncommon in these simulations, but observationally it is), implying that metals are more homogeneously distributed than observed.
\item In the FOGGIE simulations, outflowing absorbers with the highest metallicities tend to be at or within the virial radius of these central halos and probe active or recent galaxy outflows, while outflowing absorbers with the lowest metallicity are found well outside the virial radius of the central galaxies and therefore the remnants of ejected gas. On the other hand, inflowing absorbers can be found at all radii, with their metallicity generally increasing the closer they are to their central galaxy. Very metal-poor absorbers with $\xh<-2.4$ are excellent probe inflowing gas in these simulations. 
\end{enumerate}

\section*{Acknowledgements}
We are grateful to Anna Wright for providing the satellite catalogs used in Fig.~\ref{f-met_vs_nh1_sim_dist} based on the methods of \citep{pontzen18}. The main support for this research was made by NASA through the Astrophysics Data Analysis Program (ADAP) grant NNX16AF52G. Additional support was provided by NSF grant award number 1516777. Support for the development of the Cloudy ionization models was provided by NASA through grants HST-AR-12854 and HST-AR-15634 from the Space Telescope Science Institute, which is operated by the Association of Universities for Research in Astronomy, Incorporated, under NASA contract NAS5-26555. CC and BWO acknowledge support by NSF grants no. AST-1517908 and AST-1908109 and NASA ATP grants NNX15AP39G and 80NSSC18K1105. This project has received funding from the European Research Council (ERC) under the European Union's Horizon 2020 research and innovation programme (grant agreement No 757535). This work has been supported by Fondazione Cariplo, grant No 2018-2329. All the data presented in this work were obtained from KODIAQ DR1 and DR2, which was funded through NASA ADAP grants NNX10AE84G and NNX16AF52G along with NSF award number 1516777. This research has made use of the Keck Observatory Archive (KOA), which is operated by the W. M. Keck Observatory and the NASA Exoplanet Science Institute (NExScI), under contract with the National Aeronautics and Space Administration. The authors wish to recognize and acknowledge the very significant cultural role and reverence that the summit of Maunakea has always had within the indigenous Hawaiian community. The photoionization modeling was  supported by the Notre Dame Center for Research Computing through the Grid Engine software and, together with the Notre Dame Cooperative Computing Lab, through the HTCondor software.Analysis of the FOGGIE simulations used the resources of the Michigan State University High Performance Computing Center, operated by the Institute for Cyber-Enabled Research.  The FOGGIE calculations were performed using the publicly-available Enzo code \citep{Bryan2014,Enzo_2019} and analyzed using yt \citep{turk11}, both of which are the products of the collaborative effort of many independent  scientists  from  numerous  institutions  around  the world.

\software{Astropy \citep{price-whelan18}, emcee \citep{foreman-mackey13}, Matplotlib \citep{hunter07}, PyIGM \citep{prochaska17a}}, scikit-learn \citep{pedregosa11,buitinck13}, Enzo \citep{Bryan2014,Enzo_2019},
yt \citep{turk11}, Trident \citep{hummels17}, SALSA \citep{boyd20}

\facilities{Keck(HIRES)}

\makeatletter
\renewcommand{\thefigure}{A\@arabic\c@figure}
\setcounter{figure}{0}

\appendix
\label{a-data}

[Note: owing to their sizes, supplemental files are available on request. They will be easily accessible when the paper is published.]

In this Appendix, we provide information regarding the supplemental files. For each KODIAQ-Z absorber, we produced a figure as shown in Fig.~\ref{f-spectra-ex} where we plot the normalized profiles of metals and some \hi\ transitions for which we estimated the column densities (supplemental figure fA3.pdf). The red portion in each profile shows the velocity range of the absorption over which the velocity profile was integrated to derive the column densities and average velocity. The vertical dashed lines mark the zero velocity. We also provide for each KODIAQ-Z absorber (and sometimes paired or multiple closely redshift spaced absorbers) a set of figures as Fig.~\ref{f-fit-ex} where the Voigt profile fits (in red the composite profile fit, and in blue each individual component fit) to the individual transitions of \hi\ (black spectra) are shown (supplemental figure fA4.pdf). 

\begin{figure}[tbp]
\epsscale{0.8}
\plotone{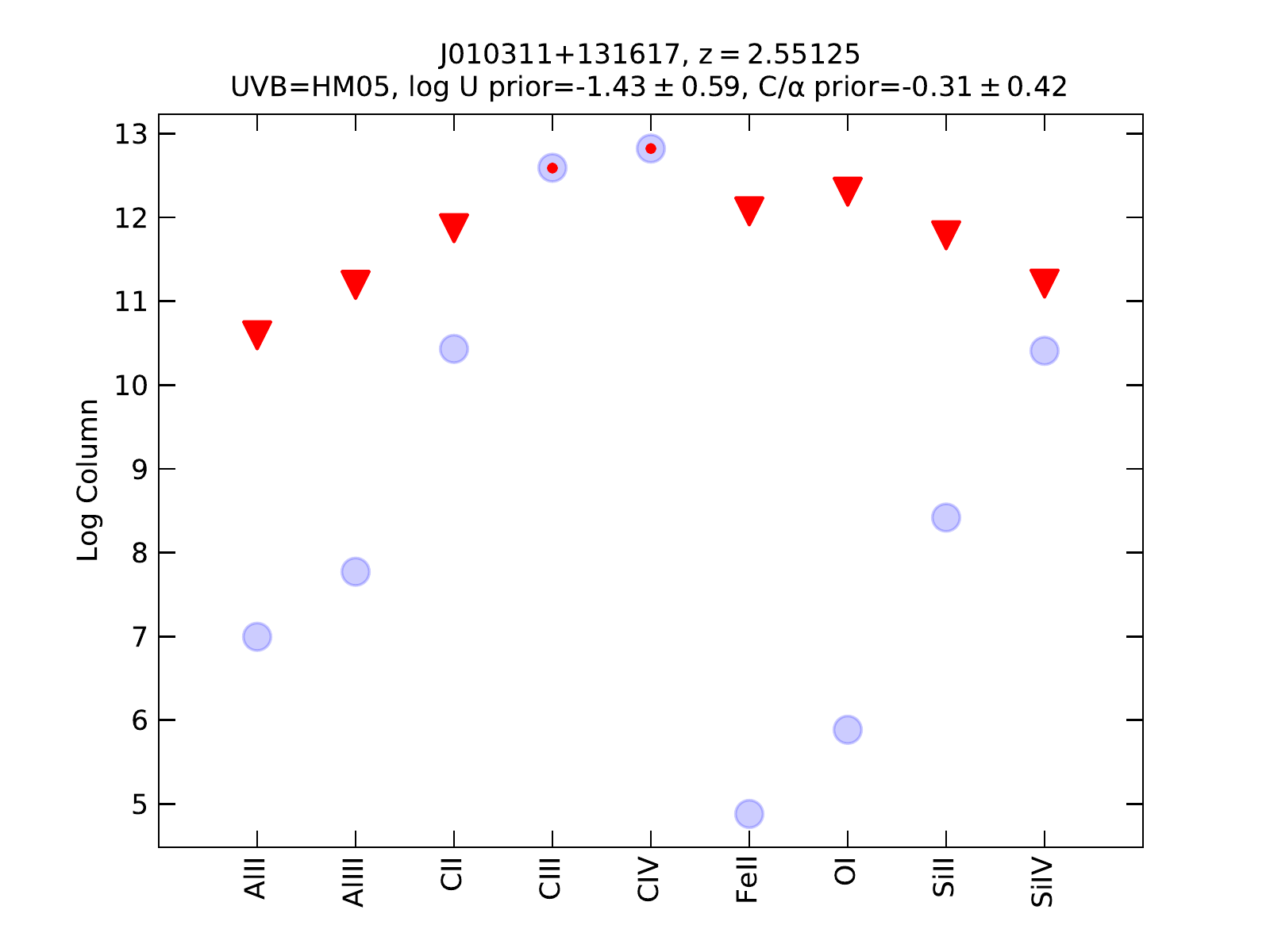}
\caption{Example of a MCMC comparison plot for the absorber at $z=2.55125$ toward J010311+131617. It shows the measured column densities for each ion (red) and the predicted column densities from the median MCMC model (blue). Downward triangles show upper limits. 
\label{f-MCMC_output-residual}}
\end{figure}

\begin{figure*}[tbp]
\epsscale{0.8}
\plotone{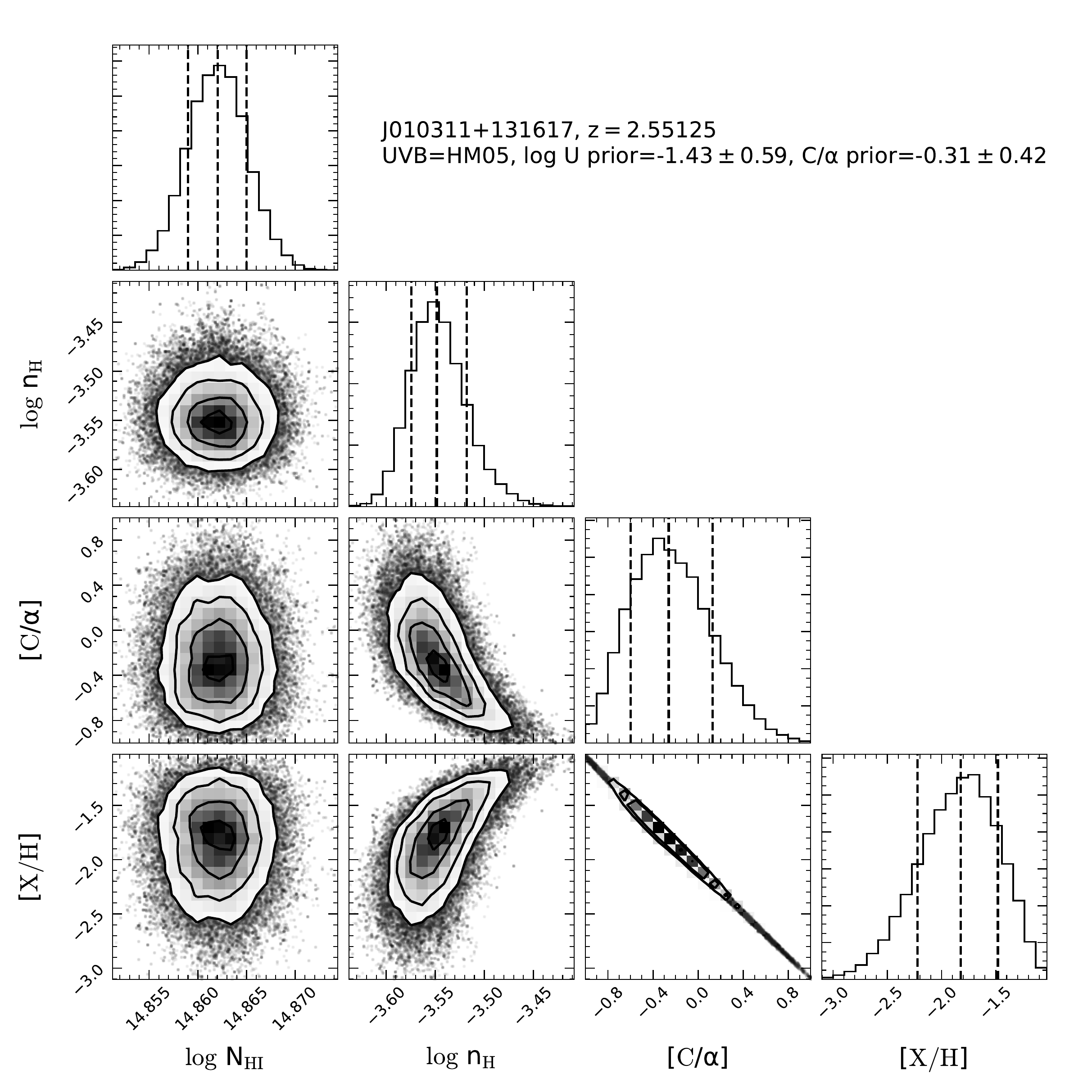}
\caption{Example of a MCMC corner plot for the absorber at $z=2.55125$ toward J010311+131617. The histograms along the diagonal show the PDFs for \nhi, hydrogen number density ($n_{\rm H}$), \ca, and metallicity, respectively. The contour plots below the diagonal show the joint posterior PDFs of the given row and column. For this absorber, there is not enough constraint on the metal column densities to derive the metallicity and \ca\ without using priors (see Fig.~\ref{f-MCMC_output-residual}).
\label{f-MCMC_output-corner}}
\end{figure*}

Next, we provide the visualization results from the MCMC Bayesian photoionization modeling. For each KODIAQ-Z absorber, we provide the comparison plots between the observations and models and corner plots as shown in Figs.~\ref{f-MCMC_output-residual} and \ref{f-MCMC_output-corner}, respectively (supplemental figures fA5.pdf and fA6.pdf). The example shown in these figures is one of the absorbers shown in Fig.~\ref{f-spectra-ex}.  In Fig.~\ref{f-MCMC_output-residual}, we compare the measured column densities for each ion (red) and the predicted column densities from the median MCMC model (blue). Triangles (when present) show lower limits (i.e., saturated transitions), while downward triangles show upper limits. Red data points with error bars (sometimes smaller than the circles) denote well-constrained column densities.  From the corner or comparison plots, one can determine readily which modeling was used: (1) if there is no entry for $\log U$ ($\log U\; {\rm prior} = {\rm False}$), then a flat prior on the ionization parameter was used; (2) if a value to $\log U$ is given, then a Gaussian prior on $\log U$ was used with the listed mean and dispersion values; (3) if $\ca$ is present, the absence of value indicates that a flat prior was used, otherwise a Gaussian prior was used on that ratio with the listed mean and dispersion values. The EUVB used in the modeling is also provided (in this case, HM05). Finally, the comparison plots show which ions were used in the ionization modeling.

\clearpage 
\startlongtable
\begin{deluxetable*}{lcccccccc}
\tabletypesize{\scriptsize}
\tablecaption{Sample Summary\label{t-sum}}
\tablehead{\colhead{Target} & \colhead{\zem} & \colhead{$\lambda_{\rm min}$} & \colhead{$\lambda_{\rm max}$} & \colhead{${\rm SNR}_{\rm 935}$} & \colhead{${\rm SNR}_{\rm 1170}$} & \colhead{${\rm SNR}_{\rm 1450}$} & \colhead{Decker}& \colhead{Obs.}\\ 
\colhead{} & \colhead{ } & \colhead{(\AA)} & \colhead{(\AA)} & \colhead{ } & \colhead{ } & \colhead{ } & \colhead{ } & \colhead{Flag}}
\startdata
J002127$-$020333 & $2.596$ & $3365$ & $6198$ & 10 & 12 & 10 & C1 & 0 \\
J003501$-$091817 & $2.418$ & $3123$ & $5979$ & 7 & 18 & 19 & C5 & 0 \\
J004530$-$261709 & $3.440$ & $3972$ & $8542$ & 10 & 17 & 15 & C1 & 0 \\
J005700+143737 & $2.643$ & $3366$ & $6198$ & 5 & 9 & 8 & C1 & 0 \\
J010311+131617 & $2.721$ & $3122$ & $8566$ & 54 & 76 & 84 & C5 & 1 \\
J010806+163550 & $2.652$ & $3047$ & $8158$ & 34 & 101 & 29 & C5 & 1 \\
J010925$-$210257 & $3.226$ & $3304$ & $6179$ & 19 & 19 & 14 & C5 & 0 \\
J011150+140141 & $2.470$ & $3132$ & $6030$ & 4 & 7 & 6 & C5 & 0 \\
\enddata
\tablecomments{
The median SNR is calculated by ${\rm SNR} = 1.0/\sigma$ per pixel within $\pm 5 \AA$ (where $\sigma$ is the error on the normalized flux) of three rest wavelengths. A $-1$ in one of the SNR column indicates there is no wavelength coverage where the SNR is estimated. The pixels are a constant 2.1 \km\ for pre-2004 observations or 2.6 \km\ for post-2004 observations and for coaddditions of observations mixing pre- and post-2004 data (see \citealt{omeara17}). The observation flag ``1" in the last column indicates if there was more than one of run or a single run (``0") (see Table~2 in  \citealt{omeara17}). The decker setup corresponds to the following resolution: C1: $R= 45,000$; C5:$R=34,000$. 
}
\end{deluxetable*}

\startlongtable
\begin{deluxetable*}{ccccccccccccc}
\tabletypesize{\scriptsize}
\tablecaption{Summary of the Column Densities of the KODIAQ-Z Absorbers \label{t-results}}
\tablehead{\colhead{Target} & \colhead{$z_{\rm abs}$} & \colhead{Ion} & \colhead{$v_1$} & \colhead{$v_2$} & \colhead{$v$} & \colhead{$\sigma_v$} & \colhead{$\log N$} & \colhead{$\sigma^1_{\log N}$} & \colhead{$\sigma^2_{\log N}$} & \colhead{Detection} & \colhead{Reliability} & \colhead{Unique}\\ 
\colhead{} & \colhead{} & \colhead{} & \colhead{} & \colhead{} & \colhead{} & \colhead{} & \colhead{} & \colhead{} & \colhead{} & \colhead{Flag} & \colhead{Flag} & \colhead{ID}\\
\colhead{} & \colhead{} & \colhead{} & \colhead{(\km)} & \colhead{(\km)} & \colhead{(\km)} & \colhead{ } & \colhead{[cm$^{-2}$]} & \colhead{} & \colhead{} & \colhead{} & \colhead{} & \colhead{}}
\startdata
J002127$-$020333 & $2.54918$ & \ion{Al}{2} & $-50.0$ & $50.0$ & $999.0$ & 999.0 & 11.625 & 0.176 & 0.301 & $-1$ & $ 3$ & $  1$ \\
J002127$-$020333 & $2.54918$ & \ion{C}{2} & $-50.0$ & $50.0$ & $999.0$ & 999.0 & 13.379 & 0.176 & 0.301 & $-1$ & $ 3$ & $  1$ \\
J002127$-$020333 & $2.54918$ & \ion{C}{4} & $-50.0$ & $50.0$ & $ 0.4$ & 3.9 & 13.175 & 0.058 & 0.066 & $ 0$ & $ 2$ & $  1$ \\
J002127$-$020333 & $2.54918$ & \ion{Fe}{2} & $-50.0$ & $50.0$ & $999.0$ & 999.0 & 13.228 & 0.176 & 0.301 & $-1$ & $ 3$ & $  1$ \\
J002127$-$020333 & $2.54918$ & \ion{H}{1} & $-50.0$ & $50.0$ & $-2.4$ & 2.7 & 15.680 & 0.019 & 0.022 & $ 0$ & $ 2$ & $  1$ \\
J002127$-$020333 & $2.54918$ & \ion{N}{1} & $-50.0$ & $50.0$ & $999.0$ & 999.0 & 13.329 & 0.176 & 0.301 & $-1$ & $ 3$ & $  1$ \\
J002127$-$020333 & $2.54918$ & \ion{N}{5} & $-50.0$ & $50.0$ & $999.0$ & 999.0 & 12.658 & 0.176 & 0.301 & $-1$ & $ 3$ & $  1$ \\
J002127$-$020333 & $2.54918$ & \ion{O}{1} & $-50.0$ & $50.0$ & $999.0$ & 999.0 & 13.860 & 0.176 & 0.301 & $-1$ & $ 3$ & $  1$ \\
\enddata
\tablecomments{A 999.0 value corresponds to a non-detection and the upper limit on the column density is quoted at the $2\sigma$ level. The detection flag has the following definition: $0=$ detection; $-1 =$ upper limit (non-detection); $-2 =$ detection but a lower limit owing to saturation of the line. Depending if a single or several transitions were detected, a reliability flag is assigned as follows: $1 =$ most reliable: results based on several transitions or a non-detection (a non-detection always reliable since it is estimated in an uncontaminated region of the spectrum);  $2 =$ reliable: results based on at least two transitions, but where only one transition is detected at the 2$\sigma$ level and the upper limits agree with that detection; $3 =$ less reliable: results based only on a single transition or several saturated transitions. }
\end{deluxetable*}
\startlongtable
\begin{deluxetable*}{lcccccccc}
\tabcolsep=3pt
\tablecolumns{9}
\tablewidth{0pc}
\tabletypesize{\scriptsize}
\tablecaption{Summary of the \hi\ Fit Results\label{t-fit}}
\tablehead{\colhead{Target} & \colhead{\zabs} & \colhead{Comp.} & \colhead{$v$} & \colhead{$\sigma_v$} & \colhead{$b$} & \colhead{$\sigma_b$} & \colhead{$\log N$} & \colhead{$\sigma_{\log N}$}\\ 
\colhead{} & \colhead{} & \colhead{} & \colhead{(\km)} & \colhead{(\km)} & \colhead{(\km)} & \colhead{(\km)} & \colhead{[cm$^{-2}$]} & \colhead{}}
\startdata
J002127$-$020333 & 2.54918 & $ 1$ & $-5.6$ & $ 0.2$ & $32.5$ & $ 0.2$ & 15.672 & 0.016 \\
J003501$-$091817 & 2.35342 & $ 1$ & $ 9.1$ & $ 0.6$ & $29.1$ & $ 0.7$ & 15.560 & 0.030 \\
J004530$-$261709 & 2.81907 & $ 1$ & $-25.8$ & $ 1.1$ & $74.0$ & $ 0.5$ & 19.664 & 0.010 \\
J004530$-$261709 & 3.24823 & $ 1$ & $-2.1$ & $ 0.2$ & $25.1$ & $ 0.2$ & 15.081 & 0.016 \\
J004530$-$261709 & 3.36517 & $ 1$ & $-18.6$ & $ 0.4$ & $53.2$ & $ 0.6$ & 15.538 & 0.016 \\
J004530$-$261709 & 3.36517 & $ 2$ & $ 5.3$ & $ 1.4$ & $16.0$ & $ 1.3$ & 15.368 & 0.055 \\
J005700$+$143737 & 2.54721 & $ 1$ & $-30.9$ & $ 1.4$ & $55.8$ & $ 1.9$ & 15.062 & 0.018 \\
J010311$+$131617 & 2.54303 & $ 1$ & $-0.5$ & $ 0.1$ & $27.9$ & $ 0.1$ & 15.629 & 0.010 \\
\enddata
\tablecomments{The redshift between parentheses in the second column indicates the actual redshift of the absorber associated with that velocity component.  
}
\end{deluxetable*}

\startlongtable
\begin{deluxetable*}{lccccc}
\tabcolsep=3pt
\tablecolumns{6}
\tablewidth{0pc}
\tablecaption{Metallicities of the Absorbers in KODIAQ-Z Sample \label{t-met-sum}}
\tabletypesize{\scriptsize}
\tablehead{\colhead{QSO Name}   & \colhead{\zabs}   & \colhead{\lnhi}   & \colhead{$\xh $}   & \colhead{$\log U$}   & \colhead{\ca}    \\
           \colhead{ }        & \colhead{ }               & \colhead{[\cmm]} & \colhead{ }               & \colhead{ }          & \colhead{ }                    }
\startdata
J010806$+$163550       &  2.31582       & $14.574 \pm 0.004$       & $<$$-$4.738, $<$$-$3.675, $<$$-$2.475 & $-$2.117, $-$1.521:, $-$0.945          & \nodata                                \\ 
J045213$-$164012       &  2.44044       & $14.611 \pm 0.004$       & $<$$-$4.739, $<$$-$3.688, $<$$-$2.568 & $-$2.141, $-$1.540:, $-$0.910          & \nodata                                \\ 
J170100$+$641209       &  2.49647       & $14.622 \pm 0.007$       & $-$2.425, $-$2.069, $-$1.776          & $-$1.743, $-$1.470, $-$1.222           & $-$0.460, $-$0.099:, +0.264            \\ 
J023359$+$004938       &  2.48836       & $14.640 \pm 0.019$       & $<$$-$4.723, $<$$-$3.594, $<$$-$2.517 & $-$2.098, $-$1.503:, $-$0.942          & \nodata                                \\ 
J073149$+$285448       &  3.42673       & $14.653 \pm 0.008$       & $<$$-$4.743, $<$$-$3.677, $<$$-$2.418 & $-$2.165, $-$1.557:, $-$0.954          & \nodata                                \\ 
J045213$-$164012       &  2.53542       & $14.674 \pm 0.003$       & $<$$-$4.790, $<$$-$3.922, $<$$-$3.079 & $-$2.109, $-$1.492:, $-$0.914          & \nodata                                \\ 
J012156$+$144823       &  2.81222       & $14.683 \pm 0.002$       & $<$$-$4.777, $<$$-$3.941, $<$$-$3.014 & $-$2.126, $-$1.567:, $-$0.976          & \nodata                                \\ 
J105756$+$455553       &  3.69194       & $14.698 \pm 0.003$       & $<$$-$4.754, $<$$-$3.776, $<$$-$2.567 & $-$2.198, $-$1.603:, $-$1.011          & \nodata                                \\
\enddata
\tablecomments{
The lower to upper bounds for each quantity represent the 68\% CI for detections and the 80\% CI for the upper or lower limits; the middle number is the median metallicity. When a colon is present after the median value of \logU, this implies that a \logU\ Gaussian prior was used to determine the properties of the absorber. 
}
\end{deluxetable*}

\begin{deluxetable}{lcccc}
\tabcolsep=3pt
\tablecolumns{5}
\tablewidth{0pc}
\tablecaption{Metallicity Statistics at $2.2\le z \le 3.6$ \label{t-stat}}
\tabletypesize{\footnotesize}
\tablehead{\colhead{Abs. Type}        & \colhead{$N_{\rm abs}$}      & \colhead{$\widetilde{\xh}$}  & \colhead{$\overline{\xh}$}  & \colhead{IQR}  }
\startdata
\multicolumn{5}{c}{KODIAQ-Z sample}\\
\hline
SLFSs    	  & 155 & $ -2.42 $ & $ -2.62 \pm 1.23$ & $[-3.61, -1.76]$ \\
pLLSs    	  &  24 & $ -2.09 $ & $ -2.27 \pm 1.04$ & $[-2.91, -1.45]$ \\
LLSs     	  &  16 & $ -2.16 $ & $ -2.17 \pm 0.90$ & $[-2.75, -1.58]$ \\
pLLSs+LLSs	  &  40 & $ -2.13 $ & $ -2.23 \pm 0.99$ & $[-2.78, -1.51]$ \\
SLFSs+pLLSs+LLSs  & 195 & $ -2.36 $ & $ -2.54 \pm 1.20$ & $[-3.37, -1.69]$ \\
SLLSs     	  &   7 & $ -2.09 $ & $ -2.11 \pm 1.29$ & $[-2.70, -0.87]$ \\
\hline
\multicolumn{5}{c}{HD-LLS sample}\\
\hline
LLSs    	  &  46 & $ -2.22 $ & $ -2.28 \pm 0.78$ & $[-2.77, -1.71]$ \\
SLLS     	  &  73 & $ -1.89 $ & $ -1.91 \pm 0.84$ & $[-2.46, -1.33]$ \\
\hline
\multicolumn{5}{c}{KODIAQ-Z+HD-LLS sample}\\
\hline
LLSs     	   &  62 & $ -2.20 $ & $ -2.25 \pm 0.81$ & $[-2.75, -1.69]$ \\
pLLSs+LLSs	   &  86 & $ -2.19 $ & $ -2.26 \pm 0.88$ & $[-2.77, -1.66]$ \\
SLFSs+pLLSs+LLSs & 241 & $ -2.34 $ & $ -2.49 \pm 1.13$ & $[-3.23, -1.70]$ \\
SLLSs     	   &  80 & $ -1.90 $ & $ -1.93 \pm 0.89$ & $[-2.47, -1.33]$ \\
\hline
\multicolumn{5}{c}{R12-DLA sample}\\
\hline
DLAs     	   & 101 & $ -1.41 $ & $ -1.41 \pm 0.50$ & $[-1.70, -1.05]$ \\
\hline
\enddata
\tablecomments{The tilde and bar above \xh\ represent the median and mean (with the standard deviation), respectively. To derive the mean, median, standard deviation, and IQR we directly use the walkers for the KODIAQ-Z and HD-LLS absorbers. Note that these are means, standard deviations, and IQRs in logarithmic space.}
\end{deluxetable}

\begin{deluxetable}{lcccc}
\tabcolsep=4pt
\tablecolumns{5}
\tablewidth{0pc}
\tablecaption{Fraction of very and extremely metal-poor absorbers at $2.2\le z \le 3.6$ \label{t-frac}}
\tabletypesize{\footnotesize}
\tablehead{\colhead{Abs.}        & \colhead{$N_{\rm abs}$}      & \colhead{Fraction with}  & \colhead{Fraction with} & \colhead{Fraction with} \\
\colhead{Type}        & \colhead{}      & \colhead{$\xh<-2.4$ }  & \colhead{$\xh<-3.0$} & \colhead{$\xh<-3.5$} 
}
\startdata
\hline
SLFSs    	     & 155 & $>$41.2\%--54.3\% 	&  $>$14.7\%--25.1\%&  $>$2.9\%--8.9\% \\
pLLSs    	     &  24 & 23.3\%--54.2\% 	&  10.5\%--37.1\%   &  5.1\%--27.5\%    \\
LLSs     	     &  62 & 26.3\%--45.9\% 	&  8.6\%--23.3\%    &  2.0\%--11.5\%     \\
SLLSs     	     &  80 & 20.1\%--36.5\% 	&  3.1\%--12.3\%    &  0.3\%--5.4\%     \\
DLAs     	     & 101 & 0.2\%--4.3\% 	&  0.0\%--2.6\%     &  0.0\%--2.6\%      \\
\hline
\enddata
\tablecomments{The SLFSs and pLLSs are from KODIAQ-Z; the LLSs and SLLSs are from KODIAQ-Z+HD-LLS; the DLAs are from R12-DLA. The percentage ranges are estimated using the Wilson score with 90\% confidence intervals. For upper limits, we use the $90^{\rm th}$ percentile to define the metallicity values. The SLFSs have an important fraction of upper limits, which is the reason that we mark them with a $>$ sign in that table. }
\end{deluxetable}

\startlongtable
\begin{deluxetable*}{lcccccc}
\tabcolsep=3pt
\tablecolumns{7}
\tablewidth{0pc}
\tablecaption{Physical Properties of the KODIAQ-Z Sample\label{t-physical_properties}}
\tabletypesize{\footnotesize}
\tablehead{\colhead{QSO Name}   & \colhead{\zabs}   & \colhead{\lnhi}   & \colhead{$\mlnh$}  & \colhead{$\log n_{\rm H}$}  & \colhead{$\log T$}   & \colhead{$l$}    \\
           \colhead{ }        & \colhead{ }               & \colhead{[\cmm]} & \colhead{[\cmm]}       & \colhead{[\cmmm]}         & \colhead{[K]}          & \colhead{(kpc)}   }
\startdata
J010806$+$163550       &  2.31582       & $14.574 \pm 0.004$       & 17.935, 18.635, 19.305       & $-$3.737, $-$3.161:, $-$2.565          &  4.387,  4.521,  4.644       &     0.1,     2.0,    35.7              \\ 
J045213$-$164012       &  2.44044       & $14.611 \pm 0.004$       & 17.944, 18.649, 19.378       & $-$3.785, $-$3.155:, $-$2.554          &  4.381,  4.516,  4.645       &     0.1,     2.1,    47.2              \\ 
J170100$+$641209       &  2.49647       & $14.622 \pm 0.007$       & 18.409, 18.725, 19.013       & $-$3.479, $-$3.231, $-$2.958           &  4.456,  4.513,  4.566       &     0.8,     2.9,    10.1              \\ 
J023359$+$004938       &  2.48836       & $14.640 \pm 0.019$       & 18.025, 18.721, 19.371       & $-$3.758, $-$3.197:, $-$2.602          &  4.391,  4.524,  4.638       &     0.1,     2.7,    43.8              \\ 
J073149$+$285448       &  3.42673       & $14.653 \pm 0.008$       & 17.955, 18.665, 19.333       & $-$3.880, $-$3.277:, $-$2.670          &  4.363,  4.502,  4.584       &     0.1,     2.8,    52.9              \\ 
J045213$-$164012       &  2.53542       & $14.674 \pm 0.003$       & 18.047, 18.769, 19.434       & $-$3.792, $-$3.214:, $-$2.597          &  4.389,  4.527,  4.640       &     0.1,     3.1,    54.5              \\ 
J012156$+$144823       &  2.81222       & $14.683 \pm 0.002$       & 18.034, 18.687, 19.368       & $-$3.765, $-$3.174:, $-$2.616          &  4.382,  4.508,  4.621       &     0.1,     2.4,    44.0              \\ 
J105756$+$455553       &  3.69194       & $14.698 \pm 0.003$       & 17.959, 18.656, 19.307       & $-$3.865, $-$3.273:, $-$2.678          &  4.352,  4.485,  4.564       &     0.1,     2.8,    48.1              \\
\enddata
\tablecomments{
The lower  and upper  bounds for each quantity represent the 68\% CI, while the middle value is the median value. When a colon is present after the median value of $\mlnnh$, this implies that a \logU\ Gaussian prior was used to determine the properties of the absorber (see \S\ref{s-photo-model}). }
\end{deluxetable*}

\begin{deluxetable*}{lcccccccccc}
\tabcolsep=3pt
\tablecolumns{11}
\tablewidth{0pc}
\tablecaption{Statistical Physical Properties of the Absorbers \label{t-phys-stat}}
\tabletypesize{\footnotesize}
\tablehead{\colhead{Absorber}   & \colhead{$\widetilde{\mlnhi}$}  & \colhead{$\overline{\mlnhi}$} & \colhead{$\widetilde{\log n_{\rm H}}$}  & \colhead{$\overline{\log n_{\rm H}}$}    & \colhead{$\widetilde{\log N_{\rm H}}$}  & \colhead{$\overline{\log N_{\rm H}}$}  & \colhead{$\widetilde{\log l}$}  & \colhead{$\overline{\log l}$}       & \colhead{$\widetilde{\log T}$}  & \colhead{$\overline{\log T}$}     \\           
           \colhead{Type}       & \colhead{[\cmm]}  & \colhead{[\cmm]}   & \colhead{[\cmmm]}  & \colhead{[\cmmm]}    & \colhead{[\cmm]}  & \colhead{[\cmm]}  & \colhead{[kpc]}  & \colhead{[kpc]}       & \colhead{[K]}  & \colhead{[K]}                     
}
\startdata
\multicolumn{11}{c}{Entire sample}\\
\hline
SLFSs                  & $15.36$ & $15.41 \pm 0.42$ &  $-3.23$ & $-3.25 \pm 0.57$  & $19.59$ & $19.47 \pm 0.68$ & $1.31$ & $1.23 \pm 1.19$ & $4.46$ & $4.46 \pm 0.14$   \\
pLLSs                  & $16.40$ & $16.52 \pm 0.27$ &  $-2.60$ & $-2.69 \pm 0.54$  & $19.97$ & $19.95 \pm 0.70$ & $1.33$ & $1.15 \pm 1.21$ & $4.36$ & $4.37 \pm 0.11$   \\
LLSs                   & $18.10$ & $17.97 \pm 0.43$ &  $-2.56$ & $-2.55 \pm 0.50$  & $20.78$ & $20.80 \pm 0.64$ & $1.86$ & $1.85 \pm 1.12$ & $4.29$ & $4.29 \pm 0.09$   \\
SLLSs                  & $19.58$ & $19.50 \pm 0.31$ &  $-2.25$ & $-2.26 \pm 0.76$  & $20.69$ & $20.80 \pm 0.68$ & $1.41$ & $1.57 \pm 1.43$ & $4.19$ & $4.19 \pm 0.15$   \\
\hline
\multicolumn{11}{c}{Restricted sample without the \logU\ constraint}\\
\hline
SLFSs                  & $15.30$ & $15.35 \pm 0.42$ &  $-3.20$ & $-3.22 \pm 0.44$  & $19.41$ & $19.38 \pm 0.58$ & $1.05$ & $1.12 \pm 0.94$ & $4.50$ & $4.48 \pm 0.11$   \\
pLLSs                  & $16.44$ & $16.55 \pm 0.28$ &  $-2.62$ & $-2.70 \pm 0.53$  & $19.99$ & $19.99 \pm 0.60$ & $1.14$ & $1.20 \pm 1.19$ & $4.38$ & $4.38 \pm 0.11$   \\
\enddata
\tablecomments{The tilde and bar above each parameter represent the median and mean (with the standard deviation), respectively, of the logarithmic values of the parameter. The entire sample consists of 321 absorbers: 155 SLFSs, 24 pLLSs, 62 LLSs, and 80 SLLSs. The restricted sample without the \logU\ constraint  contains 243 absorbers (i.e., absorbers that were modeled with a flat prior on \logU): 85 SLFSs, 21 pLLSs, 59 LLSs, and 78 SLLSs. Note that \nhi\ is estimated directly from the data, while all other quantities are estimated from our photoionization modeling. We include the results from the HD-LLS survey for the LLSs and SLLSs.}
\end{deluxetable*}
~

\begin{deluxetable}{lccc}
\tabcolsep=3pt
\tablecolumns{4}
\tablewidth{0pc}
\tablecaption{Metallicity statistics of the FOGGIE absorbers at $2\le z \le 3$ \label{t-stat-foggie}}
\tabletypesize{\footnotesize}
\tablehead{\colhead{Abs. Type}        & \colhead{$\widetilde{\xh}$}  & \colhead{$\overline{\xh}$}  & \colhead{IQR}  }
\startdata
\hline
SLFSs        & $ -2.43 $ & $ -2.87 \pm 1.68$ & $[-3.22, -1.86]$ \\
pLLSs        & $ -1.98 $ & $ -2.13 \pm 1.09$ & $[-2.49, -1.56]$ \\
LLSs         & $ -1.62 $ & $ -1.73 \pm 0.92$ & $[-2.10, -1.18]$ \\
SLLSs        & $ -1.41 $ & $ -1.45 \pm 0.76$ & $[-1.87, -0.96]$ \\
DLA         & $ -0.96 $ & $ -1.19 \pm 1.00$ & $[-1.57, -0.60]$ \\
\hline
\enddata
\tablecomments{The tilde and bar above \xh\ represent the median and mean (with the standard deviation), respectively.}
\end{deluxetable}

\makeatletter
\renewcommand{\thetable}{A\@arabic\c@table}
\setcounter{table}{0}

\phantom{This is needed for the table to fully resolve.}

\end{document}